\documentclass[times,preprint,final,nopreprintline]{elsarticle}
\pdfoutput=1

\usepackage{xifthen}
\newboolean{anonymous}
\setboolean{anonymous}{False} 

\newboolean{review}
\setboolean{review}{False} 

\newboolean{arxivpreprint}
\setboolean{arxivpreprint}{True} 

\newboolean{comments}
\setboolean{comments}{False}

\ifthenelse{\boolean{arxivpreprint}}{%
  \setboolean{review}{False} 
}{}

\ifthenelse{\boolean{review}}{%
  \usepackage{lineno}
}{}

\ifthenelse{\boolean{arxivpreprint}}{%
  \usepackage{ccicons}

  \newcommand\blfootnote[1]{%
  \begingroup
  \renewcommand\thefootnote{}\footnote{#1}%
  \addtocounter{footnote}{-1}%
  \endgroup
}
}{}

\usepackage[activate={true,nocompatibility},kerning=true,spacing=true,tracking=true,final]{microtype}
\microtypecontext{spacing=nonfrench} 

\usepackage[british]{babel}
\usepackage[babel=true]{csquotes}

\usepackage{booktabs}

\usepackage{xspace}
\DeclareRobustCommand{\hairspn}{\hspace{1pt}\nolinebreak}

\DeclareRobustCommand{\ie}{{i.\hairspn{}e.,~}}
\DeclareRobustCommand{\eg}{{e.\hairspn{}g.,~}}

\DeclareRobustCommand{\egX}{{e.\hairspn{}g.,\hspace{0pt}}}

\usepackage{siunitx}

\DeclareSIUnit{\pixel}{{\kern-1pt}pixel}
\DeclareSIUnit{\slice}{{\kern-1pt}section} 
\DeclareSIUnit{\image}{{\kern-0.5pt}image}
\DeclareSIUnit{\voxel}{{\kern-0.9pt}voxel}

\usepackage{tikz}
\usetikzlibrary{arrows}
\usetikzlibrary{calc}
\usetikzlibrary{through}
\usetikzlibrary{shapes.symbols}
\usetikzlibrary{decorations.pathmorphing}
\usetikzlibrary{fadings}
\usetikzlibrary{positioning}
\usetikzlibrary{fit}

\usepackage{natbib}

\providecommand{\tightlist}{%
}

\usepackage[]{subfig}

\usepackage[outline]{contour}
\contourlength{0.1em}

\DeclareRobustCommand{\afollsheath}{\ifthenelse{\boolean{false}}{\citep{follikel-anon,unpublished-spleen-anon}}{\citep{steiniger_capillary_2018,steiniger_locating_2018}}\xspace}
\DeclareRobustCommand{\akm}{\ifthenelse{\boolean{false}}{\citep{km16own-anon}}{\citep{km16own}}\xspace}
\DeclareRobustCommand{\afoll}{\ifthenelse{\boolean{false}}{\citep{follikel-anon}}{\citep{steiniger_capillary_2018}}\xspace}
\DeclareRobustCommand{\asheath}{\ifthenelse{\boolean{false}}{\citep{unpublished-spleen-anon}}{\citep{steiniger_locating_2018}}\xspace}

\newcommand{\hl}[1]{\textcolor{red}{#1}}
\ifthenelse{\boolean{review} \OR \boolean{arxivpreprint} \OR \NOT\boolean{comments}}{%
\newcommand{\comm}[2]{}
}{%
\newcommand{\comm}[2]{\hl{\(\spadesuit\){\bfseries #1: }}{\sffamily #2}\hl{\(\spadesuit\)}}%
}
\newcommand{\olcomment}[1]{\comm{OL}{#1}}

\usepackage[]{hyperref}

\usepackage{url}

\usepackage{floatpag}
\floatpagestyle{plain} 

\begin{document}

\begin{frontmatter}

\title{Inspection of histological 3D reconstructions in virtual~reality}

\ifthenelse{\boolean{anonymous}}{%
  \author{Anonymous}
}{%
\author[1,2,3]{Oleg Lobachev\corref{cor1}}
\ead{oleg.lobachev@leibniz-fh.de}
\ead[url]{https://orcid.org/0000-0002-7193-6258}
\author[1]{Moritz Berthold\fnref{fn1}}
\author[4]{Henriette Pfeffer}
\author[1]{Michael~Guthe}
\author[4]{Birte~S.~Steiniger}
\cortext[cor1]{Corresponding author}

\address[1]{University of Bayreuth, Visual Computing, Bayreuth, Germany}
\address[2]{Hannover Medical School, OE~4120, Carl-Neuberg-Straße~1, 30625 Hannover, Germany}
\address[3]{Leibniz-Fachhochschule School of Business, Expo Plaza~11, 30539 Hannover, Germany}
\fntext[fn1]{Present address: BCM Solutions GmbH, Rotebühlpl.~23, 70178 Stuttgart, Germany}
\address[4]{Philipps-University Marburg, Anatomy and Cell Biology, Marburg, Germany}
}

\begin{abstract}
3D reconstruction is a challenging current topic in medical research. We perform 3D reconstructions from serial sections stained by immunohistological methods. This paper presents an immersive visualisation solution to quality control (QC), inspect, and analyse such reconstructions. QC is essential to establish correct digital processing methodologies. Visual analytics, such as annotation placement, mesh painting, and classification utility, facilitates medical research insights. We propose a visualisation in virtual reality (VR) for these purposes. In this manner, we advance the microanatomical research of human bone marrow and spleen. Both 3D reconstructions and original data are available in VR. Data inspection is streamlined by subtle implementation details and general immersion in VR.
\end{abstract}  




\end{frontmatter}

  \ifthenelse{\boolean{arxivpreprint}}{%
    \blfootnote{\ccbyncndeu~
      Licensed under  Creative Commons Attribution-NonCommercial-NoDerivatives 4.0 International licence.}}{}

\ifthenelse{\boolean{arxivpreprint}}{%
}{%
  \journal{XXXXX}
}




\ifthenelse{\boolean{review}}{%
  \linenumbers
}{}

\hypertarget{introduction}{%
\section{Introduction}\label{introduction}}

\olcomment{Identify research contribution clearly!}

\begin{figure*}[tb]
\centering
\linespread{1}\selectfont
\begin{tikzpicture}[align=center,node distance=3cm and 2cm,
]

\node (rw) [draw=blue!90] {\contour[60]{white}{specimen}};
\node (dc) [draw=blue!90,fill=blue!5,right of=rw, below of=rw] {\contour[60]{white}{digital copy}}; 
\node (ar) [draw=blue!90,fill=blue!95,below of=rw,left of=dc] {\contour[60]{white}{VR}};
\node (vd) [draw=blue!90,fill=blue!45,left of=rw, below of=rw] {\contour[60]{white}{visual decisions}};

\node (vc) [below of=rw,text width=3cm] {\Large visual\\computing}; 

\draw [->] (rw.east) to [out=0,in=90] node (aq) [auto] {acquisition} (dc.north);
\draw [->] (dc.south) to [out=-90,in=0] node (si) [auto,inner color=blue!65,outer color=white,shading=radial] {~~\contour[60]{white}{simulation}~~} (ar.east);
\draw [->] (ar.west) to [out=180,in=-90] node (va) [auto,inner color=blue!95,outer color=white,shading=radial] {\contour[60]{white}{visual analytics}} (vd.south);
\draw [->] (vd.north) to [out=90,in=180] node (de) [auto,inner color=blue!15,outer color=white,shading=radial] {\contour[60]{white}{conclusions}} (rw.west);


\node (exp-s-p) [below of=si,yshift=1cm] {};
\node (exp-va-p) [below of=va,yshift=1cm] {};
\node (exp-vr-p) [below of=ar,yshift=1.5cm] {};
\node (exp-vd-p) [below of=vd,xshift=-1.5cm,yshift=2cm] {};
\node (exp-de-p) [above of=de,xshift=0cm,yshift=-1.8cm] {};
\node (exp-aq-p) [above of=aq,xshift=0cm,yshift=-1cm] {};
\node (exp-rw-p) [above of=rw,xshift=0cm,yshift=-1.7cm] {};
\node (exp-dc-p) [above of=dc,xshift=1cm,yshift=-1.7cm] {};

\draw [color=blue,->] (exp-s-p) -- (si);
\draw [color=blue,->] (exp-va-p) -- (va);
\draw [color=blue,->] (exp-vr-p) -- (ar);
\draw [color=blue,->] (exp-vd-p) -- (vd);
\draw [color=blue,->] (exp-de-p) -- (de);
\draw [color=blue,->] (exp-aq-p) -- (aq);
\draw [color=blue,->] (exp-rw-p) -- (rw);
\draw [color=blue,->] (exp-dc-p) -- (dc);

\node (exp-s) at (exp-s-p) [color=blue,text width=2.5cm,fill=white] {real-time rendering, tracking, etc.};
\node (exp-va) at (exp-va-p) [color=blue,fill=white,text width=3cm] {quality~control, annotations, mesh painting};
\node (exp-va) at (exp-vr-p) [color=blue,fill=white] {the medium};
\node (exp-vd) at (exp-vd-p) [color=blue,fill=white,text width=3cm] {coloured mesh,\\decision on quality};
\node (exp-de) at (exp-de-p) [color=blue,text width=2.5cm,fill=white] {microanatomical insights};
\node (exp-aq) at (exp-aq-p) [color=blue,text width=4cm,fill=white] {biological processing, registration, mesh construction, etc.};
\node (exp-rw) at (exp-rw-p) [color=blue,text width=1.8cm,fill=white] {sample of a lymphatic organ};
\node (exp-dc) at (exp-dc-p) [color=blue,fill=white,text width=2cm,] {reconstructed meshes};

\draw [->] (rw.east) to [out=0,in=90]  (dc.north);
\draw [->] (dc.south) to [out=-90,in=0]  (ar.east);
\draw [->] (ar.west) to [out=180,in=-90]  (vd.south);
\draw [->] (vd.north) to [out=90,in=180]   (rw.west);

\end{tikzpicture}
\caption{The general life cycle of visual computing (black) and our specific case (blue). The intensity of blue shading in the circle highlights the major focus of this paper. Blue arrows show facets of our implementation.}
\label{fig:0}
\end{figure*}
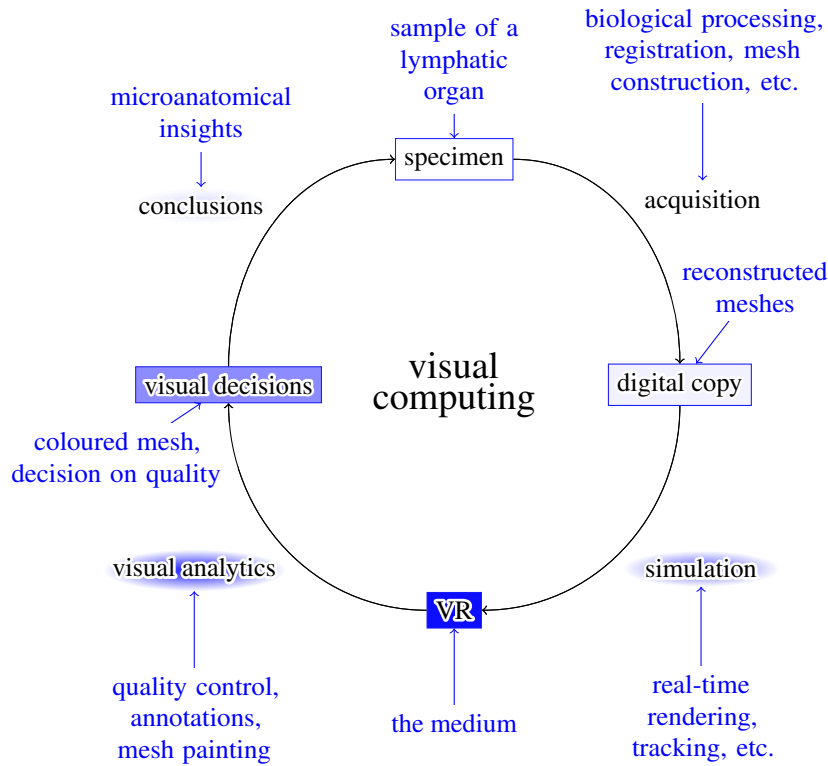

Visualisation is a significant part of modern research. It is important
not only to obtain images, but to be able to grasp and correctly
interpret data. There is always the question, whether the obtained model
is close enough to reality. Further, the question arises how to discern
important components and derive conclusions from the model. The method
we present in this paper is a typical example of this very generic
problem statement, but it is quite novel in the details. We use virtual
reality for quality control (QC) and visual analytics (VA) of our 3D
reconstructions in medical research. Figure~\ref{fig:0} positions our
tool in the visual computing methodology.

3D reconstruction from histological serial sections closes a gap in
medical research methodology. Conventional MRI, CT, and ultrasound
methods do not have the desired resolution. This is also true for
micro-CT and similar methods. Even worse, there is no way to identify
the structures of interest in human specimens under non-invasive imaging
techniques. In contrast, immunohistological staining provides a reliable
method to mark specific kinds of molecules in the cells as long as these
molecules can be adequately fixed after specimen removal. It is possible
to unequivocally identify, \eg  the cells forming the walls of small
blood vessels, the so-termed endothelial cells. Only a thin section of
the specimen---typically about \SI{7}{\micro\meter} thick---is
immunostained to improve recognition in transmitted light microscopy. If
larger structures, such as microvessel networks, are to be observed,
multiple sections in a series (\enquote{serial sections}) need to be
produced. These series are necessary, because paraffin sections cannot
be cut at more than about \SI{30}{\micro\meter} thickness due to
mechanical restrictions. The staining solution can penetrate more than
\SI{30}{\micro\meter} of tissue. It is not possible up to now to
generate focused \(z\)-stacks from immunostained thick sections in
transmitted light. Hence, the information gathered from single sections
is limited. Thus, registration is a must.

For 3D reconstruction, serial sections are digitally processed after
obtaining large images of each section with a special optical scanning
microscope. The resolution is typically in the range
0.11--0.5~\si{\micro\meter\per\pixel}, the probe may cover up
to~\SI{1}{cm^2}. With a our registration method \citep{media16} we
produce stacks of serial sections that are spatially correct. After some
post-processing \citep[\eg  morphological operations, but also an
interpolation][]{sccg17}, a surface polygon structure (a mesh) is
obtained from volume data with the marching cubes algorithm
\ifthenelse{\boolean{anonymous}}{\citep{Lorensen1987}}{\citep{Lorensen1987,mc14}}.
Both the actual mesh construction and the post-processing operations
feature some subjective decisions, most prominently, choice of an
iso-value for mesh construction. Thus, it is necessary to demonstrate
that the 3D reconstruction is correct.

\begin{figure}[t]
\centering%
\includegraphics[width=1\linewidth]{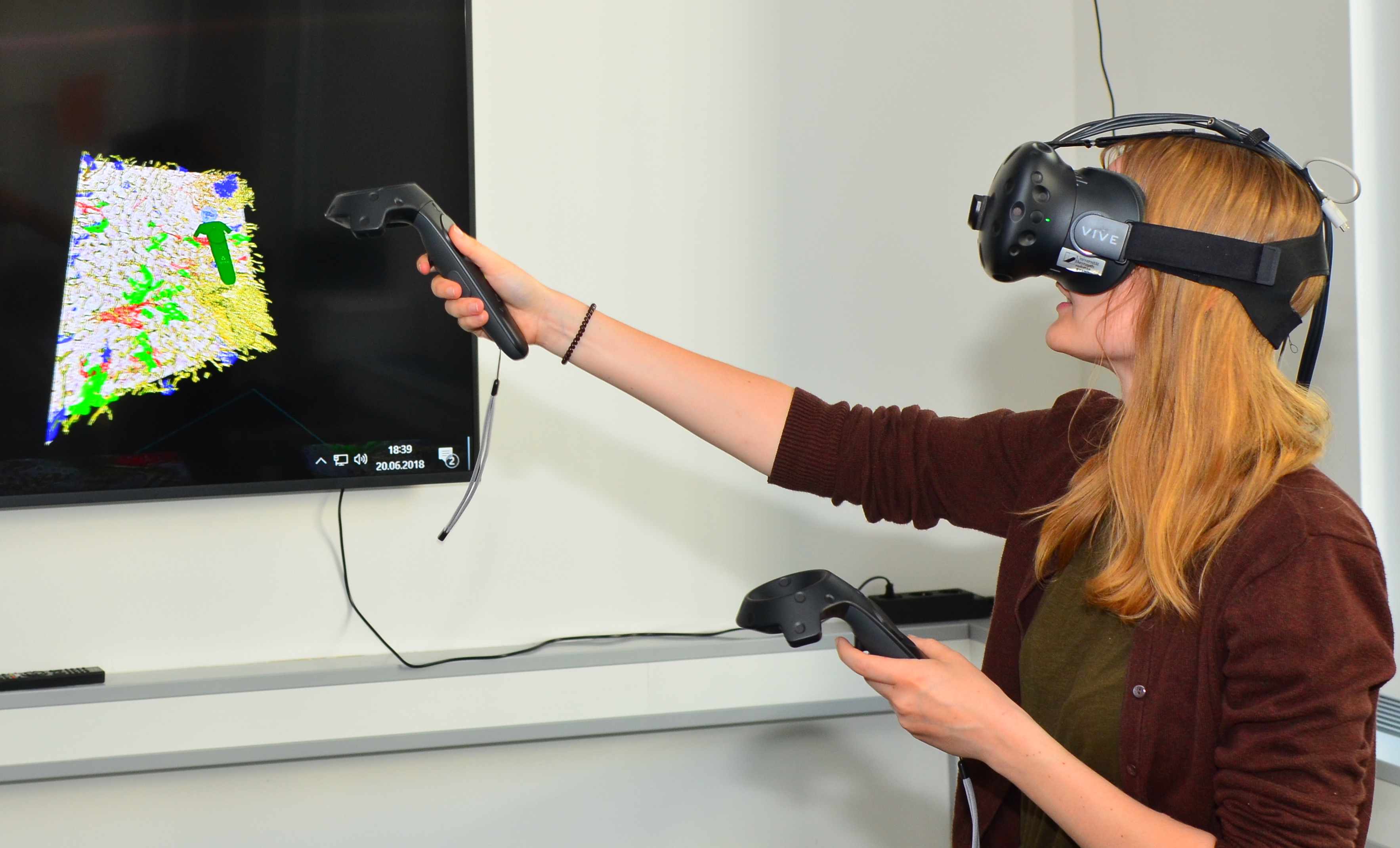}
\caption{A~user in VR. The large display mirrors the headset image.}
\label{fig:new-user}
\end{figure}

We now present a method for controlling that 3D reconstructions tightly
correspond to the original immunostained sections by directly comparing
the reconstructions to the original serial sections. This method
accelerates QC and mesh colouring. QC is facilitated by showing single
sections in the visualised mesh, without volume rendering. We inspect,
annotate, and colour 3D models (directly compared to original data, the
serial sections) in virtual reality (VR). Figure~\ref{fig:new-user}
documents a QC session from \enquote{outside}. The presented method has
been extensively used in microanatomical research
\citep{steiniger_capillary_2018, steiniger_locating_2018, habil, poster, steiniger_150}.

Our domain experts are much better trained in distinguishing and
analysing details in stained histological sections than in reconstructed
meshes. However, only 3D reconstructions provide an overview of
continuous structures spanning multiple sections, \eg  blood vessels.

\begin{figure*}[tb]
\subfloat[]{\label{fig:volren-a}\includegraphics[width=0.498\linewidth]{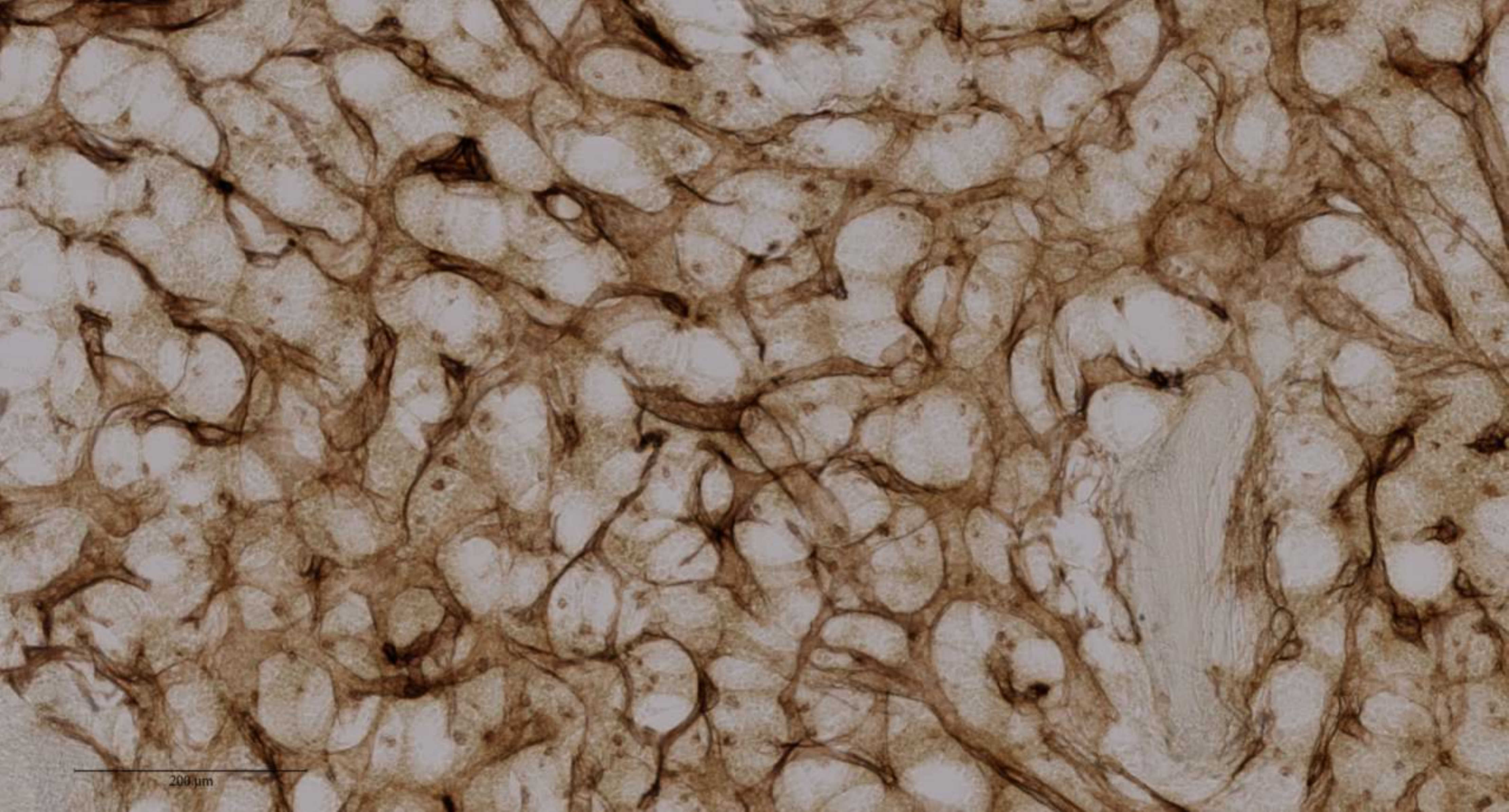}}\hfill
\subfloat[]{\label{fig:volren-b}\includegraphics[width=0.498\linewidth]{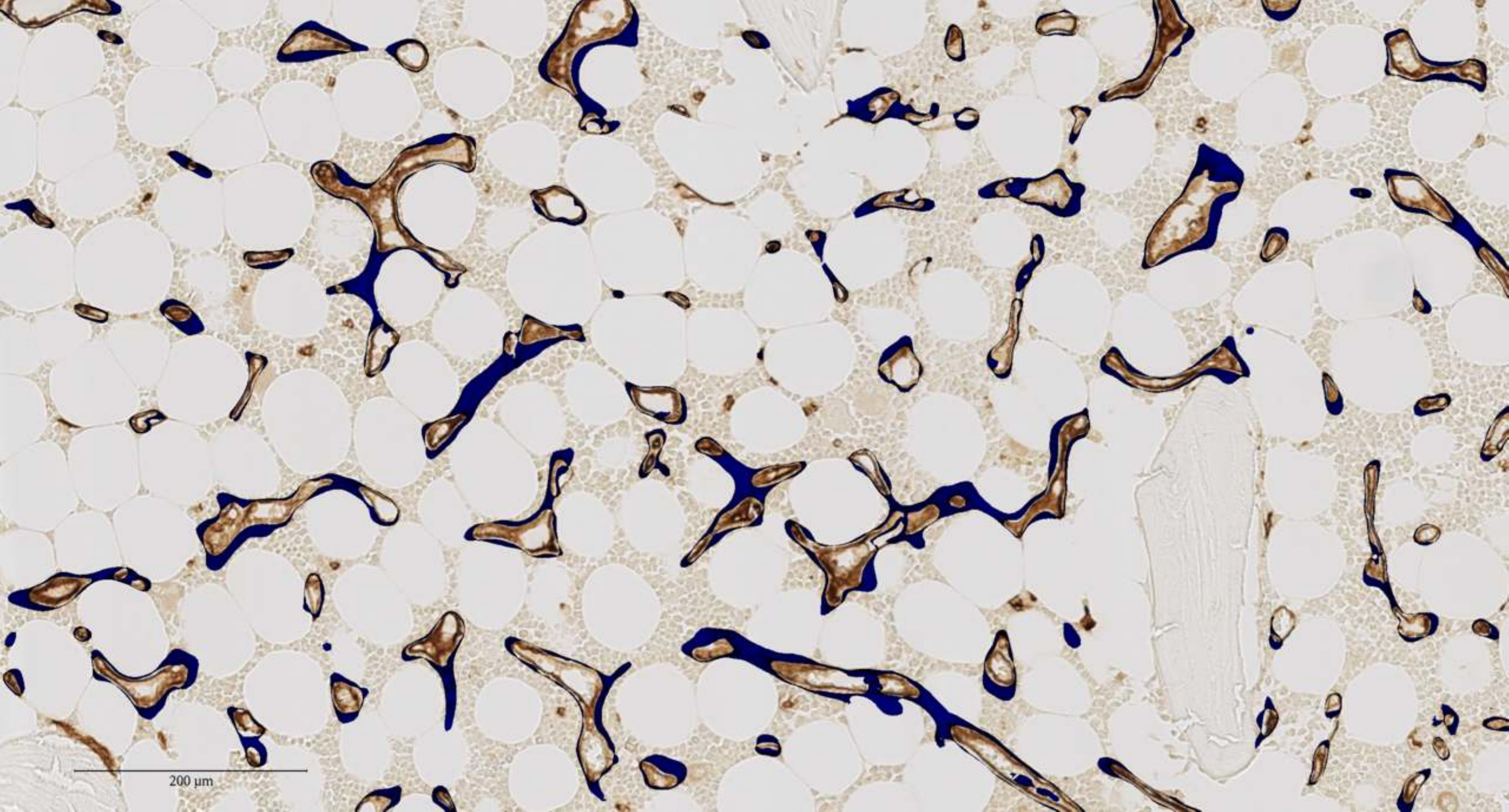}}
\caption{Showcasing our non-VR volume renderer. 
  Endothelia of blood vessels are stained
  brown in \enquote{bone marrow} data set. The blended-in mesh is blue. The volume renderer played an important
  role in data verification for our publication~\akm. \protect\subref{fig:volren-a} shows volume data representation, \protect\subref{fig:volren-b} presents the visualisation of the final, filtered mesh vs.\ corresponding single section.}
\label{fig:volren}
\end{figure*}

Further, the reconstructed mesh permits novel findings. In our prior
experience, domain experts often encounter problems when trying to
understand 3D arrangements using conventional mesh viewers. For this
reason, we previously showed pre-rendered videos to the experts to
communicate the reconstruction results and to enable detailed
inspection. Videos are, however, limited by the fixed direction of
movement and the fixed camera angle. Our experience with non-immersive
interactive tools has been negative, both with standard \citep{km39} and
with custom (Fig.~\ref{fig:volren}) software. Further, our data suffer
from a high degree of self-occlusion. In VR the user can move freely and
thus intuitively control the angle of view and the movement through the
model. In our experience, immersion allows for much easier and more
thorough inspection of visualised data. Occlusion of decisive structures
in the reconstruction does no longer pose a problem.

\hypertarget{contributions}{%
\subsection*{Contributions}\label{contributions}}
\addcontentsline{toc}{subsection}{Contributions}

We present a modern, immersive VR approach for inspection, QC, and VA of
histological 3D reconstructions. Some of the reconstructed meshes are
highly self-occluding, we are able to cope with this problem. For QC,
the original data is simultaneously displayed with the reconstruction.
User annotations and mesh painting facilitate VA. With our application,
novel research results concerning the micoanatomy of human spleens
became viable for the first
time\olcomment{verstehe ich nicht ganz: "novel ... for the first time"?};
our findings have been established and published in a series of papers
\citep{steiniger_capillary_2018, steiniger_locating_2018, steiniger_150}.

\begin{figure}[p]%
\centering%
\ifthenelse{\boolean{review} \OR \boolean{arxivpreprint}}{%
\hfill\subfloat[][]{\includegraphics[height=0.3\linewidth]{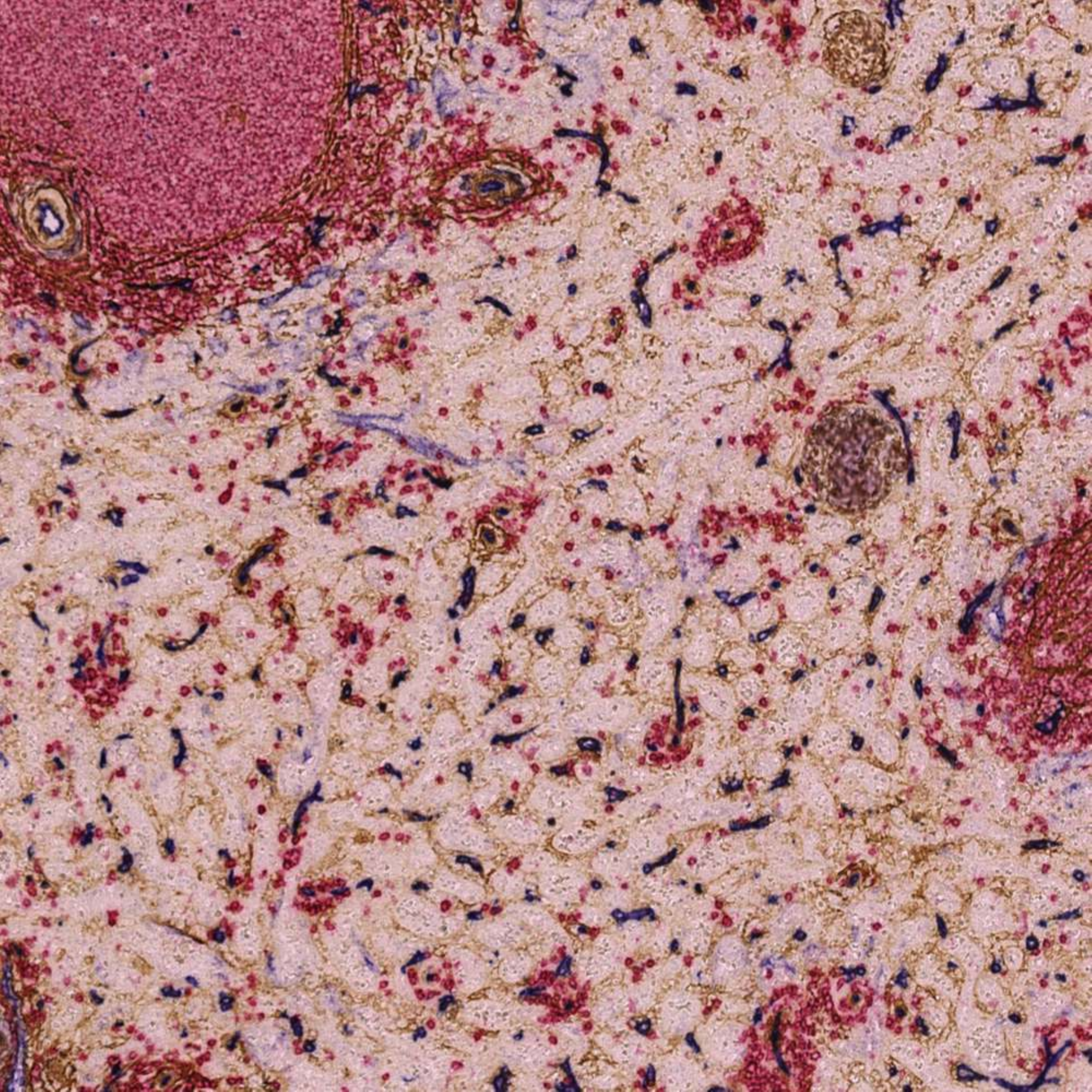}\label{fig:overv-f-a}}\hfill%
\subfloat[][]{\includegraphics[height=0.3\linewidth]{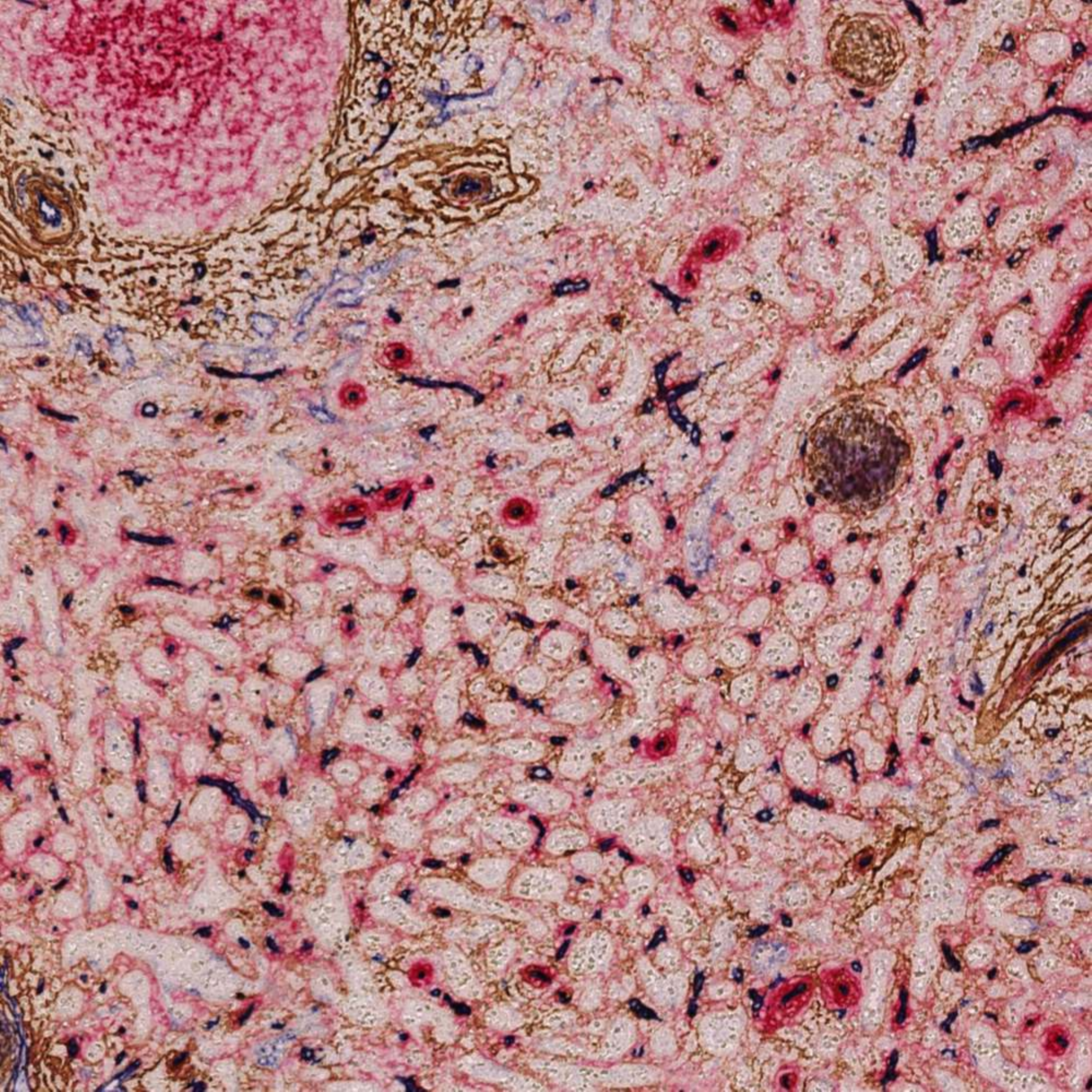}\label{fig:overv-f-b}}\hfill~

\hfill\subfloat[][]{\includegraphics[height=0.3\linewidth]{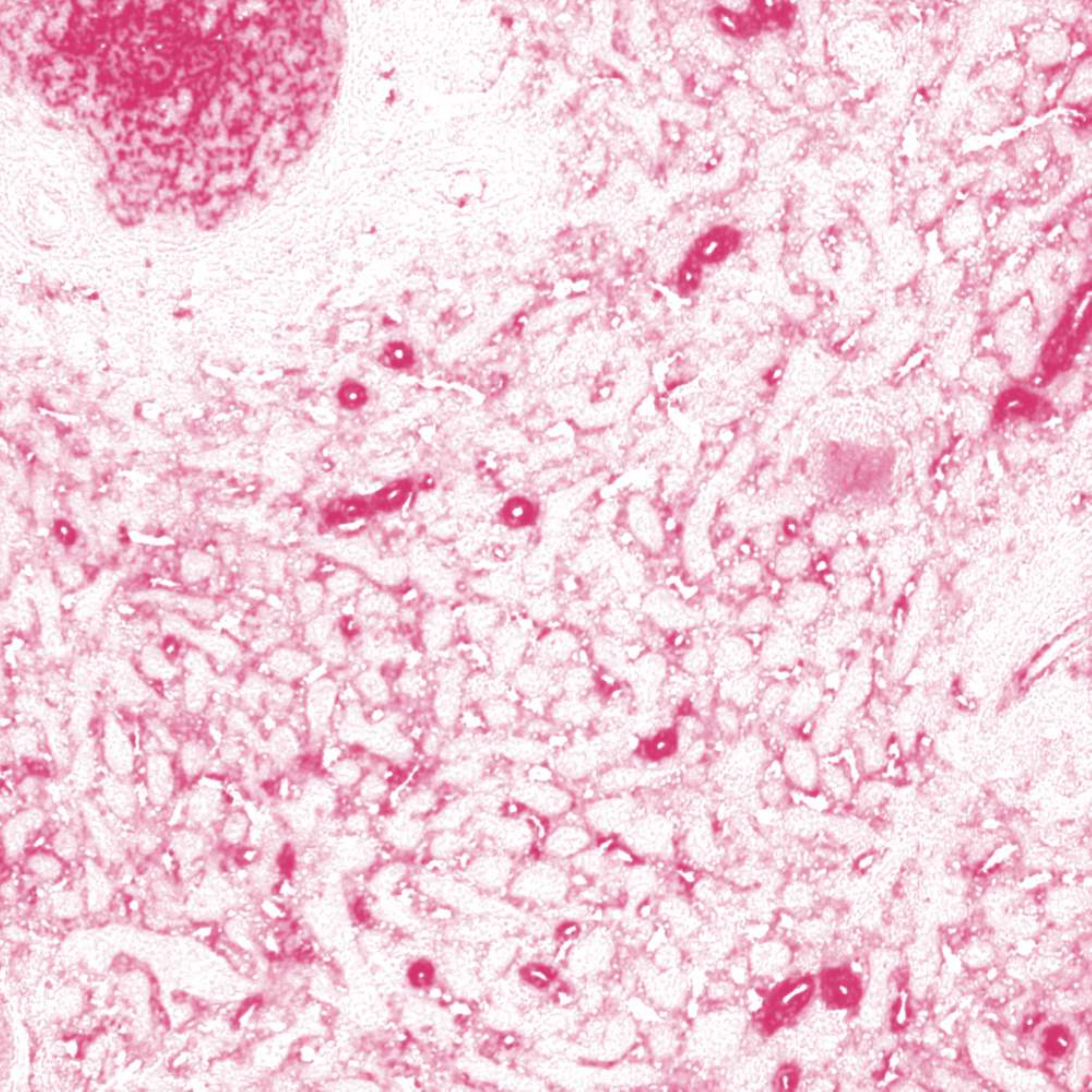}\label{fig:overv-f-deconv1}}\hfill%
\subfloat[][]{\includegraphics[height=0.3\linewidth]{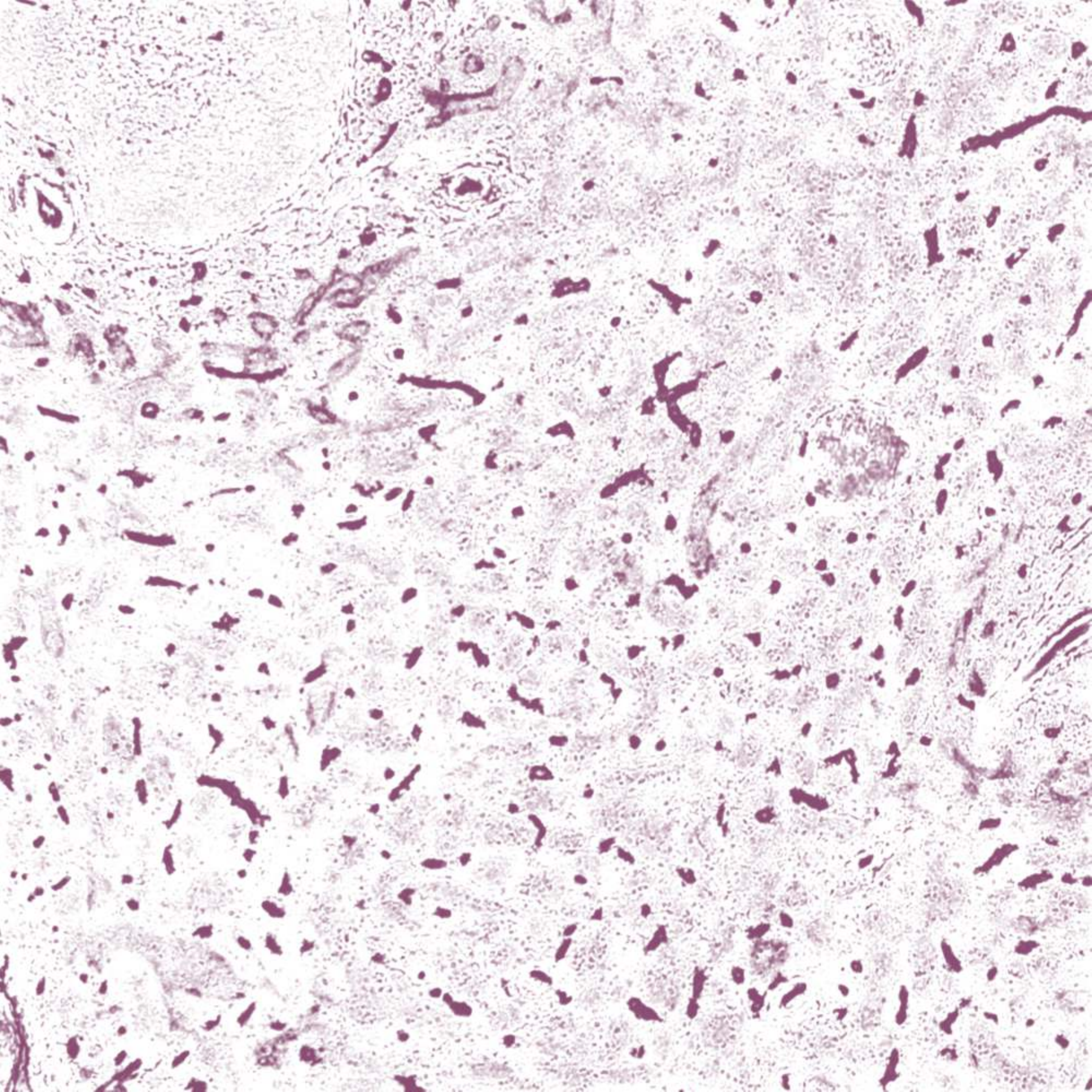}\label{fig:overv-f-deconv2}}\hfill~

\hfill\subfloat[][]{\includegraphics[height=0.3\linewidth]{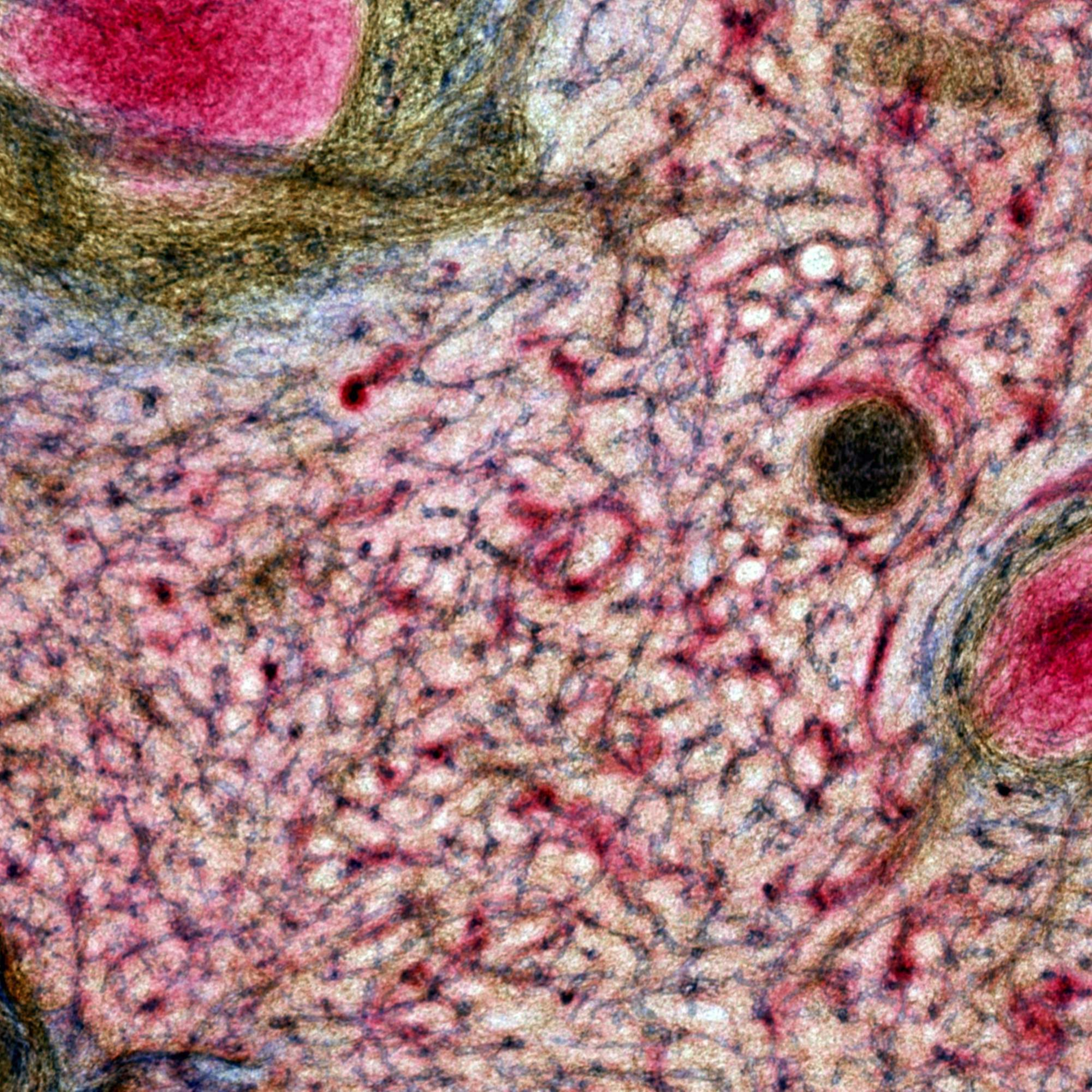}\label{fig:overv-f-c}}\hfill%
\subfloat[][]{\includegraphics[height=0.3\linewidth]{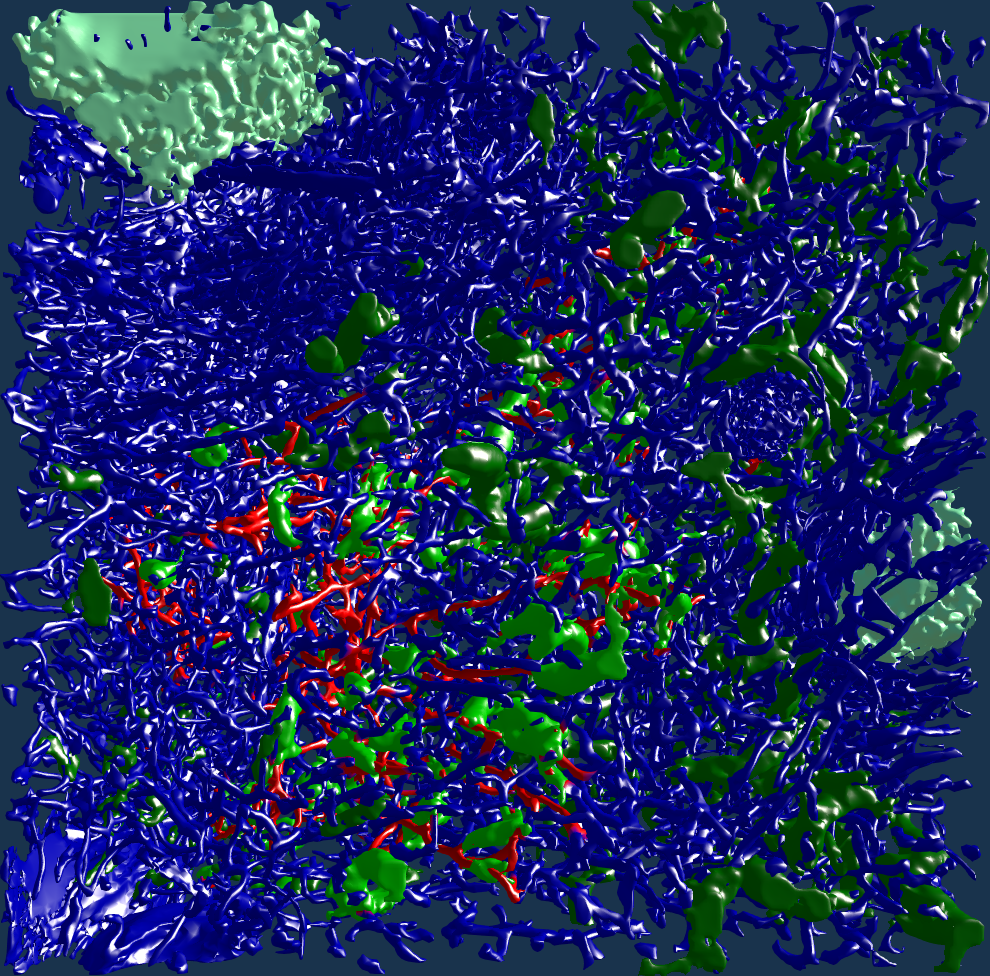}\label{fig:overv-f-d}}\hfill~%

}{%
\subfloat[][]{\includegraphics[height=0.498\linewidth]{images/new/r32_redo-2019-08-17_sorted_001.pdf}\label{fig:overv-f-a}}\hfill%
\subfloat[][]{\includegraphics[height=0.498\linewidth]{images/new/r32_redo-2019-08-17_sorted_003.pdf}\label{fig:overv-f-b}}

\subfloat[][]{\includegraphics[height=0.498\linewidth]{images/new/r32_redo-2019-08-17_sorted_003-_Colour_1_.pdf.}\label{fig:overv-f-deconv1}}\hfill%
\subfloat[][]{\includegraphics[height=0.498\linewidth]{images/new/r32_redo-2019-08-17_sorted_003-_Colour_3_.pdf.}\label{fig:overv-f-deconv2}}

\subfloat[][]{\includegraphics[height=0.498\linewidth]{images/new/mult_sh-only_01-30_8_6_levels_auto_8bit.pdf}\label{fig:overv-f-c}}\hfill%
\subfloat[][]{\includegraphics[height=0.498\linewidth]{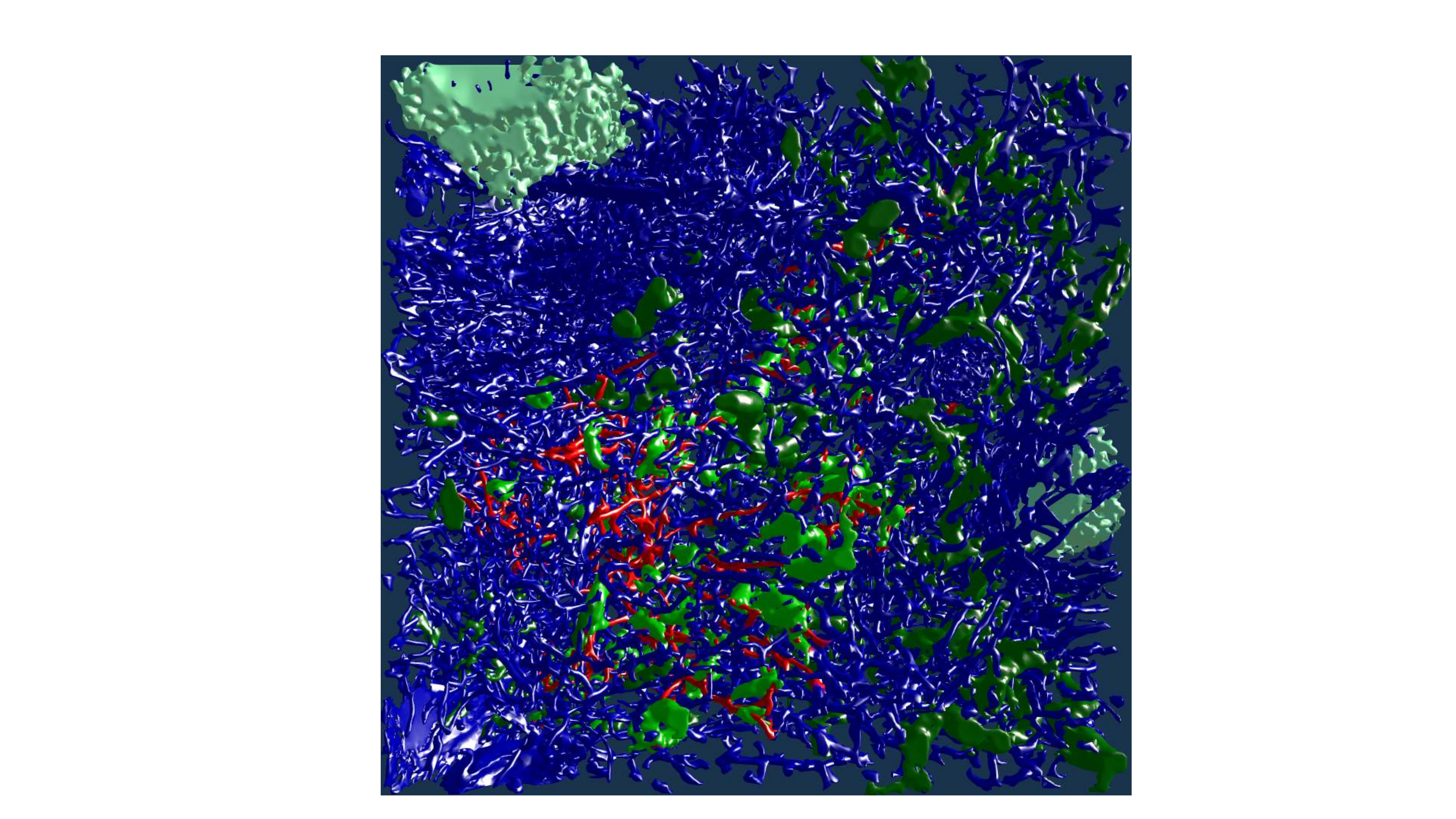}\label{fig:overv-f-d}}%
}
\caption{From section images to final results: A~human spleen section is stained for SMA (brown), CD34 (blue), and either CD271 (red) or CD20 (red), this is the \enquote{Sheaths alternating} data set. \protect\subref{fig:overv-f-a}: The region of interest (ROI), staining of B-lymphocytes (CD20) in red. \protect\subref{fig:overv-f-b}: The ROI, staining of capillary sheaths (CD271) in red. \protect\subref{fig:overv-f-deconv1}: Result of colour deconvolution for CD271 of~\protect\subref{fig:overv-f-b}, a single image. \protect\subref{fig:overv-f-deconv2}: Same, but for CD34. \protect\subref{fig:overv-f-c}: A~volume rendering of the first 30 sheath-depicting sections of the ROI. \protect\subref{fig:overv-f-d}:  Final meshes. The colours highlight the different functions. The arterial blood vessels are in blue and red. The red colour is highlighting a specific tree of blood vessels. The sheaths related to this tree are green, the unrelated sheaths are dark green. The follicular dendritic cells (that are also weakly CD271\textsuperscript{+}) are depicted in light green. The SMA mesh was used for a heuristics to find arterioles among blood vessels. SMA and B-lymphocytes are not shown in the rendering.}
\label{fig:2}
\end{figure}

\hypertarget{related-work}{%
\section{Related work}\label{related-work}}

\hypertarget{immersive-visualisation}{%
\subsection{Immersive visualisation}\label{immersive-visualisation}}

Immersive visualisation is not at all a new idea, \citet{799723} quotes
a vision of
\citet{sutherland_ultimate_1965,sutherland_head-mounted_1968,sutherland_computer_1970}.
However, an immersive scientific visualisation was quite hard to obtain
in earlier years, if a multi-million-dollar training simulation was to
be avoided \citep{888006}. The availability of inexpensive hardware such
as Oculus Rift or HTC Vive head-mounted displays (HMD) has massively
changed the game recently. This fact (and the progress in GPU
performance) allows for good VR experiences on commodity hardware.
Immersive visualisation has been previously suggested for molecular
visualisation \citep{Stone2010}, for medical volumetric data
\citep{4618615, scholl_direct_2018}, for dentistry
\citep{694195, xia_virtual_2013}, and for computational fluid dynamics
\citep{vr-fluid}. More relevant to our approach are the visualisations
of the inside of large arterial blood vessels
\citep{Forsberg:2000:IVR:375213.375297, egger_comprehensive_2020}. There
is a trend is to utilise VR in medical education and training
\citep{walsh_virtual_2012, doi:10.1227/NEU.0b013e3182750d26, 7223437, moro_effectiveness_2017, bouaoud_diva_2020, lopez_chavez_comparative_2020, pieterse_design_2020};
\citet{rea_augmented_2020} and \citet{duarte_learning_2020} provide an
overview. The availability of head-mounted displays has sparked some new
research \citep{Chen2015124, 7814821, doi:10.1121/2.0000381} in addition
to already covered fields. \citet{mann2018all} present a taxonomy of
various related approaches (virtual reality, augmented reality, etc.).
\citet{checa_review_2020} review VR applications in the area of serious
games.

A~radically different approach is to bypass mesh generation altogether
and to render the volumes directly. \citet{scholl_direct_2018} do so in
VR, although with quite small volumes (with largest side of 256 or 512
voxels). \citet{faludi_direct_2019} also use direct volume rendering and
add haptics. In this case the largest volume side was 512 voxels. The
key issue in such direct volume renderings is the VR-capable frame rate,
which is severely limited in volume rendering of larger volumes. The
volumes we used for the mesh generation had 2000 pixels or more at the
largest side.

\citet{zoller_force_2020} use volume rendering for haptic feedback in
VR. Their goal was to simulate pedicle screw tract palpation. This work
is not concerned with haptics.

\citet{wismann_accelerated_2020} use reprojection to improve binocular
ray tracing. The idea is to use the image from one eye to create the
image for the other eye faster. We do not use ray tracing in this paper,
our visualisations are conventional OpenGL-based rasterizations in VR.
However, future improvements to our work both in the visuals and in
rendering times, might base on the results by
\citeauthor{wismann_accelerated_2020}.

In the next section we discuss additional modern medical applications of
VR, however, VR visualisation is a much broader topic
\citep{berg_industry_2017, misiak_islandviz:_2018, rizvic_virtual_2019, slater_ethics_2020}.

\hypertarget{virtual-reality-in-medical-research}{%
\subsection{Virtual reality in medical
research}\label{virtual-reality-in-medical-research}}

Although there have been precursors long before the \enquote{VR boom},
\eg \citet{tomikawa_real-time_2010}, most relevant publications on the
use of VR in medical research, training, and in clinical applications
appeared after 2017. This section focuses on medical research.

\citet{stets_visualization_2017} work with a point cloud. We work with
surface meshes. \citet{esfahlani_rehabgame:_2018} reported on
non-immersive VR in rehab. We use immersive VR in medical research.
\citet{uppot_implementing_2019} describe VR and AR for radiology, we use
histological sections as our input data. \citet{knodel_virtual_2018}
discuss the possibilities of VR in medical imaging.
\citet{stefani_confocalvr_2018} show confocal microscopy images in VR.
We use images from transmitted light microscopy.
\citet{cali_method_2019} visualise glial and neuronal cells in VR. We
visualise blood vessels and accompanying cell types in lymphatic organs,
mostly in the spleen.

A~visualisation support for HTC Vive in popular medical imaging toolkits
has been presented before \citep{10.1371/journal.pone.0173972}. Unlike
our approach, this method is tied into existing visualisation libraries.
Our method is a stand-alone application, even if easily usable in our
tool pipeline. Further, visualising both reconstructed meshes \emph{and}
original input data was a must in our use case. We also implemented a
customised mesh painting module for visual analytics. Both our approach
and the works of \citeauthor{10.1371/journal.pone.0173972}
(\citeyear{10.1371/journal.pone.0173972},
\citeyear{egger_comprehensive_2020}) generate meshes prior to the
visualisation. We discuss the differences between
\citet{egger_comprehensive_2020} and our approach on
page~\pageref{page:free-mov}.

\citet{el_beheiry_virtual_2019} analyse the consequences of VR for
research. In their opinion, VR means navigation, but also allows for
better education and provides options for machine learning. They can
place annotations in their program, but focus on (immersed) measurements
between the selected points. \citeauthor{el_beheiry_virtual_2019}
perform some segmentations in VR, but primarily work with image stacks.
Our mesh painting in VR can be seen as a form of segmentation, but we
perform it on the 3D models, not on image data. Mesh painting uses
geodesic distances, as detailed in Section~\ref{sec:geo}.

\citeauthor{daly_imaging_2019}
(\citeyear{daly_future_2018,daly_confocal_2019,daly_imaging_2019}) has
similar goals to this work, however he uses a radically different tool
pipeline, relying more on off-the-shelf software---which alleviates a
larger part of software development, but is also a limitation in the
amount of features. \citeauthor{daly_imaging_2019} \citep[and also
others, \egX ][]{preim_survey_2018} also focus a lot on teaching, we use
our system at the moment mostly for research purposes.

\citet{dorweiler_zukunftsperspektiven_2019} discusses the implications
of VR, AR, and further technologies in blood vessel surgery. We are
concerned with \emph{analysis} of microscopic blood vessels in removed
probes.

The work by \citet{liimatainen_virtual_2020} allows the user to inspect
3D reconstructions from histological sections (created in a radically
different manner from how we section, they skip a lot of tissue in an
effort to cover a larger volume). The user can view the sections and
\enquote{interact with single areas of interest}. This is elaborated to
be a multi-scale selection of the details and allowing the user to zoom
in. We stay mostly at the same detail level, but allow for more in-depth
analysis. They put histological sections of tumors in their correct
location in the visualisation, which was also one of the first
requirements to our tool, as Section~\ref{sec:req} details.

We are not aware of other implementations of advanced VR- and mesh-based
interactions, such as our mesh paining that follows blood vessels
(Section~\ref{sec:geo}). To our knowledge, annotations have never before
been implemented in the manner we use: The markers are preserved after
the VR session and can be used in a mesh space for later analysis. This
paper presents both those features. In general, most VR-based
visualisations focus on presentation and exploration of the data. We do
not stop there, but also perform a lot of visual analytics.

\hypertarget{why-non-vr-tools-do-not-suffice}{%
\subsection{Why non-VR tools do not
suffice}\label{why-non-vr-tools-do-not-suffice}}

While enough non-VR tools for medical visualisation exist, such as
3D~Slicer \citep{pieper_3d_2004, kikinis_3d_2014}, ParaView
\citep{ahrens_paraview:_2005, ayachit_paraview_2015}, or MeVisLab
\citep{silva_processing_2009, ritter_medical_2011}, we are proponents of
VR-based visualisation. Rudimentary tasks in QC can be done, \eg in
3D~Slicer or using our previous work, a custom non-VR tool (detailed
below on page~\pageref{page:volren}), but in our experience, our
VR-based QC was much faster and also easier for the user.
(\citet{bouaoud_diva_2020} and \citet{lopez_chavez_comparative_2020}
report similar experiences.) The navigation and generation of insights
are a larger problem with non-VR tools. The navigation in VR is highly
intuitive. A~lot of insight can be gathered by simply looking at the
model from various views.

The relation of implementation efforts to the usability impact was
favourable for our VR tool. The complexity of software development of
large standard components also plays a role here. We base our VR
application heavily on available components, such as Steam~VR and VCG
mesh processing library, as Section~\ref{sec:comp} details. However, our
tool is not an amalgamation of multiple existing utilities (\eg using
Python or shell as a glue), but a stand-alone application, written in
C++.

Merely paging through the registered stack of serial sections does not
convey a proper 3D perception. Single entities in individual sections
(\eg capillaries) have a highly complex shape and are entangled among
similar objects. While it is possible to trace a single entity through a
series, gaining a full 3D perception is impossible without a
full-fledged 3D reconstruction. An inspection of our reconstructions in
VR
\citep{steiniger_capillary_2018, steiniger_locating_2018, steiniger_150}
was much faster than a typical inspection of 3D data without VR \akm, as
Section~\ref{sec:impact} details.

\hypertarget{background}{%
\section{Background}\label{background}}

\hypertarget{vr-visualisation-requirements}{%
\subsection{VR visualisation:
Requirements}\label{vr-visualisation-requirements}}

\label{sec:req}

Our domain experts provided feedback on the earlier versions of the
software in order to shape our application. The following features were
deemed necessary by medical researchers:

\begin{itemize}
\tightlist
\item
  Load multiple meshes corresponding to parts of the model and to switch
  between them. This allows for the analysis of multiple
  \enquote{channels} from different stainings.
\item
  Load the original data as a texture on a plane and blend it in VR at
  will at the correct position. The experts need to discriminate all
  details in the original stained sections.
\item
  Remove the reconstructed mesh to see the original section underneath.
\item
  Provide a possibility to annotate a 3D position in VR. Such
  annotations are crucial for communication and analysis.
\item
  Adjust the perceived roles of parts of the meshes by changing their
  colour. Colour changes form the foundation of visual analytics.
\item
  Cope with very complex, self-occluding reconstructions. Otherwise it
  is impossible to analyse the microvasculature in thicker
  reconstructions (from about \SI{200}{\micro\meter} in \(z\) axis
  onward).
\item
  Free user movement. This issue is essential for long VR sessions.
  Basically, no movement control (\eg flight) is imposed on the user. In
  our experience, free user movement drastically decreases the chances
  of motion sickness.
\item
  Provide a possibility for voice recording in annotations (work in
  progress).
\item
  Design a method for sharing the view point and current highlight with
  partners outside VR (trivial with Steam~VR and its display mirroring),
  and for communicating the findings from live VR sessions as non-moving
  2D images in research papers (an open question).
\end{itemize}

\hypertarget{a-3d-reconstruction-pipeline-for-serial-sections}{%
\subsection{A~3D reconstruction pipeline for serial
sections}\label{a-3d-reconstruction-pipeline-for-serial-sections}}

In a short summary our method includes:

\begin{enumerate}
\def\labelenumi{\arabic{enumi}.}
\tightlist
\item
  Biological processing: tissue acquisition, fixation, embedding,
  sectioning, staining, coverslipping.
\item
  Digital data acquisition: serial scanning in an optical scanning
  microscope.
\item
  Coarse registration: fitting the sections to each other.
\item
  Selection of regions of interest (ROIs).
\item
  Fine-grain registration~\citep{media16}. The decisive step for
  maintaining the connectivity of capillaries.
\item
  Colour processing, \eg  channel selection or colour deconvolution.
\item
  Optional: healing of damaged regions \citep{lobachev_tempest_2020}.
\item
  Interpolation to reduce
  anisotropy\ifthenelse{\boolean{anonymous}}{ [cite omitted for review]}{ \citep{sccg17}}.
\item
  Volume filtering, \eg  a closing filter and a blur.
\item
  Mesh
  construction\ifthenelse{\boolean{anonymous}}{ [cite omitted for review]}{ \citep{mc14}}.
\item
  Mesh processing, \eg  decimation or repair \citep{km37}.
\item
  Mesh colouring, \eg  colouring of selected components
  \citep{steiniger_capillary_2018, steiniger_locating_2018, steiniger_150}
  or visualisation of shape diameter function~\akm.
\item
  QC as initial visual analytics. If the reconstruction, \eg  interrupts
  microvessels or includes non-informative components, identify the
  cause and repeat from there.
\item
  Further visual analytics, \eg mesh colouring.
\item
  Final rendering.
\end{enumerate}

Figure~\ref{fig:2} showcases some important steps of this pipeline.

\hypertarget{histological-background}{%
\subsection{Histological background}\label{histological-background}}

The human spleen is a secondary lymphatic organ which serves to
immunologically monitor the blood. In order to intensify the contact
between blood-borne molecules, which provoke immune reactions, and the
specific immunocompetent lymphocytes and macrophages, the spleen
harbours a so-termed \enquote{open circulation system}. This system is
unique to the spleen. It does not comprise continuous arteries,
arterioles, capillaries, venules and veins as in other organs, but there
is an open vessel-free space between capillaries and the beginning of
the draining venous system, which is passed by all constituents of the
blood. In addition, the initial venous vessels which re-collect the
blood into the circulation system are also organ-specific and are termed
\enquote{sinuses}.

In humans, the initial parts of the splenic capillary network is covered
by peculiar multicellular arrangements termed \enquote{capillary
sheaths}. The detailed function of these sheaths is unknown, but
comparative anatomy suggests, that they collect certain foreign
molecules from the blood and guide the immigration of special
immunocompetent lymphocytes (B-lymphocytes) into the spleen.

Prior to our works
\citep{steiniger_capillary_2018, steiniger_locating_2018}, arrangement
of capillary sheaths has never been shown in three dimensions. It has
been unknown, whether all splenic capillaries are covered by sheaths,
how long the sheaths are, what shape they have and, finally, which cell
types they consist of. In addition, the location of the sheaths with
respect to the open ends of capillaries feeding the \enquote{open
circulation} has remained enigmatic.

During recent years our research
\citep{steiniger_capillary_2018, steiniger_locating_2018, steiniger_150}
has clarified many of these questions. We now report an advanced study
comprising 3D models derived from up to 150 serial paraffin sections
stained for conventional transmitted light microscopy utilising three
different chromogens (brown, blue and red) to visualise four molecules
by immunohistological methods. In detail, we demonstrate smooth muscle
alpha-actin (SMA), CD34, CD271 and CD20 (Table~\ref{tab:histo}).

\begin{table*}
\caption{Cell distribution of the molecules detected in human spleens.}
\label{tab:histo}
\centering
\linespread{1}\selectfont
\begin{tabular}{p{0.5\linewidth}ccccc}
\toprule
\multicolumn{1}{c}{Target} & \multicolumn{5}{c}{Expression} \\
\midrule
 & SMA   & CD34 & CD271 & CD20 & CD141 \\
\multicolumn{1}{l}{\ifthenelse{\boolean{review} \OR \boolean{arxivpreprint}}{\phantom{MMM}}{\phantom{fibroblastic}}%
Colour in \citep{steiniger_capillary_2018}}
  & -- & brown & blue   & --  & -- \\
\multicolumn{1}{l}{\ifthenelse{\boolean{review} \OR \boolean{arxivpreprint}}{\phantom{MMM}}{\phantom{fibroblastic}}%
Colour in \citep{steiniger_locating_2018}}
  & brown & brown & blue   & --  & -- \\
\multicolumn{1}{l}{\ifthenelse{\boolean{review} \OR \boolean{arxivpreprint}}{\phantom{MMM}}{\phantom{fibroblastic}}%
Colour in \citep{steiniger_150}}
  & brown & blue & red\textsuperscript{1}   & red\textsuperscript{1} & -- \\
\multicolumn{1}{l}{\ifthenelse{\boolean{review} \OR \boolean{arxivpreprint}}{\phantom{MMM}}{\phantom{fibroblastic}}%
Colour in \enquote{sinus} data set}
  & -- & blue & red   & -- & brown \\  
\midrule
endothelial cells (inner vessel lining) 
  & -- & +\textsuperscript{2} & -- & -- & -- \\
smooth muscle cells (wall of arteries, arterioles)
  & +  & -- & -- & -- & --\\
fibroblasts (connective tissue cells) at surface of follicles
  & +  & -- & +/-- & -- & + \\
ubiquitous fibroblasts in red pulp 
  & +/-- & -- & +/-- & -- & -- \\
fibroblastic capillary sheath cells
  & --\textsuperscript{3} & -- & + & -- & --\\
fibroblasts in trabeculae 
  & + & +/-- & -- & -- & --\\
B-lymphocytes
  & -- & -- & -- & + & -- \\
sinus endothelia
  & -- & +/--\textsuperscript{4} & -- & -- & + \\ 
\bottomrule
\end{tabular}
\begin{minipage}{0.7\linewidth}
\raggedright
\footnotesize
\textsuperscript{1} stained in alternating serial sections\\
\textsuperscript{2} except endothelial cells of most sinuses\\
\textsuperscript{3} most cells\\
\textsuperscript{4} expression varies by proximity to a follicle\\
\end{minipage}
\end{table*}

\begin{figure*}[!t]
\centering
\subfloat[][]{\label{fig:4-a}\includegraphics[height=0.3\linewidth]{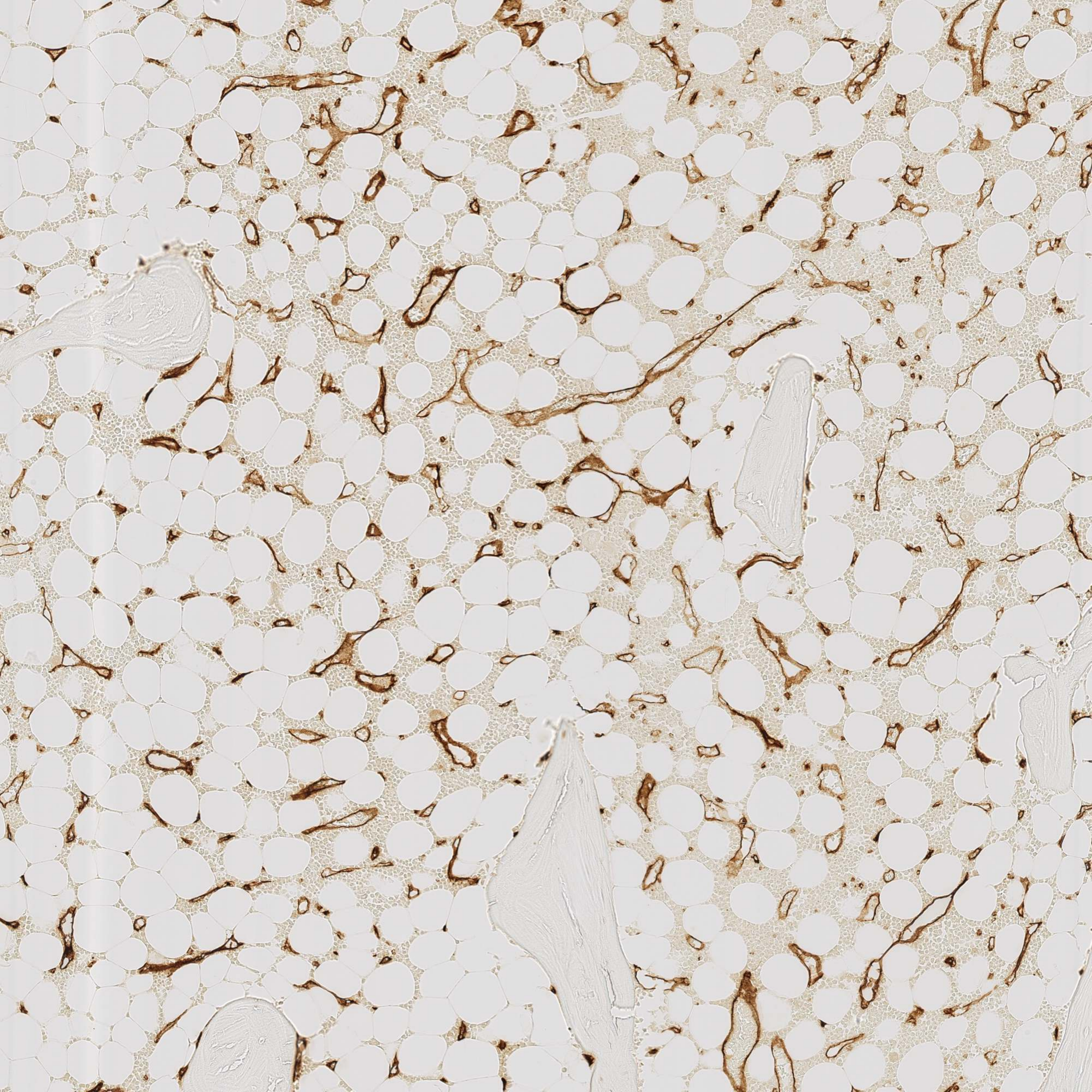}}\hfill%
\subfloat[][]{\label{fig:4-b}\includegraphics[height=0.3\linewidth]{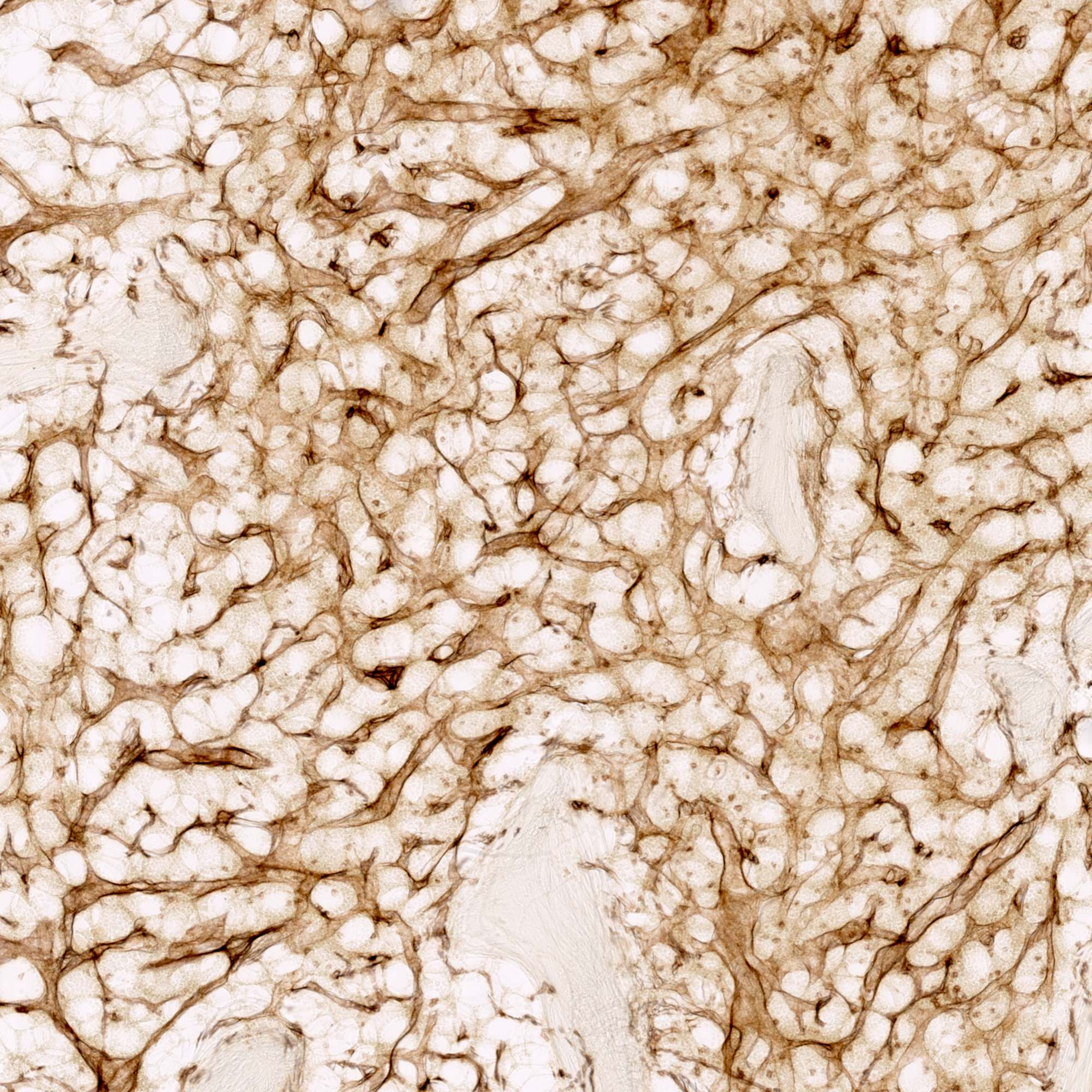}}\hfill%
\subfloat[][]{\label{fig:4-c}\includegraphics[height=0.3\linewidth]{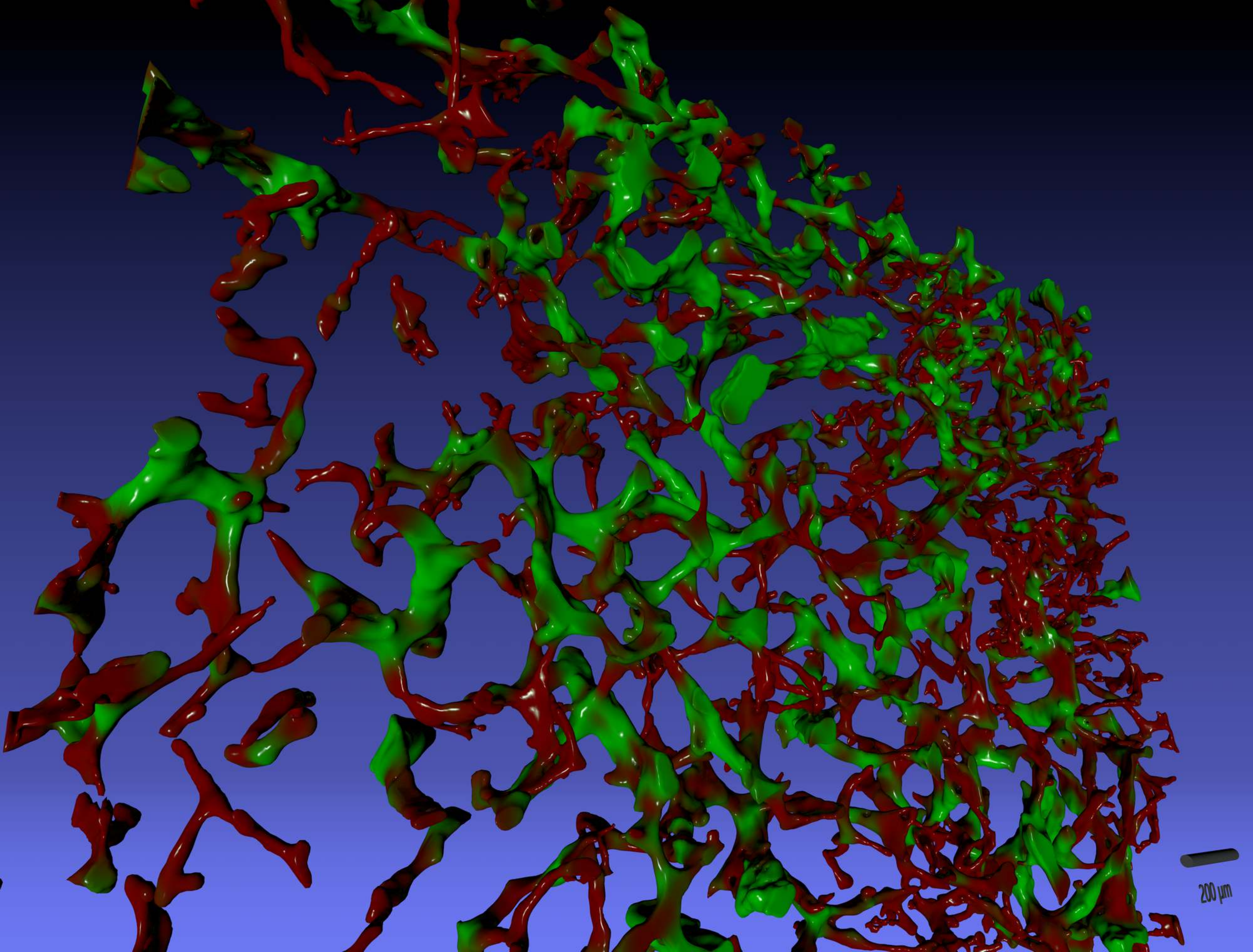}}

\subfloat[][]{\label{fig:4-d}\includegraphics[height=0.333\linewidth]{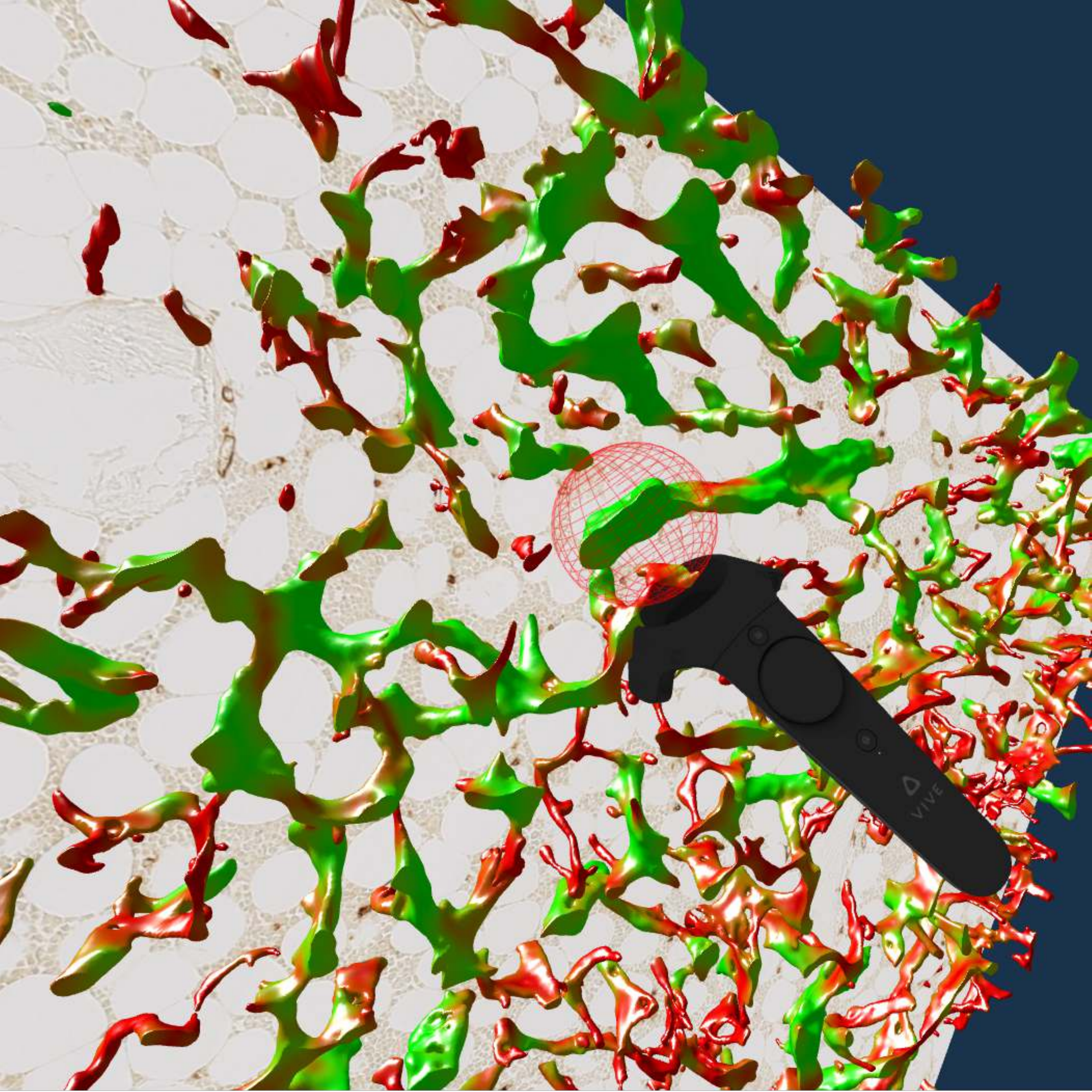}}\hfill%
\subfloat[][]{\label{fig:4-e}\includegraphics[height=0.333\linewidth]{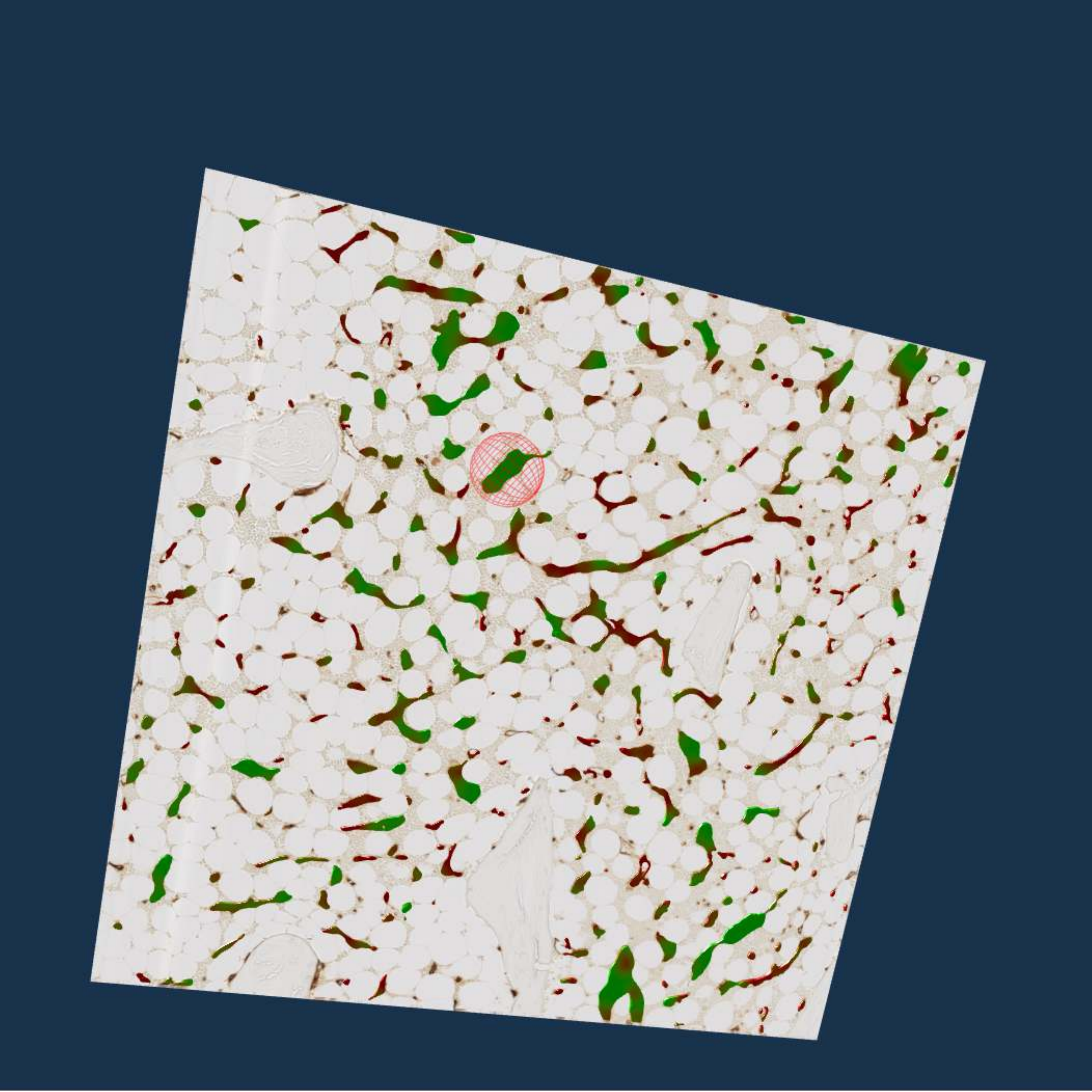}}\hfill%
\subfloat[][]{\label{fig:4-f}\includegraphics[height=0.333\linewidth]{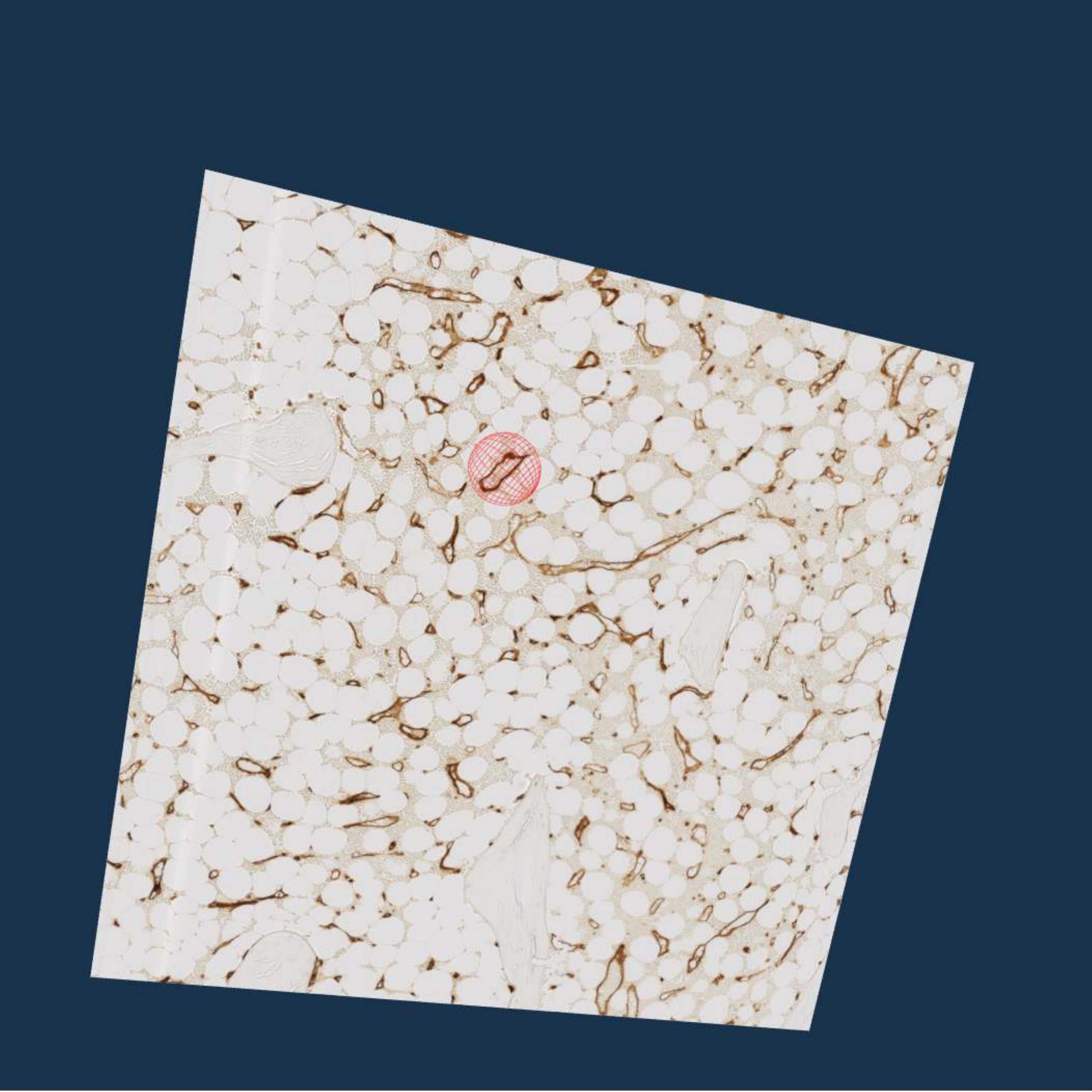}}

\caption{Images, renderings, and VR screenshots showing mesh reconstructions of blood vessels in a human bone marrow specimen, stained with anti-CD34 plus anti-CD141. \protect\subref{fig:4-a}: A~single section. The staining colour is brown for both molecules in the original data. \protect\subref{fig:4-b}: A~volume rendering of 21 consecutive serial sections. \protect\subref{fig:4-c}: The reconstructed mesh. It shows shape diameter function values, colour-coded from red to green. \protect\subref{fig:4-d}: We annotate a position of interest in the mesh in VR. An original section is seen in the background. \protect\subref{fig:4-e}: We have found the section containing the annotation, the mesh is still visible. \protect\subref{fig:4-f}: Only the section with the annotation is shown in VR. Domain experts can now reason on the stained tissue at the marked position.}
\label{fig:4}
\end{figure*}

\begin{figure*}[tbh]
\centering
\subfloat[][]{\label{fig:7-a}\includegraphics[width=0.48\linewidth,angle=90]{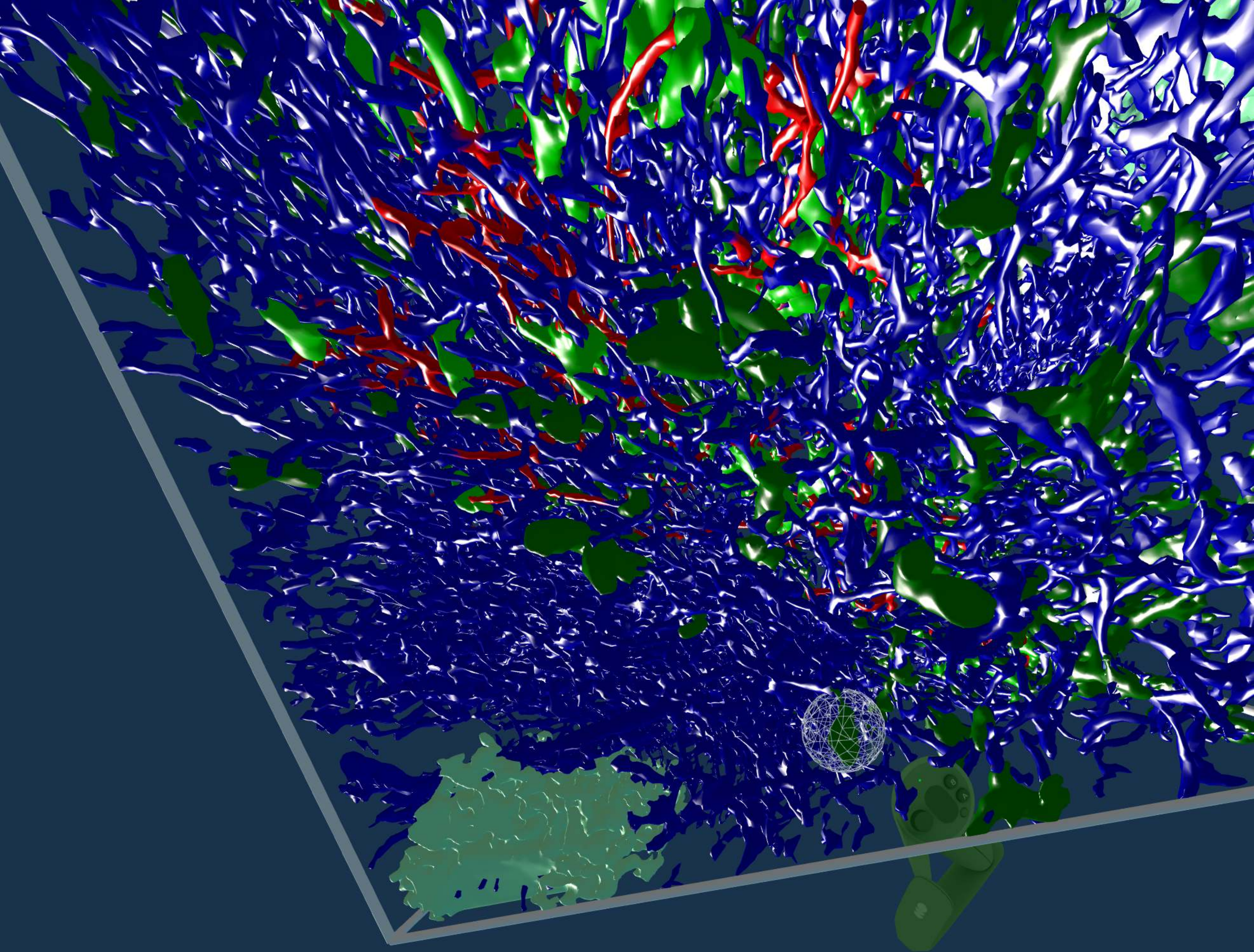}}\hfill%
\subfloat[][]{\label{fig:7-b}\includegraphics[width=0.48\linewidth,angle=90]{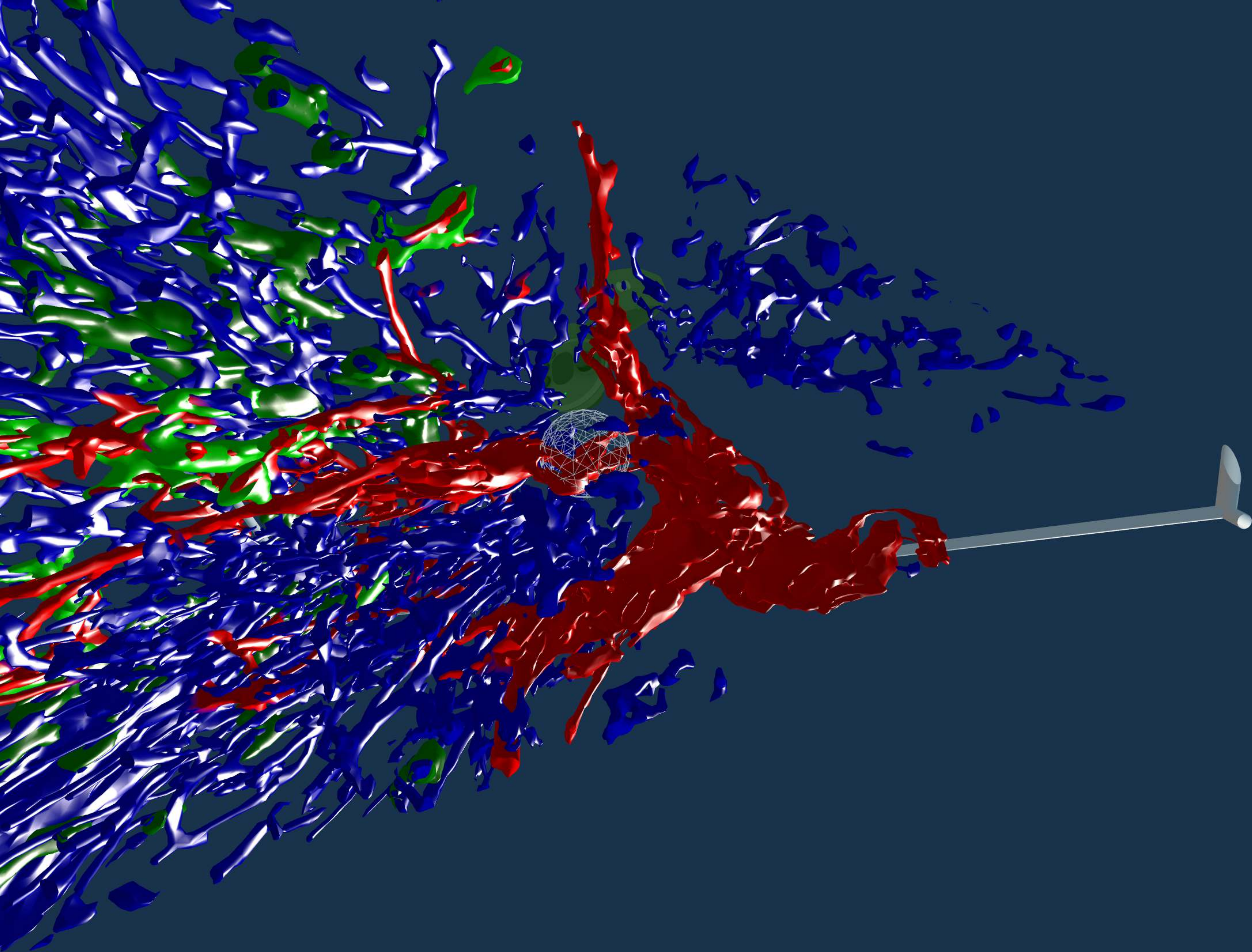}}

\olcomment{moved to front!}
\caption{Working with our annotation tool. We show VR screenshots of our application, in a human spleen \enquote{sheaths alternating} data set \citep{steiniger_150}. In Fig.~\protect\subref{fig:7-b} the front plane clipping is evident, viz.\ Fig.~\ref{fig:clipping}. Notice the Valve Index controller with a ball, showing an anatomical structure. In this manner, the VR user can clarify some morphological details or demonstrate an issue to an audience outside VR.\newline
All images are produced with our VR tool. Similar illustrations can be found in \citep{steiniger_150}.}
\label{fig:7}
\end{figure*}

\begin{figure}[tbh]
\newcommand{\Tube}[6][]%
{   \colorlet{InColor}{#4}
    \colorlet{OutColor}{#5}
    \foreach \I in {1,...,#3}
    {   \pgfmathsetlengthmacro{\h}{(\I-1)/#3*#2}
        \pgfmathsetlengthmacro{\r}{sqrt(pow(#2,2)-pow(\h,2))}
        \pgfmathsetmacro{\c}{(\I-0.5)/#3*100}
        \draw[InColor!\c!OutColor, line width=\r, #1] #6;
    }
}
\centering
\begin{tikzpicture}

    \Tube{2mm}{25}{red!40}{black}
        {(5,8) to [looseness=1,out=-70,in=80] (3,3)}

    \Tube{2mm}{25}{green!25}{black}
        {(2,7) to [looseness=1,out=50,in=180] (5,5) to [looseness=1,out=0,in=180] (8,6)}

    \Tube{2mm}{25}{green!25}{black}
        {(5,5) to [looseness=1,out=0,in=180] (7,2)}

\pgftransparencygroup
\pgfsetfillopacity{0.2}
 \fill[yellow!80!black] (4.5,5) circle (2);
\endpgftransparencygroup
 \draw[yellow!20!black] (4.5,5) circle (2);
  \fill[black] (4.5,5) circle (0.05);

\end{tikzpicture}

\caption{This figure demonstrates why we need geodesic distances for mesh painting in our VR application. The yellow circle is the painting tool. We would like to mark the green blood vessels inside the circle, but do not want to co-mark the red blood vessel, even if it is also inside the circle. Red and green blood vessels might even be connected somewhere outside the circle, but the geodesic distance from the centre of the circle (the black dot) to any vertex of the red blood vessel is too large, even if they are reachable. Conversely, a part of the green blood vessel is selected, as a vertex of the green mesh is closest to the centre of the circle. As many vertices are selected as the geodesic distance (corresponding to the radius of the circle with some heuristics) allows for.}
\label{fig:geo}
\end{figure}
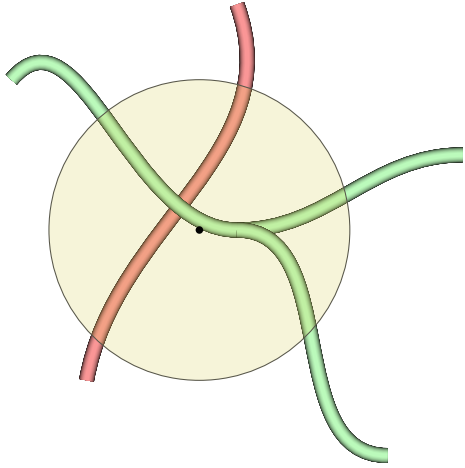

\hypertarget{input-data}{%
\subsection{Input data}\label{input-data}}

The input data was generated from the registered stack of serial
sections. Typical data volume was \(2.3k \times 2.3k \times 161\)
voxels, with \(z\)-axis interpolation. The original data typically
featured 21 to 24~sections, but we have used up to 150~sections in some
reconstructions. The final meshes had typically 1.7M--2.3M~vertices.
(Most of the GPU memory used by the application was occupied by textures
anyway.) Real time rendering was possible with Vive-native resolution at
90~fps. Our experiments with Valve Index ran at even higher frame rates.
Original data were projected as textures, typically at
\(2.3k \times 2.3k\). Although we experimented with showing all sections
at once, in a productive use only one section was shown at a time.

We quality control the following data sets derived from the bone marrow
of a 53-year-old male and from the spleen of a 22-year-old male.
Acquisition of the specimens complied with the ethical regulations of
Marburg University Hospital at the time of processing.

\begin{figure*}[tbp]
\centering
\subfloat[][]{\label{fig:5-a}\includegraphics[width=0.333\linewidth]{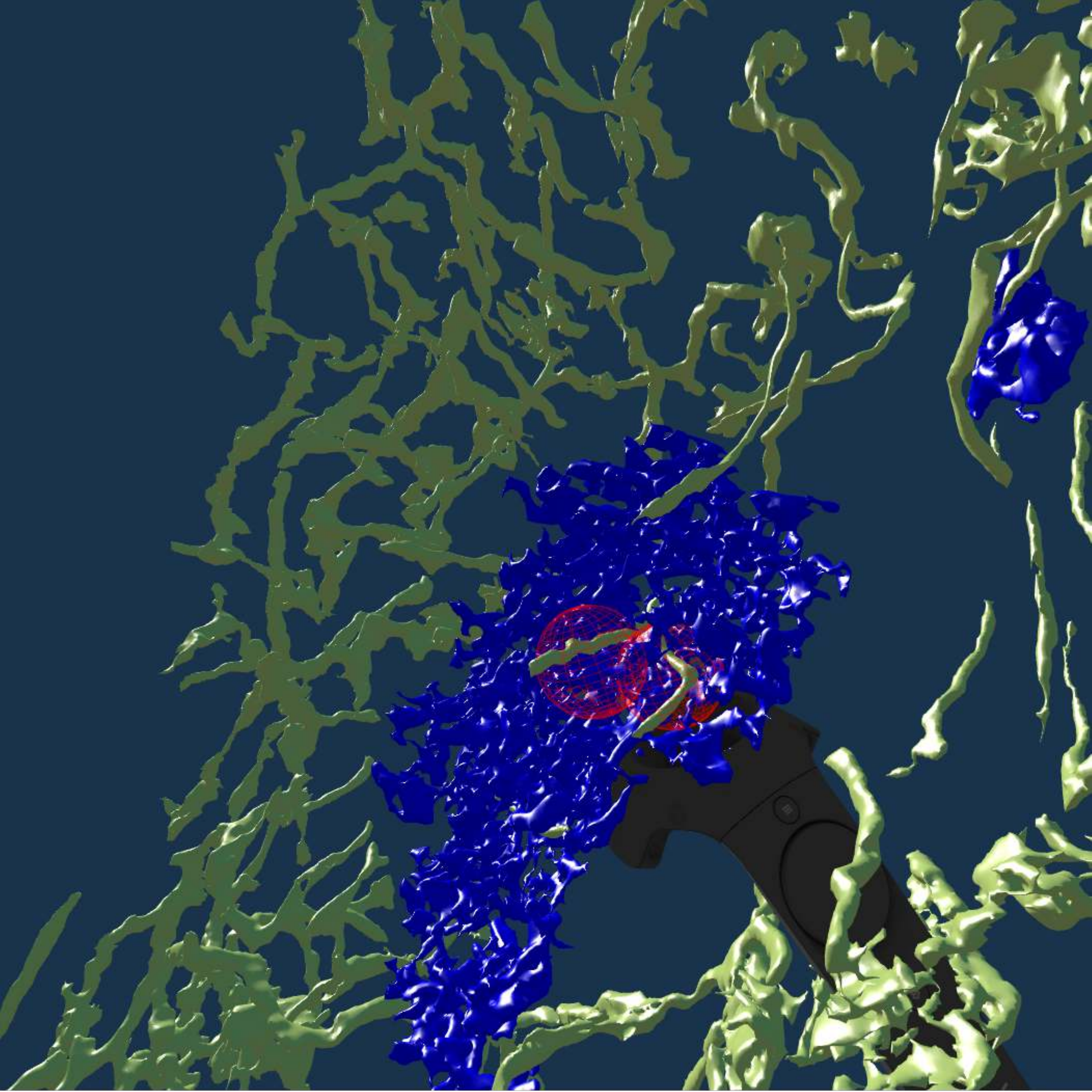}}~
\subfloat[][]{\label{fig:5-b}\includegraphics[width=0.333\linewidth]{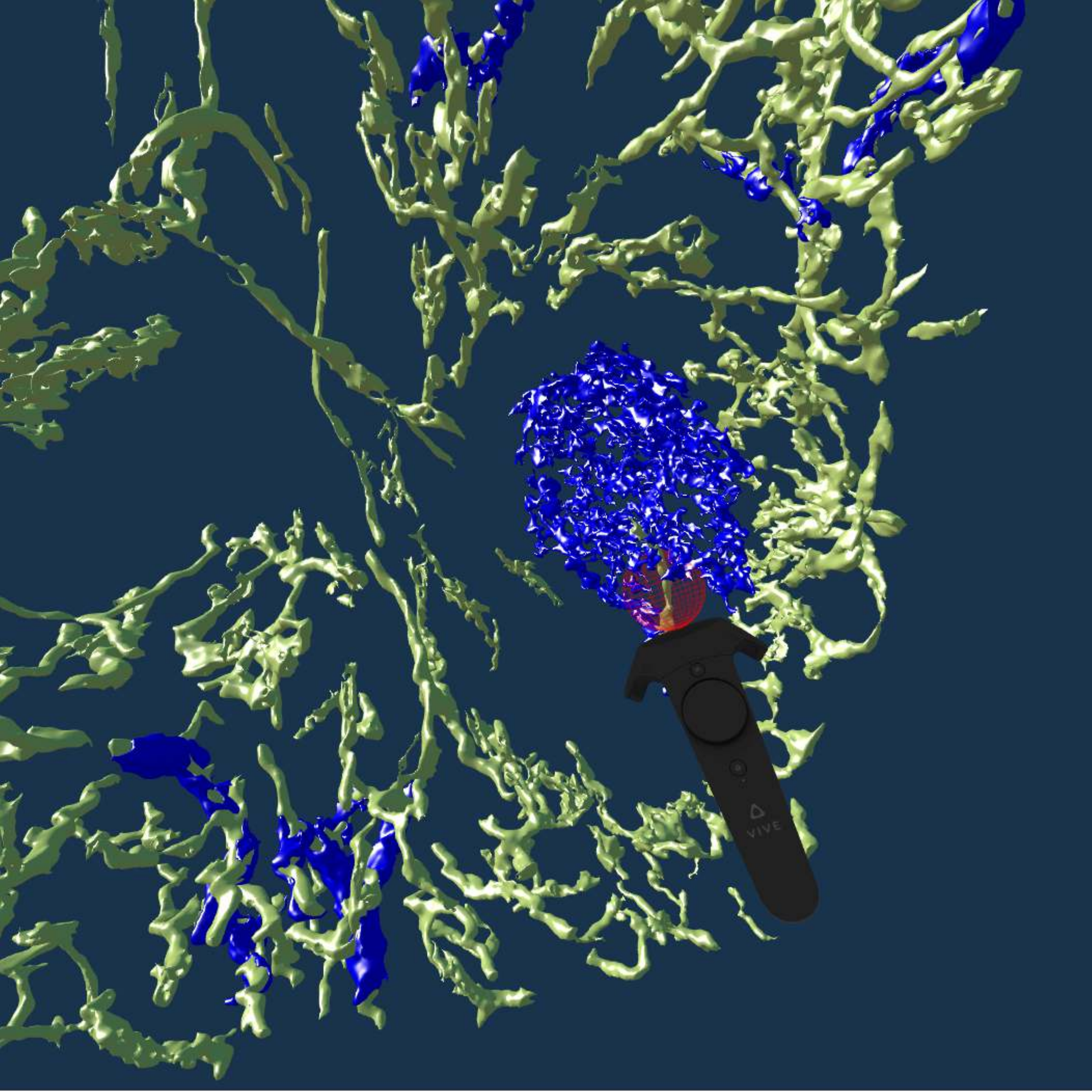}}

\subfloat[][]{\label{fig:5-c}\includegraphics[width=0.333\linewidth]{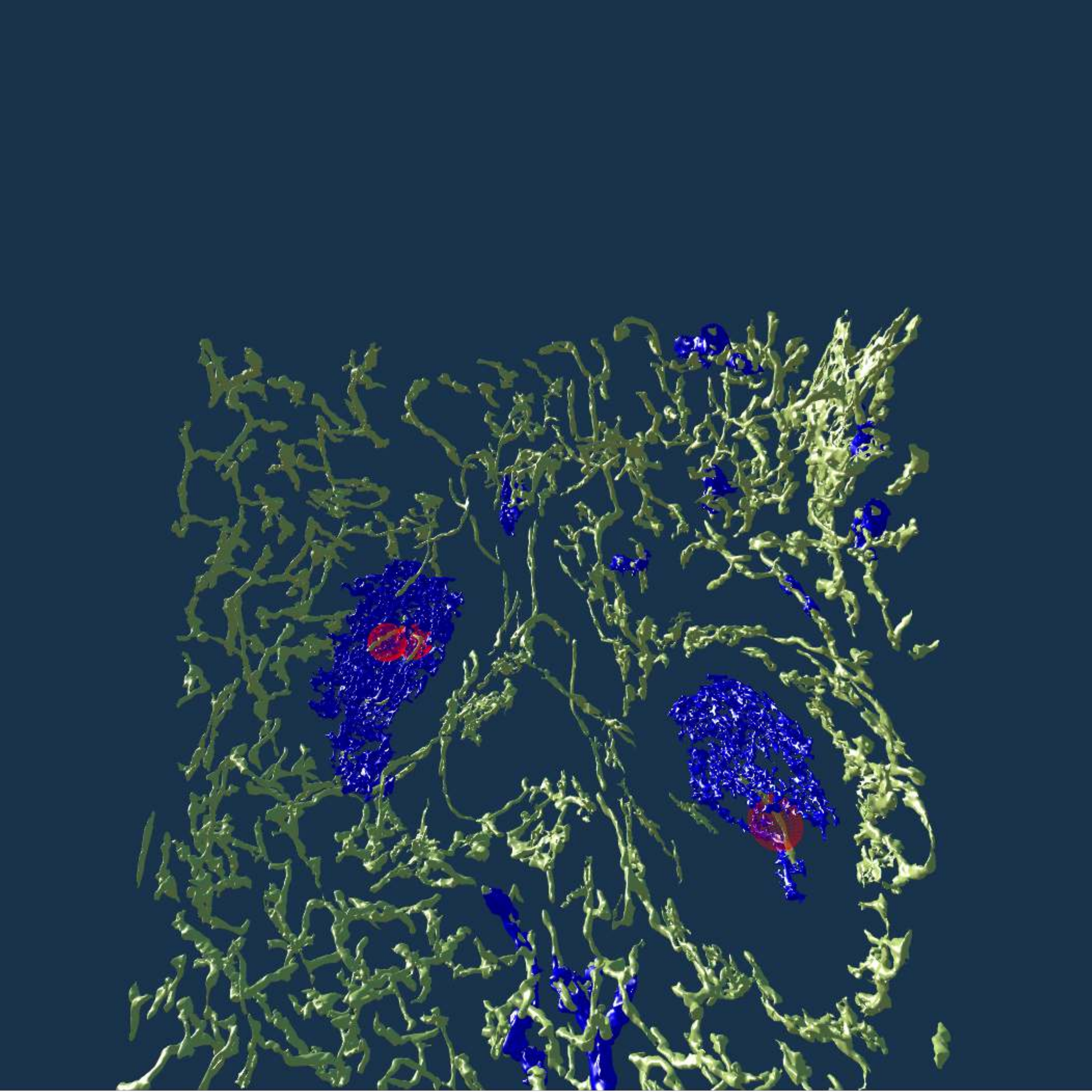}}~
\subfloat[][]{\label{fig:5-d}\includegraphics[width=0.333\linewidth]{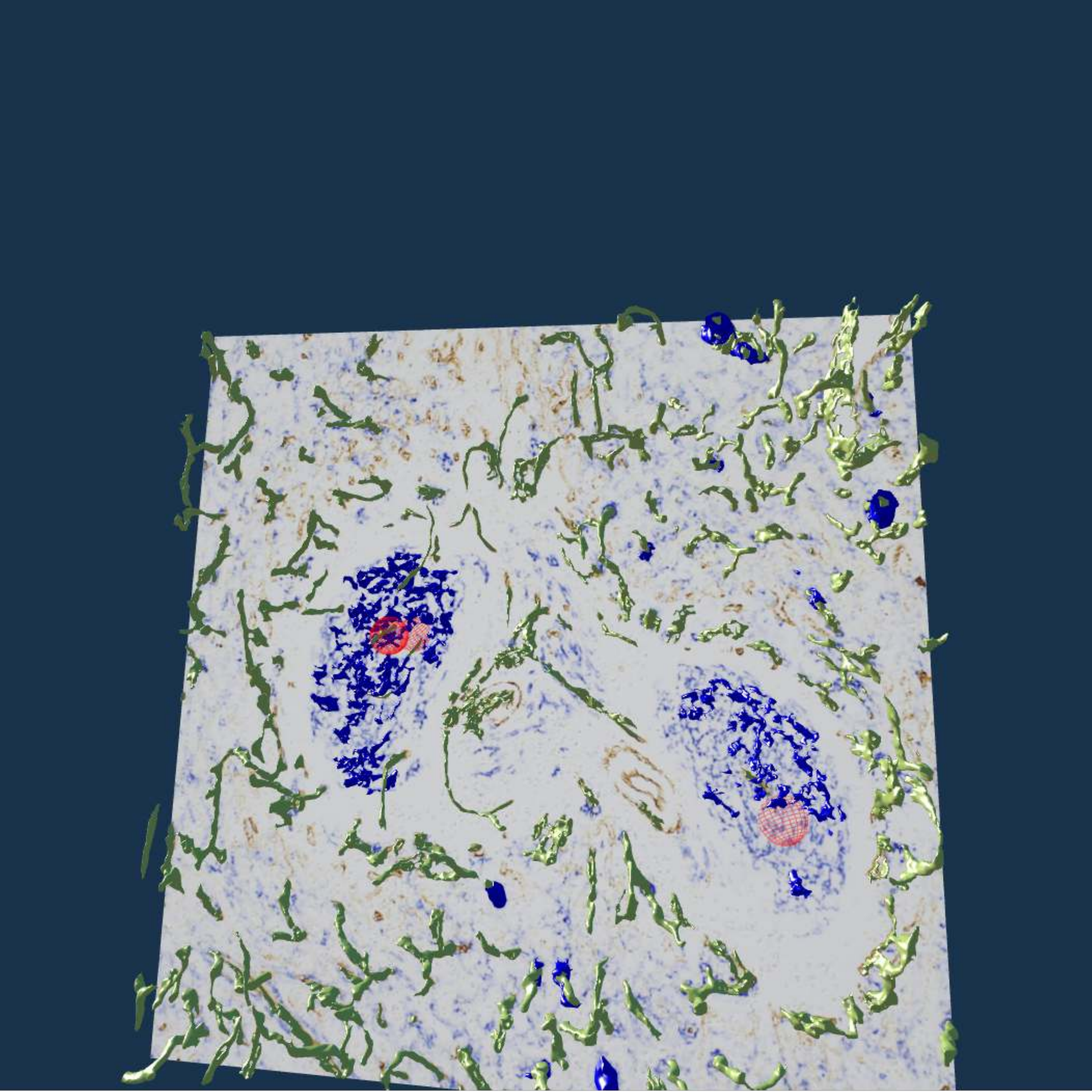}}

\subfloat[][]{\label{fig:5-e}\includegraphics[width=0.333\linewidth]{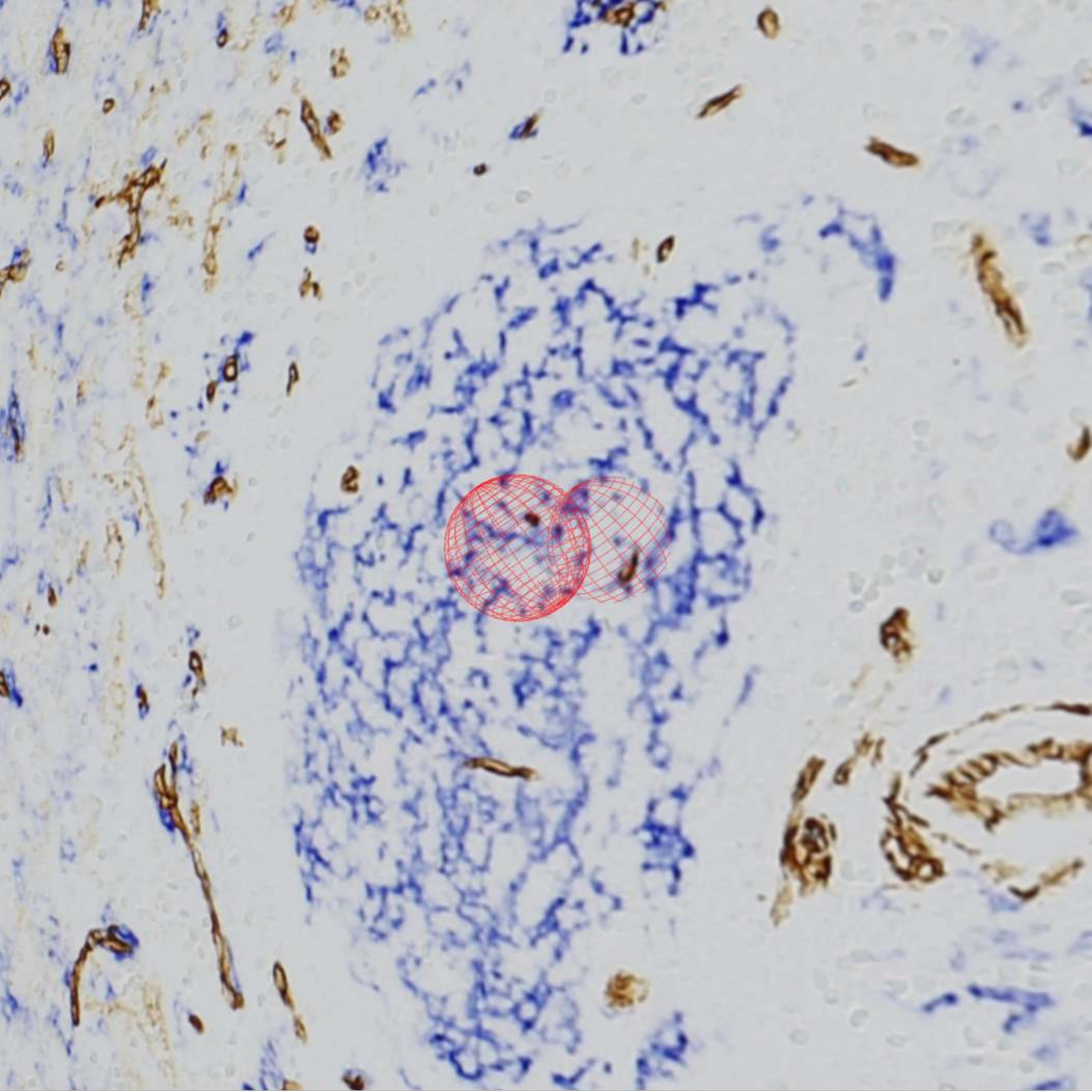}}~
\subfloat[][]{\label{fig:5-f}\includegraphics[width=0.333\linewidth]{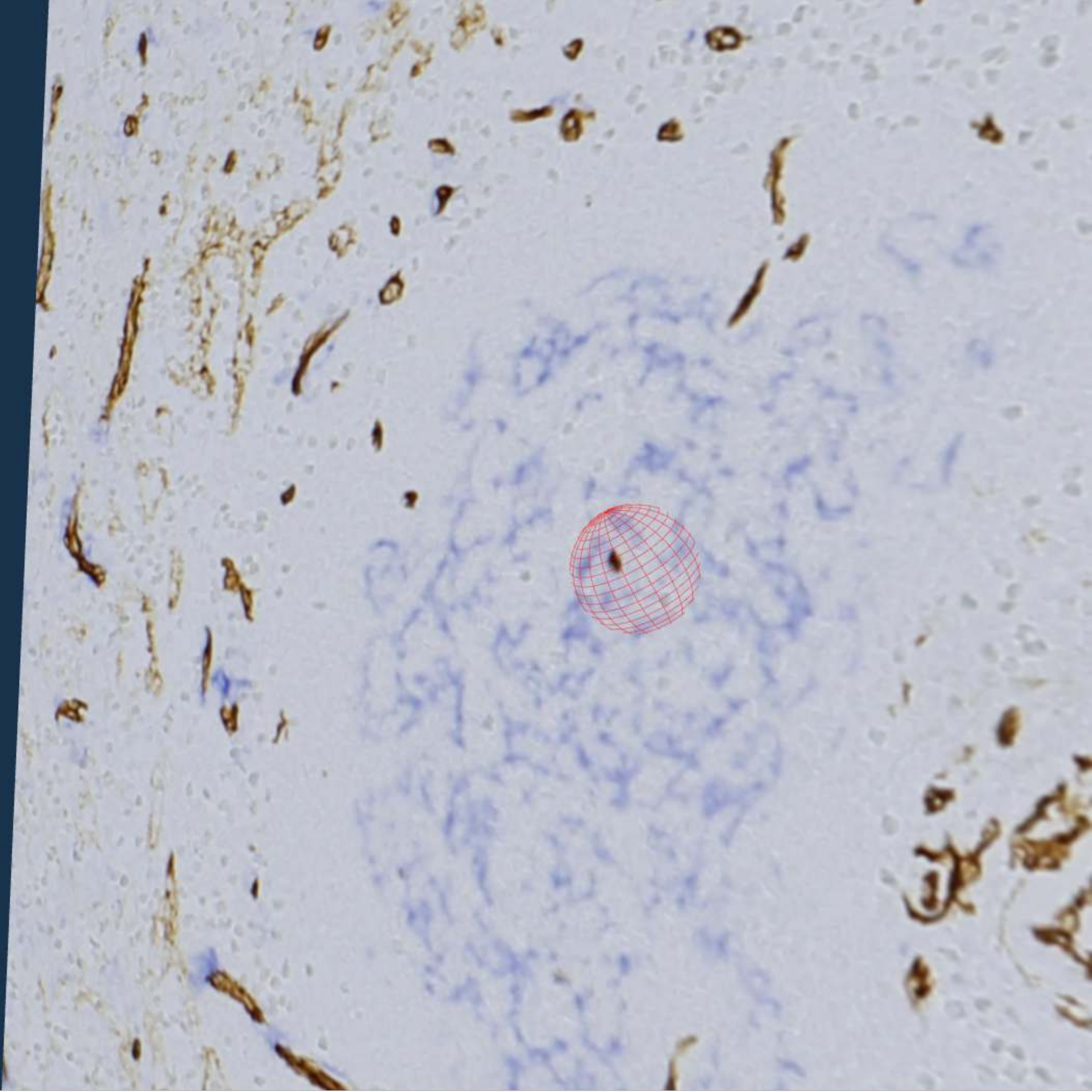}}

\caption{Real or artefact? The models are derived from human spleen sections from the \enquote{follicle-double} data set. These sections were stained for CD34 (brown in staining, yellow in the reconstruction) and for CD271 (blue). In VR we spotted and annotated putative capillaries inside  follicles (large blue structures, a, b). We can look at the meshes only \protect\subref{fig:5-c} or also show the original data~\protect\subref{fig:5-d}. A~closer view \protect\subref{fig:5-e}, \protect\subref{fig:5-f} confirms: the reconstruction is correct, these structures are CD34\textsuperscript{+} objects inside the follicle. As the structures in question continue through multiple sections, they do not represent single CD34\textsuperscript{+} cells. Hence the objects in question must be blood vessels. The reconstruction is correct, the brown structures  are real.\newline
All images in this figure are screenshots from our application. Similar results can be found in~\afoll.}
\label{fig:5}
\thisfloatpagestyle{empty} 
\end{figure*}

\begin{figure}[tbh]
\centering
\ifthenelse{\boolean{review} \OR \boolean{arxivpreprint}}{%
\includegraphics[width=0.5\linewidth]{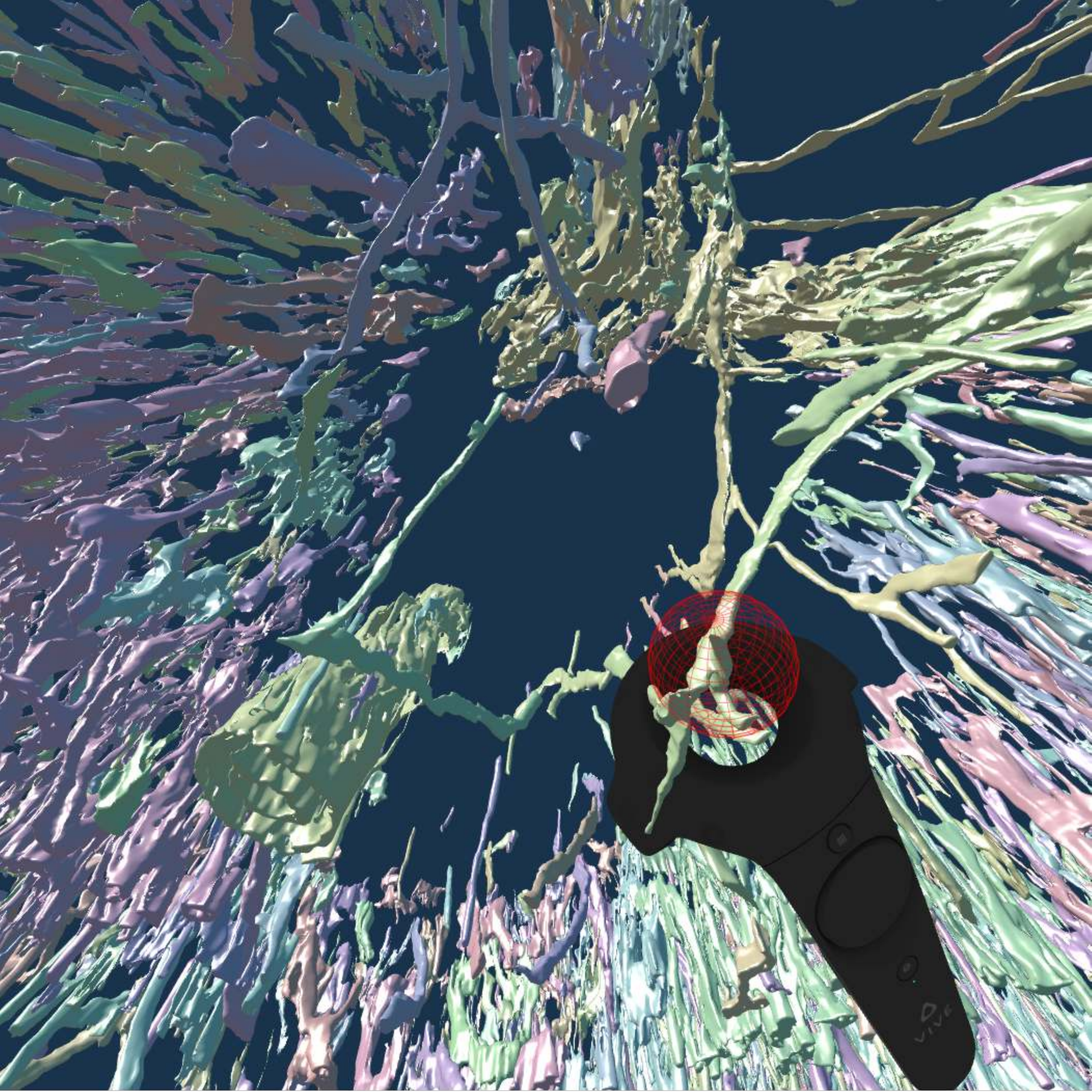}
}{%
\includegraphics[width=\linewidth]{images/fig_6_a.pdf}
}

\caption{A~VR screenshot showing mesh reconstructions of blood vessels in a human spleen specimen, anti-CD34 staining, \enquote{follicle-single} data set \afoll. Unconnected mesh components were set to distinct colours. The user is highlighting a smaller blood vessel that follows larger ones with the HTC Vive controller.}
\label{fig:6}
\end{figure}

\begin{figure*}[p]
\centering
\subfloat[][]{\label{fig:8-a}\includegraphics[width=0.35\linewidth]{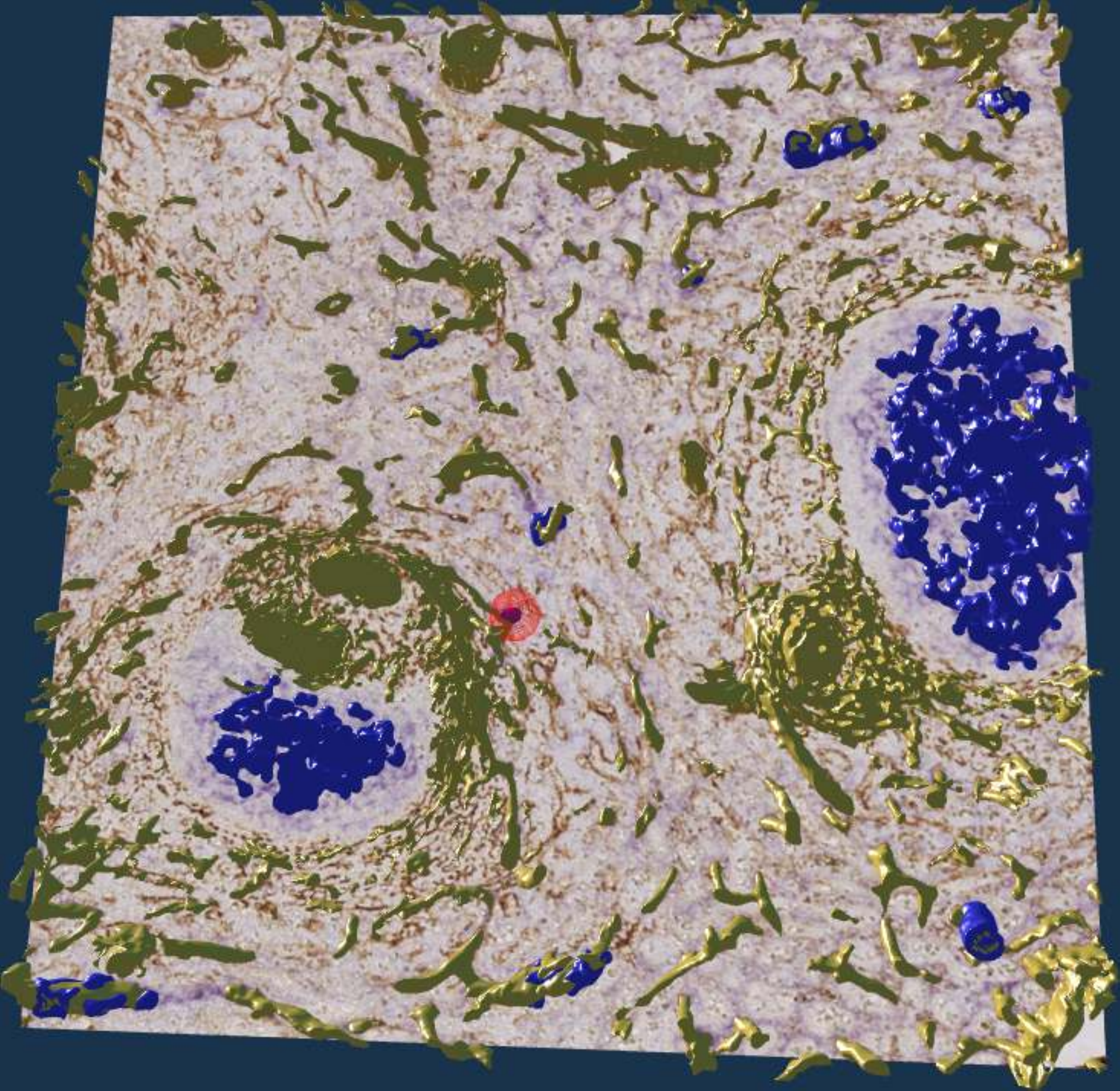}}

\subfloat[][]{\label{fig:8-b}\includegraphics[width=0.35\linewidth]{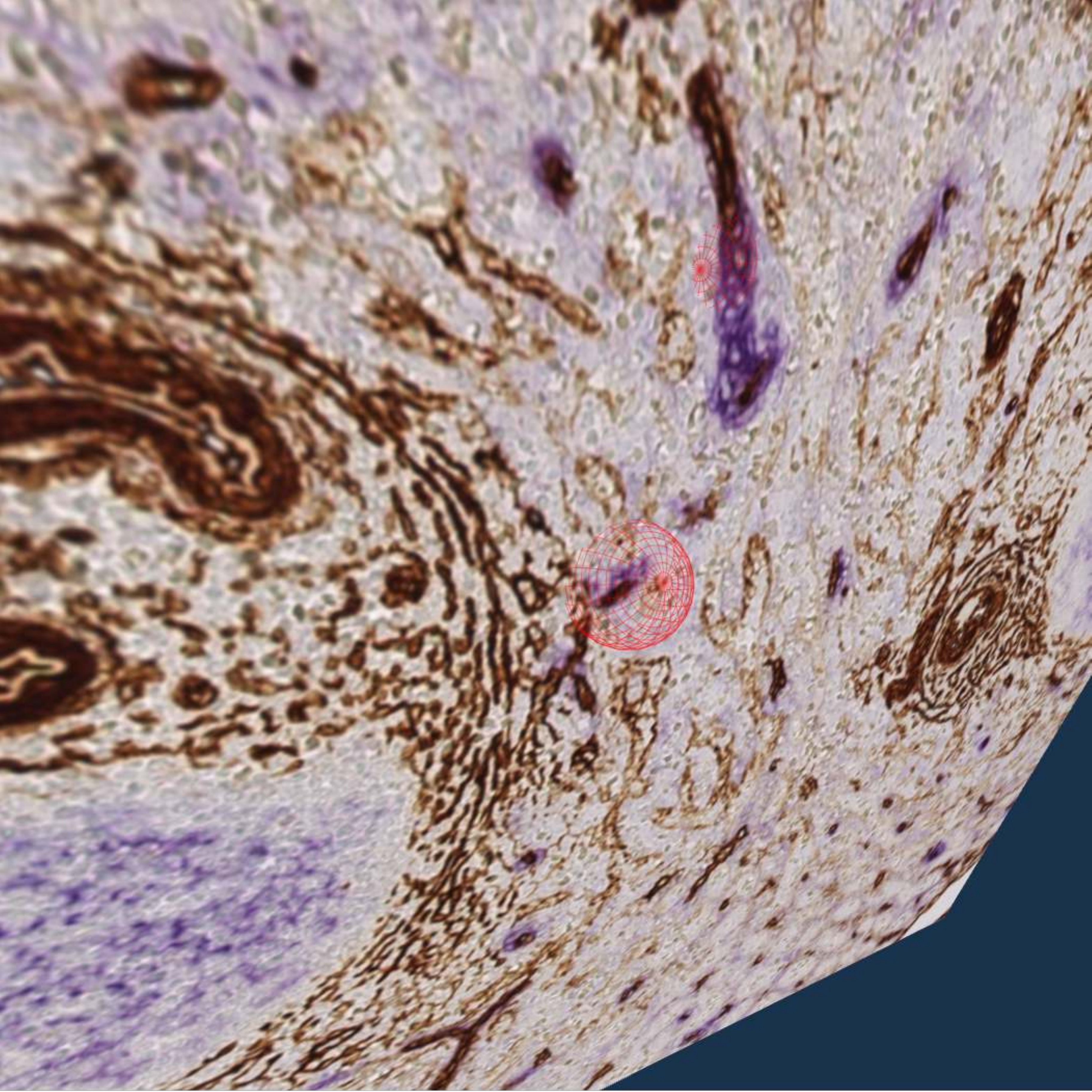}}~
\subfloat[][]{\label{fig:8-c}\includegraphics[width=0.35\linewidth]{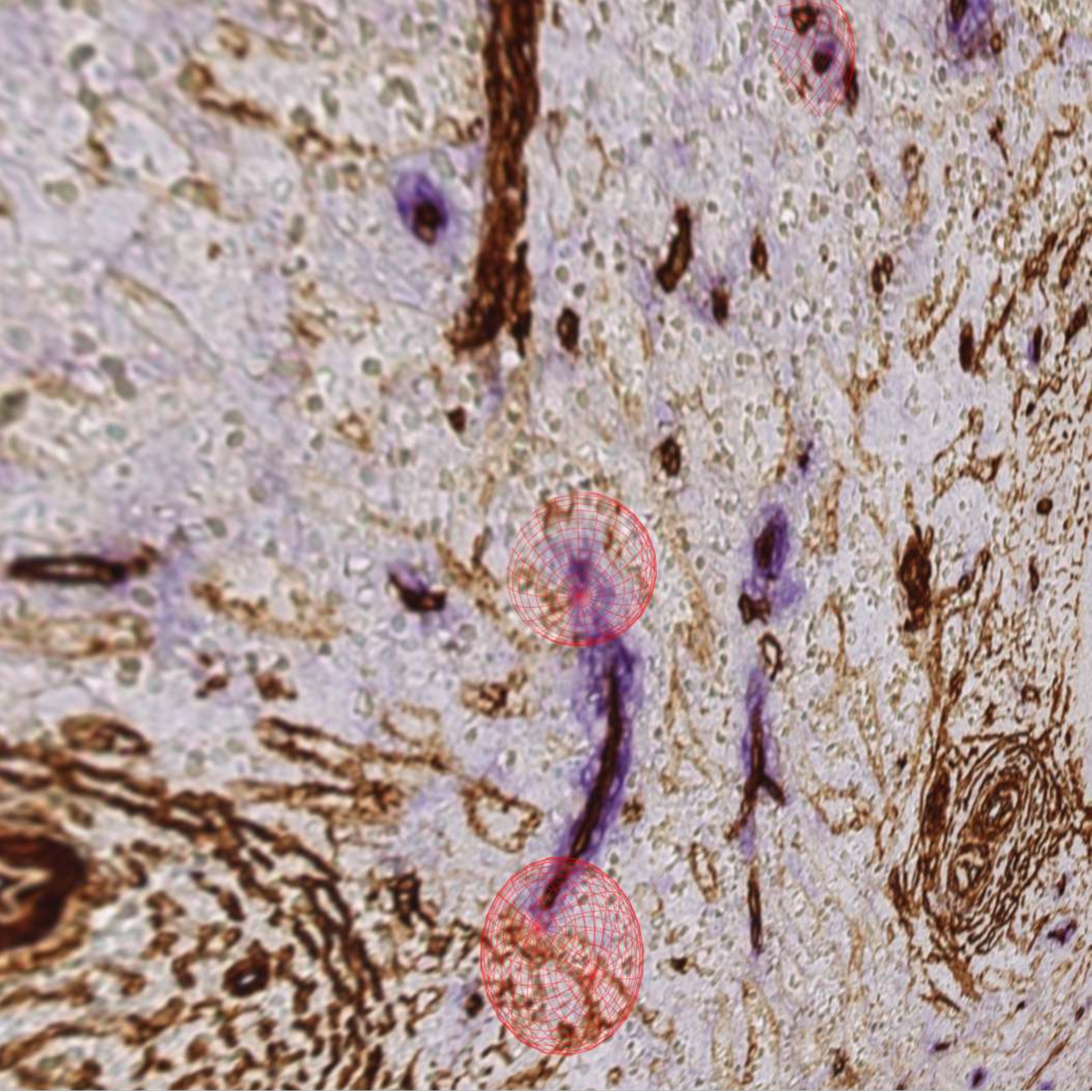}}

\subfloat[][]{\label{fig:8-d}\includegraphics[width=0.35\linewidth]{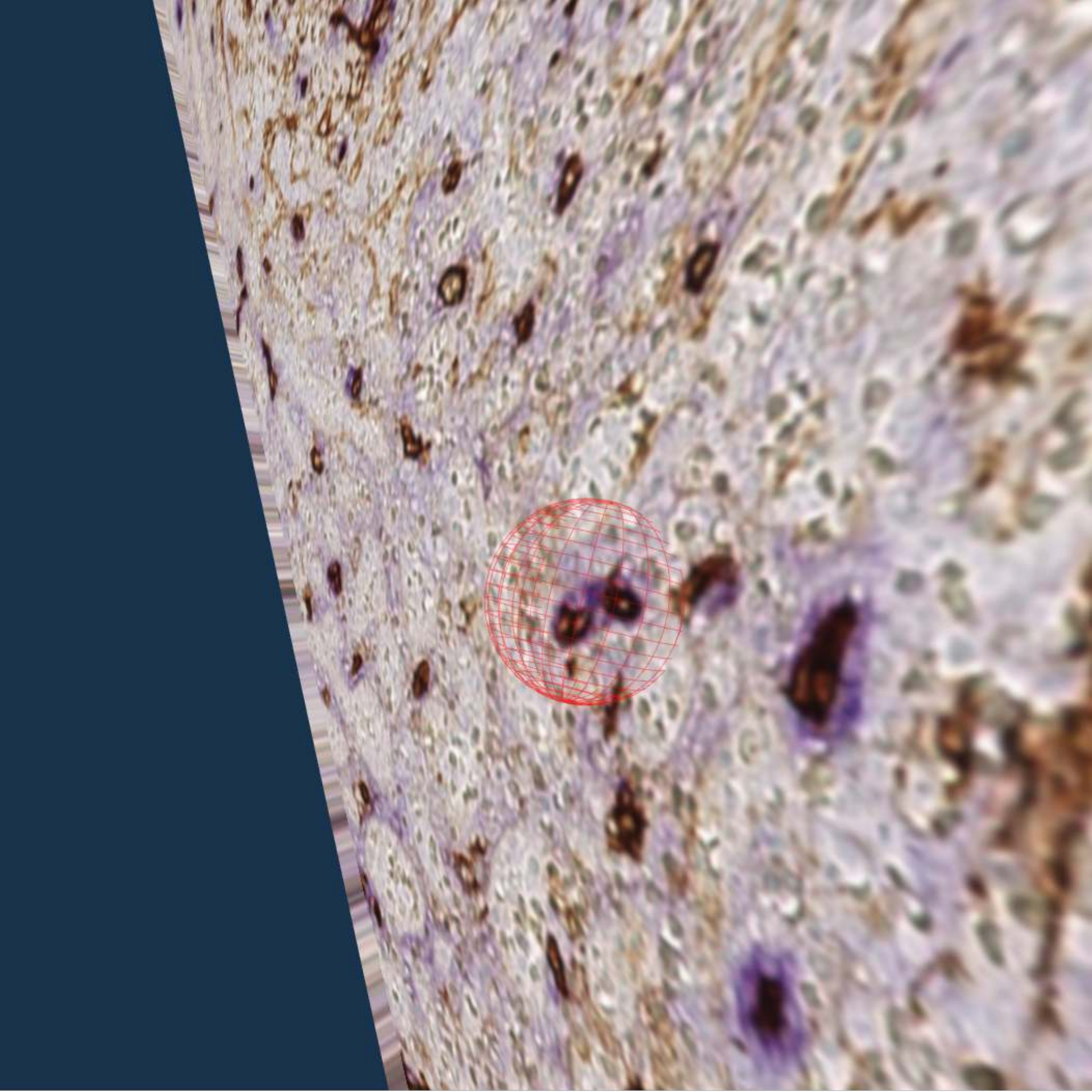}}~
\subfloat[][]{\label{fig:8-e}\includegraphics[width=0.35\linewidth]{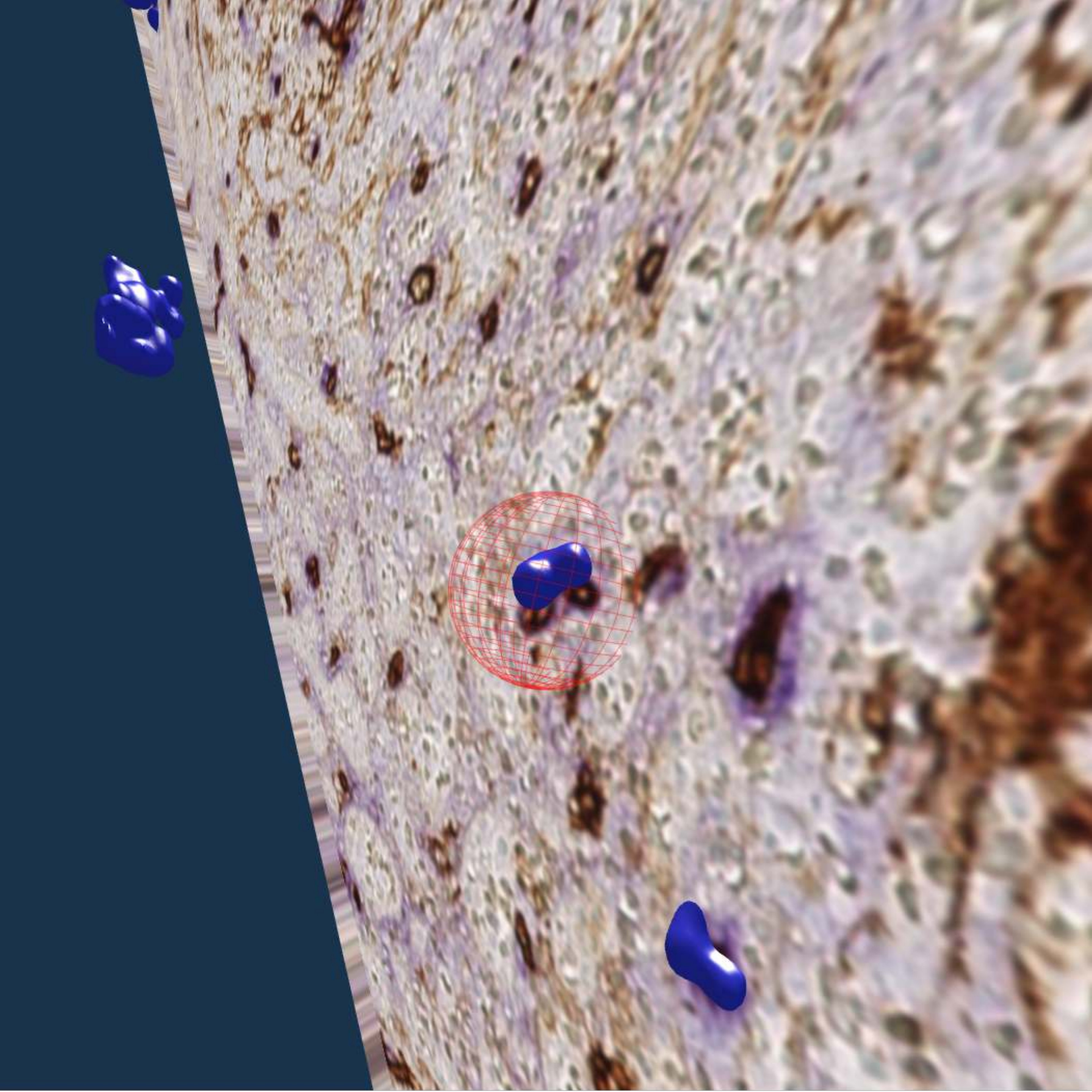}}

\caption{Investigating the annotated regions, VR screenshots of our application.The human spleen \enquote{red pulp} data set is used \asheath; we have annotated some ends of capillary sheaths in meshes reconstructed from human spleen data. \protect\subref{fig:8-a}: Overview. \protect\subref{fig:8-b}--\protect\subref{fig:8-d}: Original data and an annotation. Experts can more easily reason on such visualisations because of 3D perception and intuitive navigation. \protect\subref{fig:8-e}: The same annotation as in \protect\subref{fig:8-d}, showing additionally the mesh for sheaths.}
\ifthenelse{\boolean{review} \OR \boolean{arxivpreprint}}{%
}{
\thisfloatpagestyle{empty} 
}
\label{fig:8}
\end{figure*}

\begin{figure*}[!t]
\centering
\subfloat[][]{\label{fig:clip-a}\includegraphics[height=0.356\linewidth]{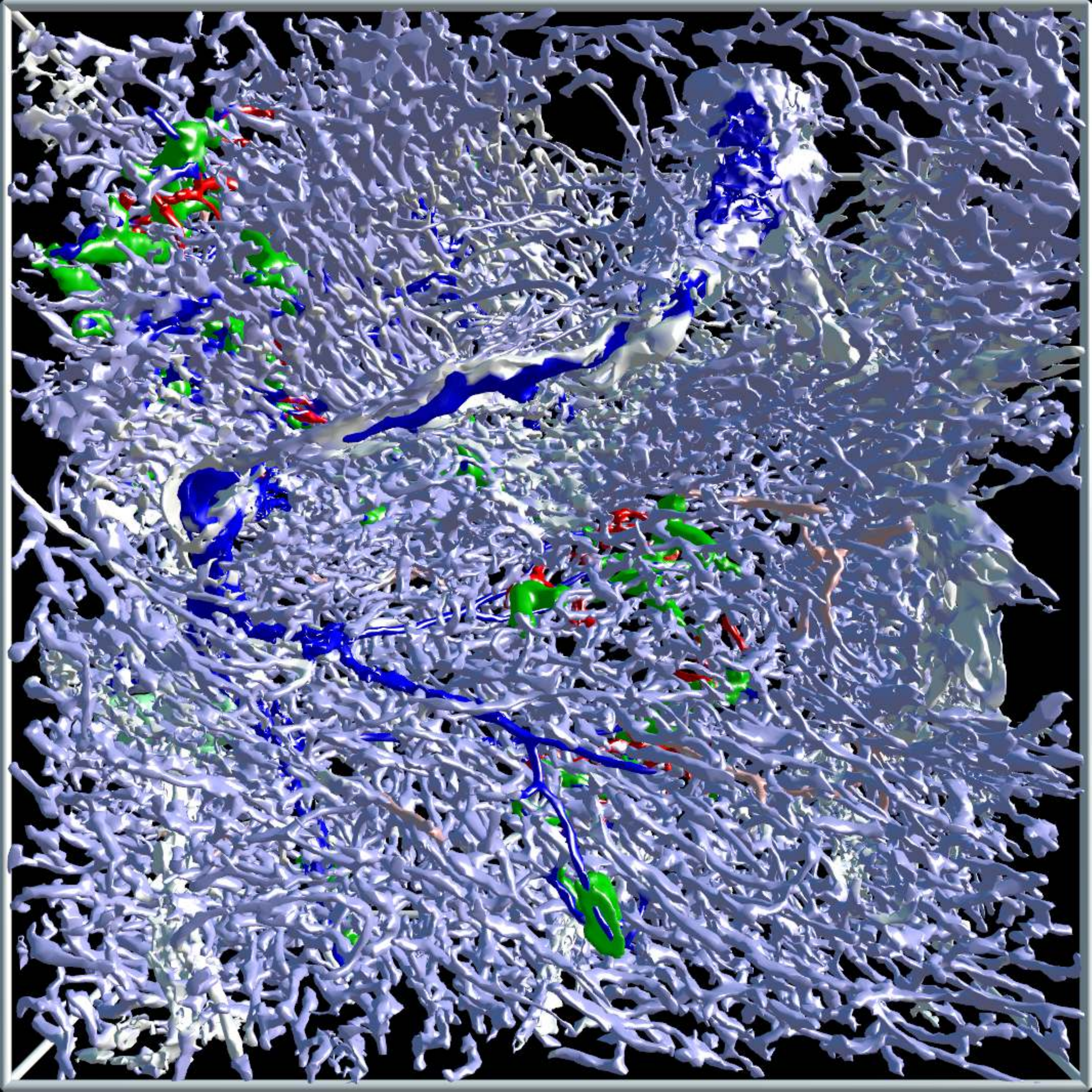}}\hfill%
\subfloat[][]{\label{fig:clip-b}\includegraphics[height=0.356\linewidth]{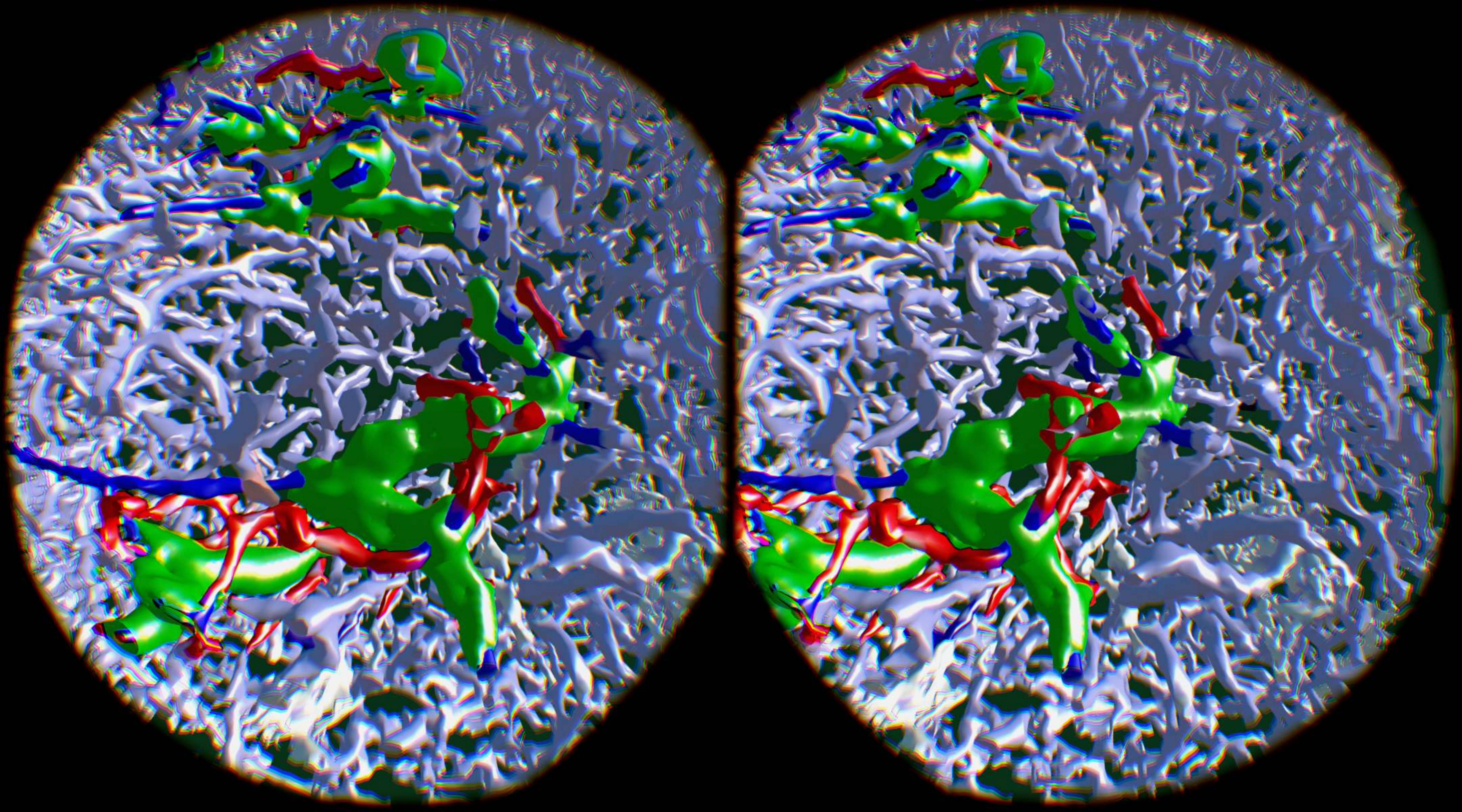}}

\subfloat[][]{\label{fig:clip-c}\includegraphics[width=0.453\linewidth,angle=90]{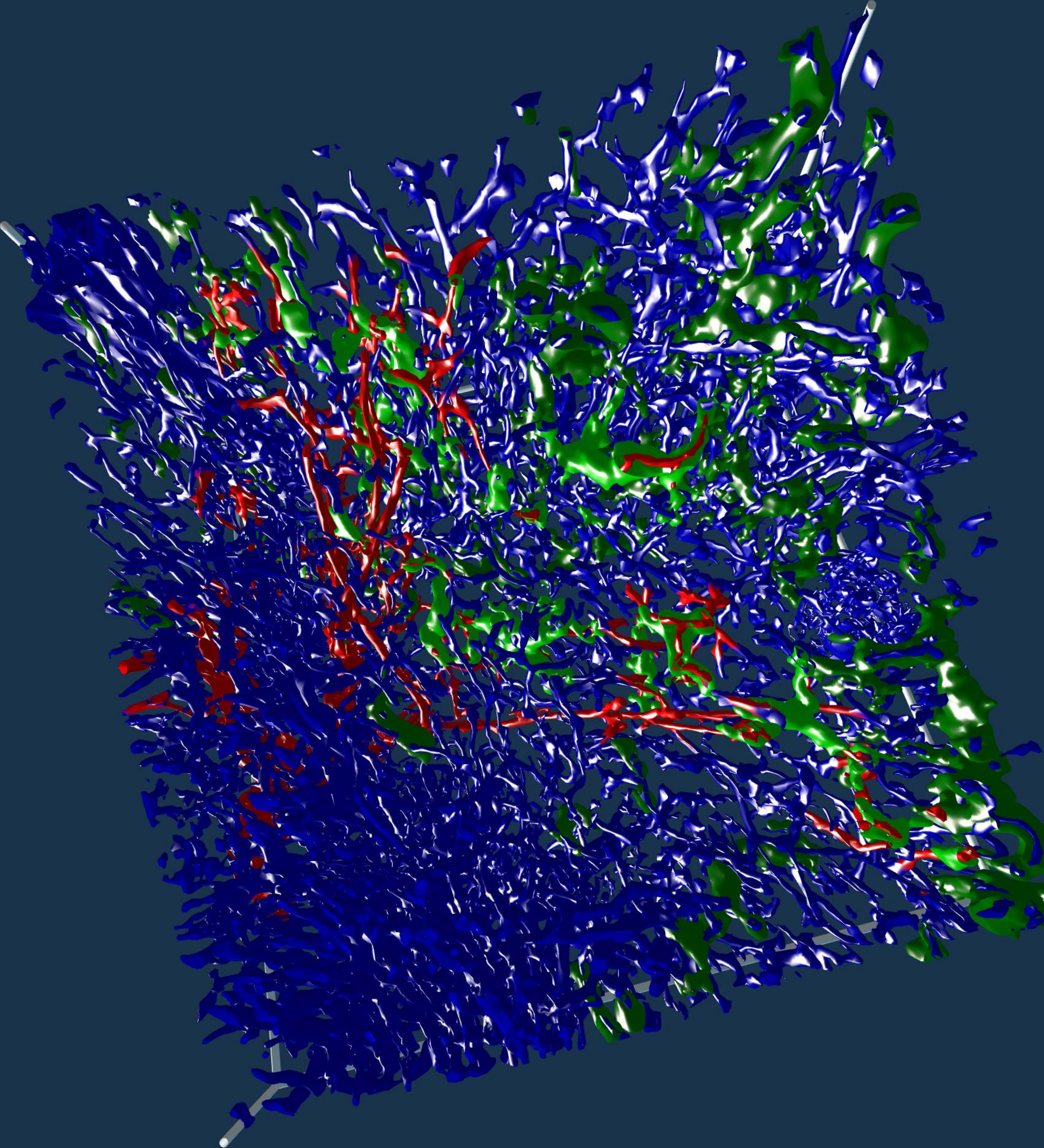}}\hfill%
\subfloat[][]{\label{fig:clip-d}\includegraphics[width=0.453\linewidth,angle=90]{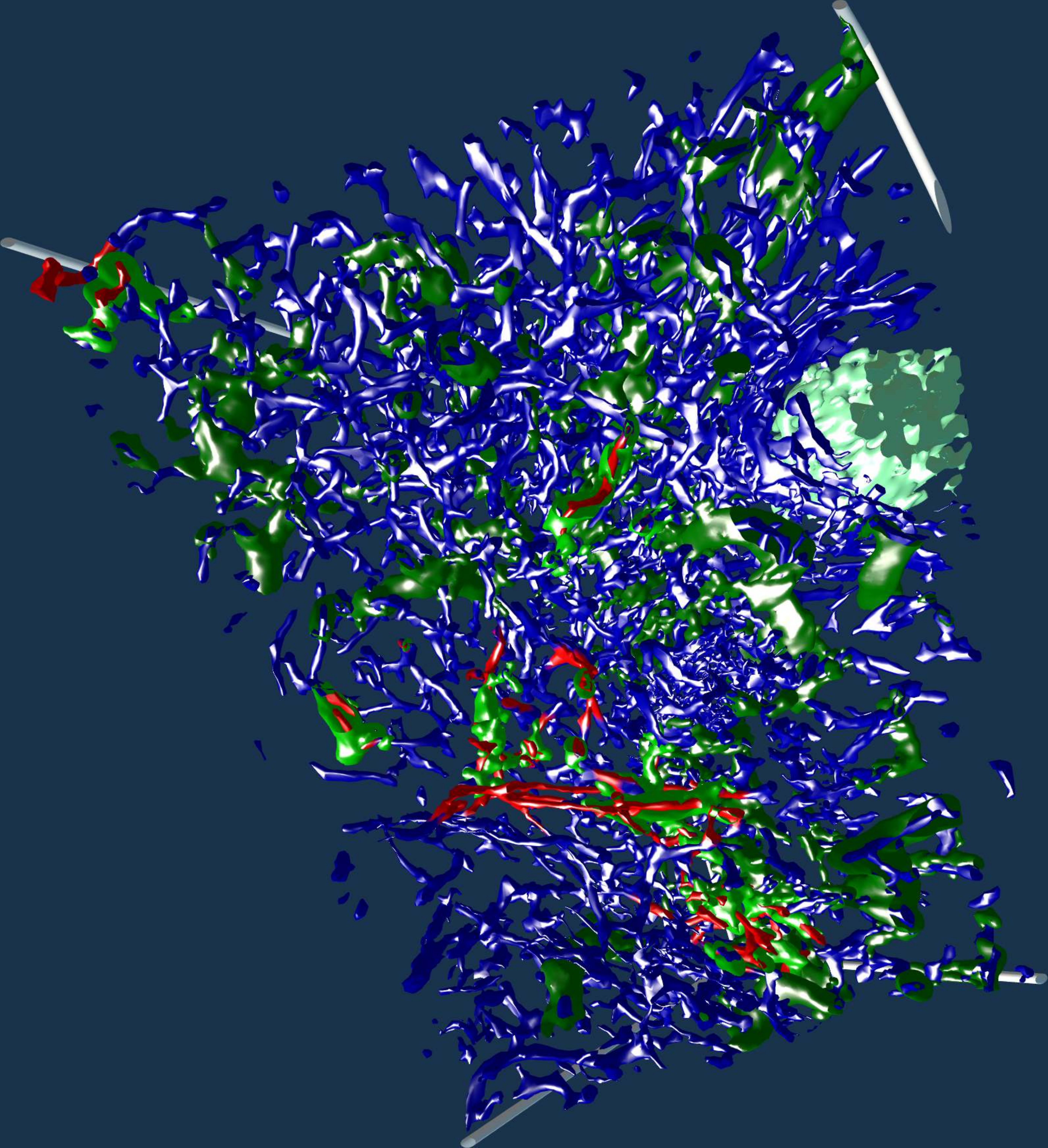}}

\caption{Showcasing front plane clipping on \enquote{sheaths alternating} data sets \citep{steiniger_150}. \protect\subref{fig:clip-a}: A~complete data set in a frontal visualisation. \protect\subref{fig:clip-b}: The user cuts into objects of interest using clipping. \protect\subref{fig:clip-c}--\protect\subref{fig:clip-d}: Utilisation of clipping during the exploration of the data set. 
\newline
All images are produced with our VR tool either directly or with Steam~VR interface. \protect\subref{fig:clip-a}, \protect\subref{fig:clip-b} were featured in a poster \citep{poster}. \protect\subref{fig:clip-a}, \protect\subref{fig:clip-c}, \protect\subref{fig:clip-d}: Similar illustrations can be found in \citep{steiniger_150}.
}
\label{fig:clipping}
\end{figure*}

\begin{figure}[tb]
\centering
\includegraphics[width=1\linewidth]{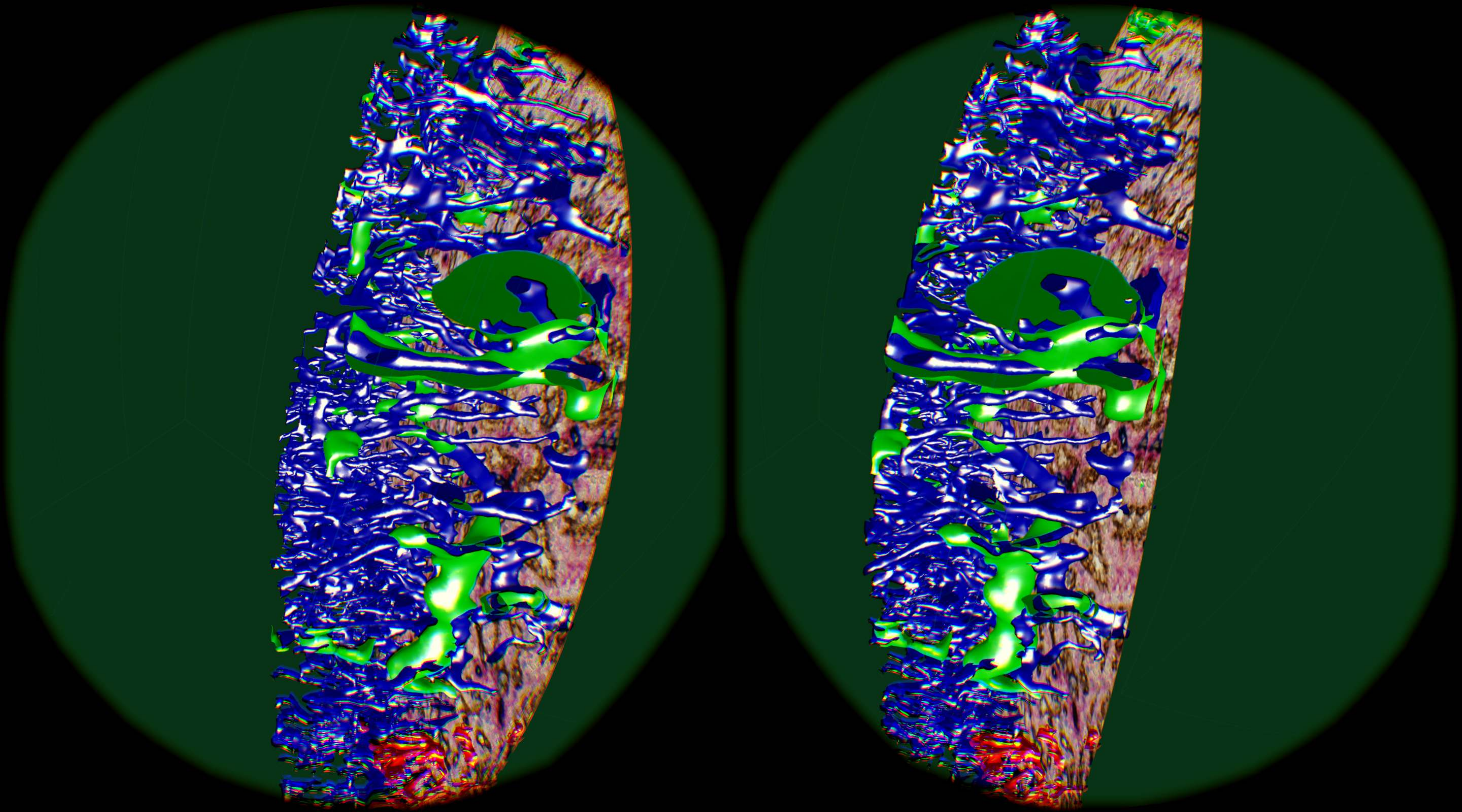}
\caption{Cutting structures open with front clipping plane, using the \enquote{sinus} data set. Capillaries are blue, capillary sheaths are green in the reconstruction. An~original section is visible on the right.}
\label{fig:sinus-cut}
\end{figure}

\begin{figure}[tb]
\centering
\includegraphics[width=1\linewidth]{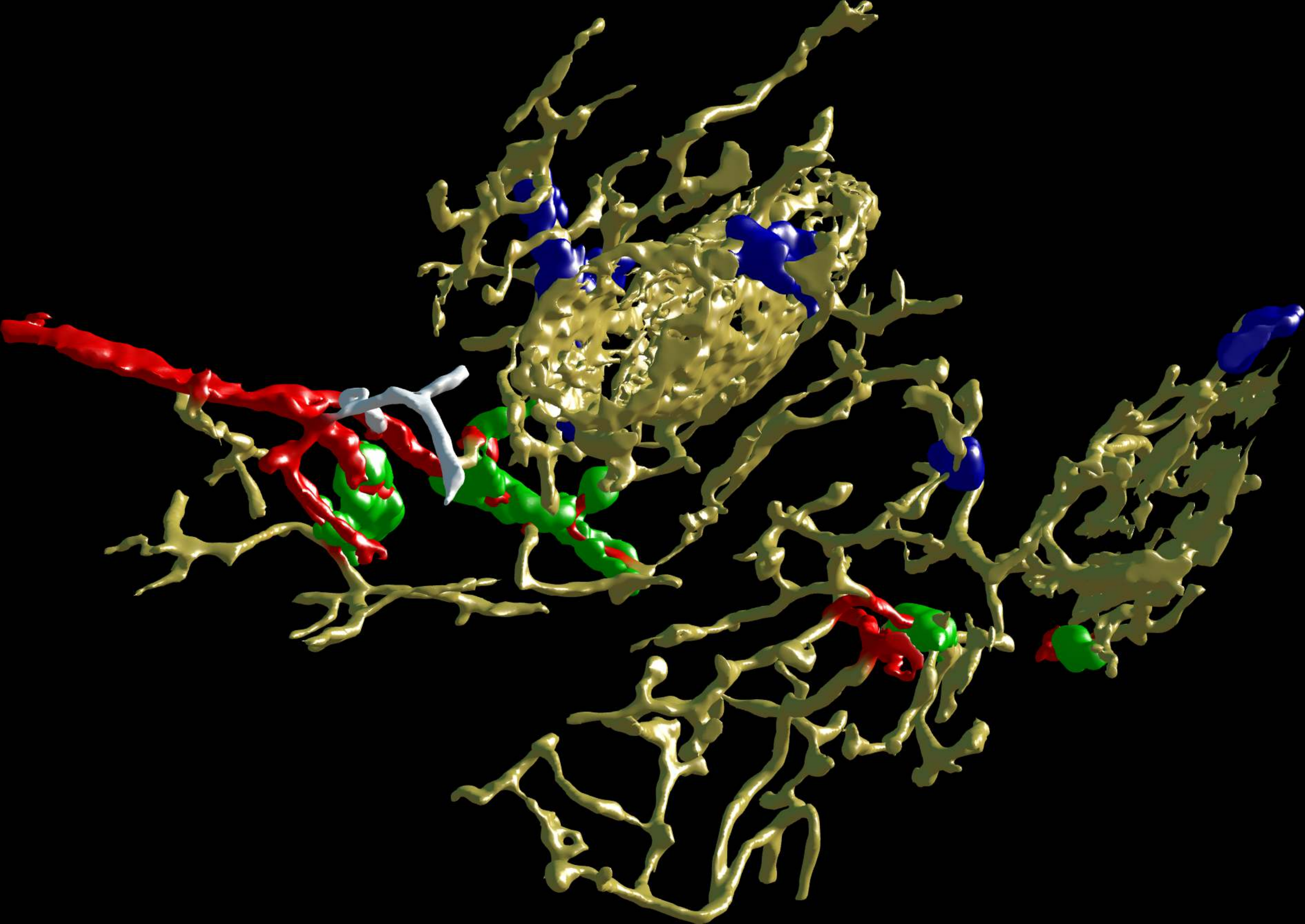}

\caption{Final result of mesh annotation and painting, \enquote{red pulp} data set \asheath. Blood vessels are yellow. Certain support structures in the spleen that feature smooth muscle actin are also reconstructed and displayed in yellow. (A~trained histologist can discern these structures from various kinds of blood vessels though.)  Some of the blood vessels (red) lead from larger blood vessels (arterioles) to capillary sheaths (green). Some sheaths are fed by arterioles not traced in the reconstruction. These sheaths are marked blue. Finally, while some capillaries are red (having green sheaths), some other capillaries, coming from the same arteriole, do not have a sheath at all. Such capillaries are coloured in white. The background is black to better discern the white colour.\newline
A~similar, but different image appeared in \asheath under CC-BY 4.0 license.}
\label{fig:sheaths}
\end{figure}

\begin{figure}[tbp]
\centering
\ifthenelse{\boolean{review} \OR \boolean{arxivpreprint}}{%
\subfloat[][]{\label{fig:complex-150-a}\includegraphics[width=0.7\linewidth]{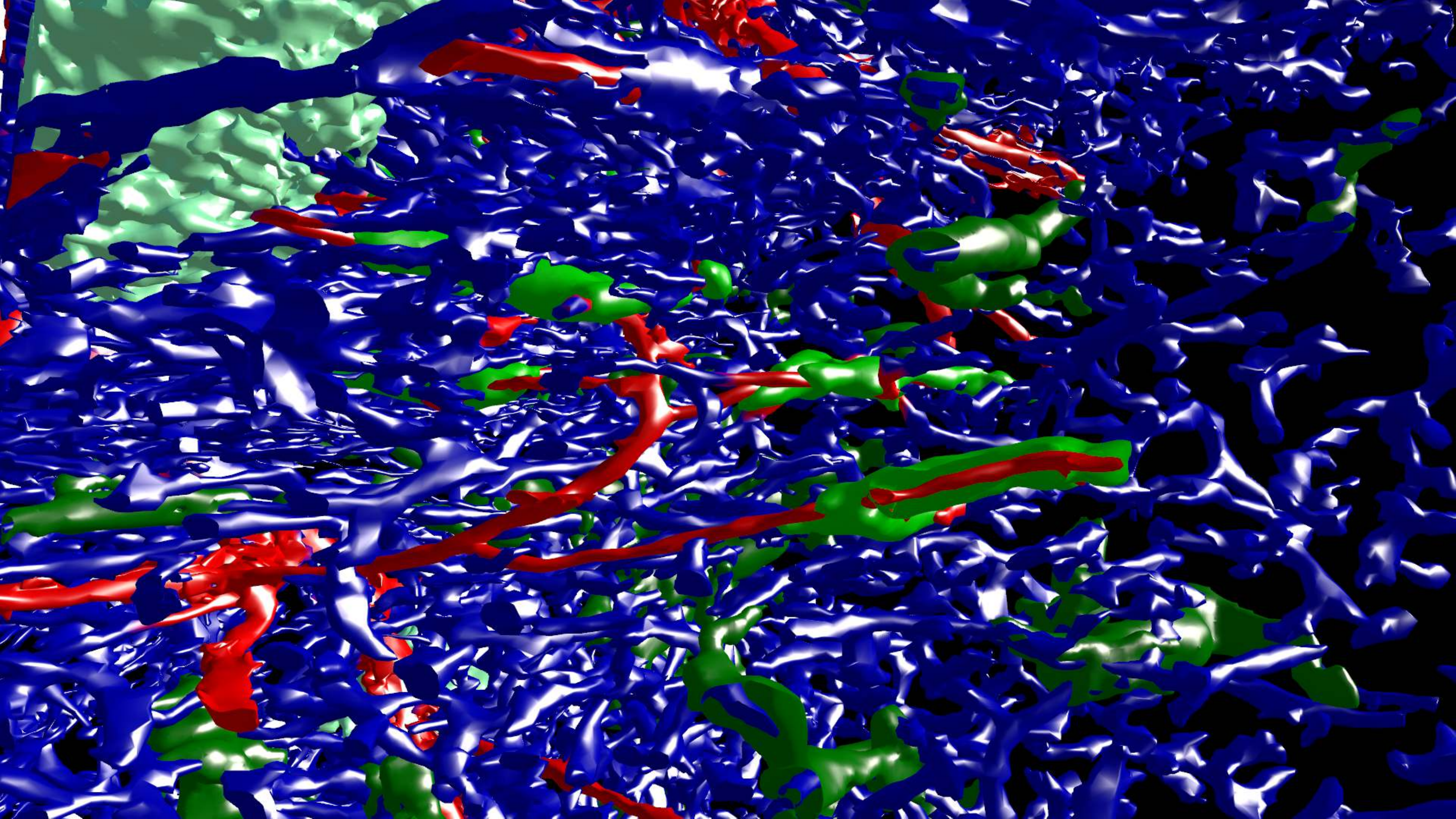}}

\subfloat[][]{\label{fig:complex-150-b}\includegraphics[width=0.7\linewidth]{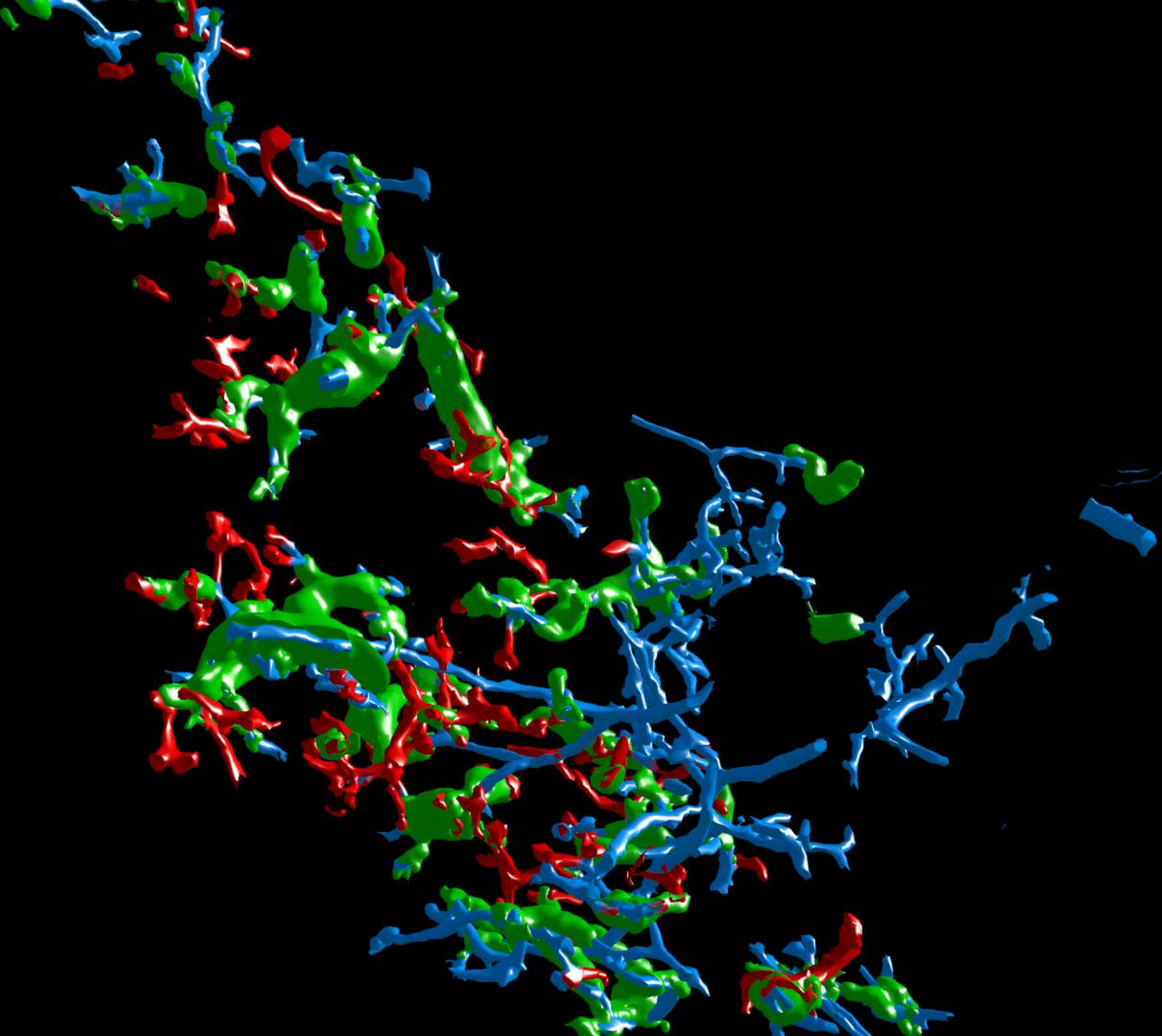}}
}{%
\subfloat[][]{\label{fig:complex-150-a}\includegraphics[width=\linewidth]{images/new/complex_1574536398.pdf}}

\subfloat[][]{\label{fig:complex-150-b}\includegraphics[width=\linewidth]{images/new/complex_1574537602.pdf}}
}
\caption{The natural complexity of the \enquote{sheaths alternating} data set \citep{steiniger_150} is very high. With mesh annotation, painting, and removal of irrelevant details we were able to keep the complexity at a tolerable level.\newline
\protect\subref{fig:complex-150-a}: The capillary network of the splenic red pulp is blue. Arterioles have been highlighted in red. Capillary sheaths are green and dark green, depending on whether they belong to the red vessels. Special supportive cells in a follicle are light green. An~arteriole (red) is entering from left, one of the branches is splitting up into capillaries (still red) that immediately enter capillary sheaths. One such sheath (green) is is cut open, showing the capillary inside.\newline
\protect\subref{fig:complex-150-b}: Arterioles and sheathed capillaries are light blue, capillary sheaths are green. The open-ended side branches of sheathed capillaries are red. This figure shows a single system, starting with an arteriole. It has been separated from other arterial vessel systems in the surroundings. Front plane clipping opens the capillary sheath and shows the blue capillary inside.  We see some open green capillary sheaths with light blue \enquote{main line} blood vessels inside.
\newline
Similar, but different figures can be found in \citet{steiniger_150}.}
\label{fig:complex-150}
\ifthenelse{\boolean{review} \OR \boolean{arxivpreprint}}{%
\thisfloatpagestyle{empty}
}{}
\end{figure}

\begin{figure}[tb]
\centering
\includegraphics[width=\linewidth]{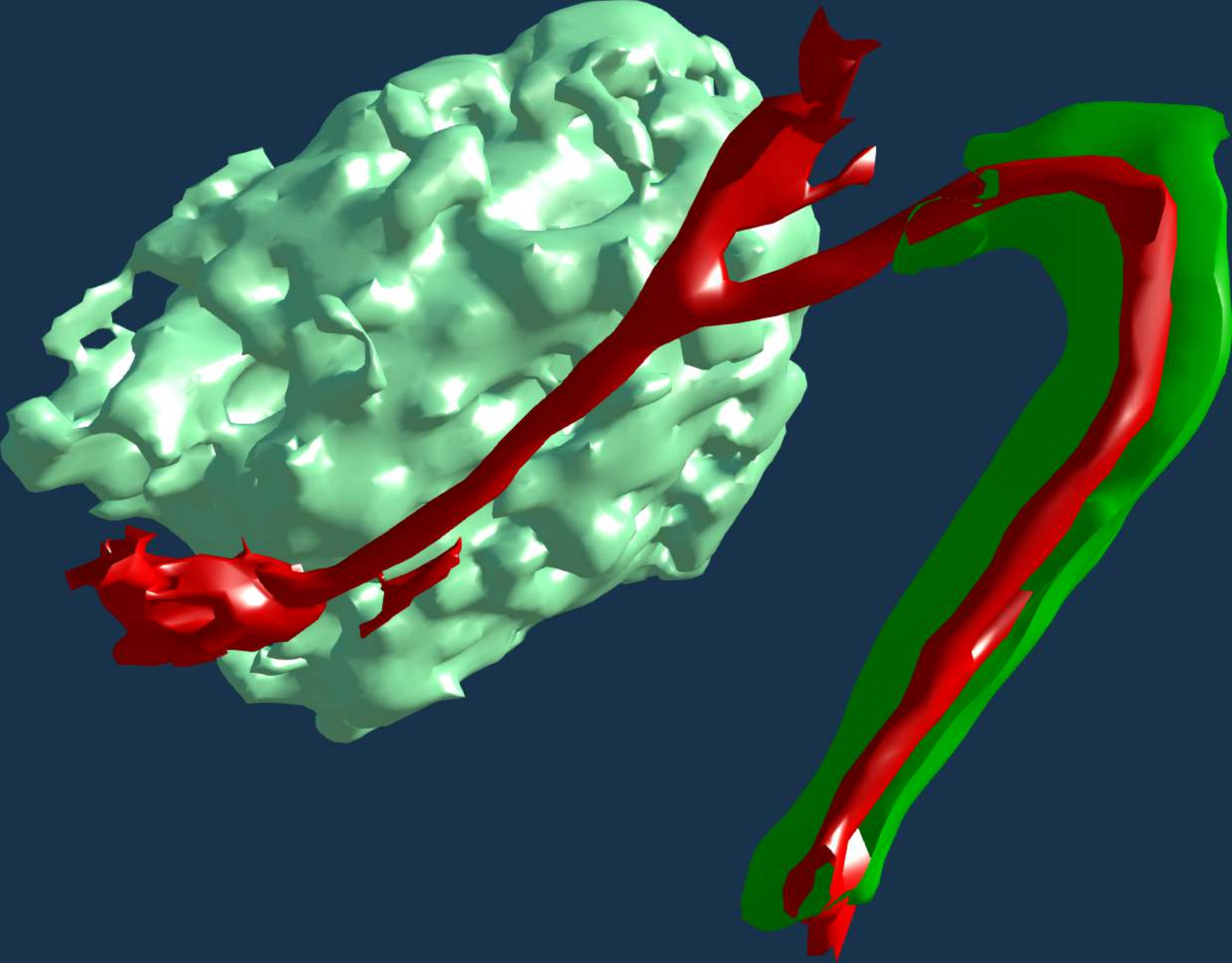}
\caption{Final result of mesh annotation, painting, and removal of irrelevant details. The complexity is now greatly reduced. This is the \enquote{sheaths alternating} data set \citep{steiniger_150}. The meshes are cut open by the front clipping plane. Blood vessels are red, capillary sheaths are green, cells in follicle are light green. An~arteriole (red) is entering from left in the proximity of a follicle, this arteriole splits further, one of the branches is splitting up into capillaries (still red) that immediately enter capillary sheaths. One such sheath (green) is curved around the follicle. The sheath is cut open, showing the capillary inside.\newline
A~similar, but different figure, depicting a different sheath, can be found in \citet{steiniger_150}.}
\label{fig:sep-150}
\end{figure}

\begin{enumerate}
\def\labelenumi{\arabic{enumi}.}
\tightlist
\item
  \enquote{Bone marrow}: stained with anti-CD34 plus anti-CD141 (both
  brown), 4~ROI \(3500\times 3500\)~pixel at
  \SI{0.5}{\micro\meter\per\pixel}, 21 serial sections (\citet{km16own};
  Fig.~\ref{fig:4}, \ref{fig:volren}).
\item
  \enquote{Follicle-double}: spleen sections stained with anti-CD34
  (brown) followed by anti-CD271 (blue), ROI \(2300\times 2300\) pixel
  at \SI{0.416}{\micro\meter\per\pixel}, 24~serial sections
  (\citet{steiniger_capillary_2018}; Fig.~\ref{fig:5});
\item
  \enquote{Follicle-single}: spleen sections stained with anti-CD34
  (brown), a ROI \(4k \times 4k\) pixel,
  \SI{0.3}{\micro\meter\per\pixel}, 24 serial sections
  (\citet{steiniger_capillary_2018}; Fig.~\ref{fig:6});
\item
  \enquote{Red pulp}: spleen sections stained with anti-CD34 plus
  anti-SMA (both brown), followed by anti-CD271 (violet-blue, different
  pigment than above), 11~ROI \(2k\times 2k\) pixel at
  \SI{0.5}{\micro\meter\per\pixel}, 24~serial sections
  (\citet{steiniger_locating_2018}; Figs.~\ref{fig:8},
  \ref{fig:sheaths});
\item
  \enquote{Sheaths alternating}: 148~immunostained sections plus 2~HE
  sections stained with anti-SMA, anti-CD34 and anti-CD271. In every
  other section CD271 was replaced by CD20. We processed 4~ROI at
  \(2k\times 2k\) pixel at \SI{0.5}{\micro\meter\per\pixel}, of which
  from 82 to 148~sections were used
  \citep{poster, steiniger_150, lobachev_tempest_2020}, see also
  Figs.~\ref{fig:2}, \ref{fig:7}, \ref{fig:clipping},
  \ref{fig:complex-150}, \ref{fig:sep-150}.
\item
  \enquote{Sinus}: 21~spleen sections immunostained with anti-CD141
  (brown, not shown here), anti-CD34 (blue), anti-CD271 (red), work in
  progress with currently two ROI \(2k \times 2k\) pixel at
  \SI{0.44}{\micro\meter\per pixel}, shown in Figs.~\ref{fig:sinus-cut},
  \ref{fig:sinus-geo}. A~phenotypical investigation was performed in
  \citet{steiniger_phenotypic_2007}, but without serial sections and
  capillary sheaths.
\end{enumerate}

\hypertarget{system-architecture-and-features}{%
\section{System architecture and
features}\label{system-architecture-and-features}}

\hypertarget{components}{%
\subsection{Components}\label{components}}

\label{sec:comp}

Our application makes use of existing software libraries. We load meshes
with the VCG library. Multiple meshes with vertex colours are supported.
We utilise Open~GL~4.2. We enable back face culling, multisampling, and
mipmaps. The section textures are loaded with the FreeImage library. The
Steam~VR library is used for interaction with Vive controllers and the
headset.

With respect to hardware, the system consists of a desktop computer with
a Steam~VR-capable GPU and a Steam~VR-compatible headset with
controllers.

\hypertarget{controls}{%
\subsection{Controls}\label{controls}}

For control, a simple keyboard and mouse interface (for debugging
outside VR), XBox One controller, and Steam~VR-compatible controllers
can all be used. Our initial idea was to use an XBox~360 or an XBox One
controller, as such controllers provide a simple control metaphor.
However, the expert users were not acquainted to gaming controllers and
could not see the XBox One controller in VR. Thus, initial error rates
were high when they \eg tried to simultaneously use an \enquote{X} key
and a \enquote{D-Pad} in blind.

Hence, a more intuitive approach with the native Vive controllers was
targeted. We have kept the keyboard-and-mouse and the XBox controller
options, but duplicated required input actions with Vive controllers.
Native HTC Vive controllers proved their benefits. Although the
metaphors were much more complicated, the intuitive control payed off
immediately. Further, the visibility of the tracked controllers in VR
helped a lot. Later on, we extended the application support to Valve
Index knuckle controllers.

Further spectators can follow the domain expert from \enquote{outside}
of the immersion, as the HMD feed is mirrored on a monitor. Further,
annotations significantly improve communication
(Figs.~\ref{fig:5}--\ref{fig:8}). The main controller actions of the
domain expert are:

\begin{itemize}
\tightlist
\item
  Blending the original data (the stained sections) in or out;
\item
  Blending the mesh(es) in or out;
\item
  Advancing the currently displayed section of original data;
\item
  Placing an annotation;
\item
  Mesh painting.
\end{itemize}

The most significant user interaction happens with intuitive movements
of the immersed user around (and through) the displayed entities in VR.

\hypertarget{communication}{%
\subsection{Communication}\label{communication}}

Without a beacon visible in VR it is almost impossible to understand
what the expert tries to show. With a VR controller and our annotation
tool, interesting areas in the visualisation can be shown to the outside
spectators in real time.

\hypertarget{annotation-markers}{%
\subsection{Annotation markers}\label{annotation-markers}}

We designed a spherical selection tool for marking points in space
(Fig.~\ref{fig:7}). The sphere is located at the top front of a Vive or
Index controller and can be seen in the virtual space (and, by proxy,
also in the mirrored display, Fig.~\ref{fig:new-user}). We need to note,
however, that the annotation sphere appears much more vivid to the VR
user than it appears on screenshots. User's movement and live feedback
are in our opinion a major reason for such a difference in perception.
Figures~\ref{fig:4-d}--\ref{fig:4-f}, \ref{fig:7}, \ref{fig:5-b},
\ref{fig:6}, show our annotation tool in images captured from VR
sessions.

The annotations and mesh modifications are saved for further analysis.
For example, after the domain expert has marked suspicious areas, the 3D
reconstruction expert can inspect them in a later VR session.
Reconstruction improvements can be deduced from this information.

\hypertarget{anti-aliasing}{%
\subsection{Anti-aliasing}\label{anti-aliasing}}

If a \enquote{direct} rendering approach is used, there is a very
dominant aliasing effect at certain points. We used multisampling (MSAA)
on meshes and mipmaps on textures to alleviate this problem.

\hypertarget{front-face-culling}{%
\subsection{Front face culling}\label{front-face-culling}}

Consider the interplay between the model and original serial sections.
A~section is not an infinitely thin plane. We show the original data as
an opaque cuboid that is one section thick in the \(z\) direction and
spans over the full surface in the \(xy\) plane. The actual data points
of the mesh, corresponding to the displayed section, are \emph{inside}
the opaque block. Decisive parts of the mesh are occluded by the front
face of the cuboid. On the one hand, this is, of course, not desired,
and requires correction. On the other hand, the behaviour of the model
spanning for multiple sections \emph{in front} of the current section is
best studied when it ends at the front face of the cuboid. The solution
is to enable or disable front face culling of the original data display
at will.

With front face culling enabled, the user can look inside the opaque
block with original section texture. This is well suited for the
inspection of lesser details and small artefacts. (Figure~\ref{fig:8},
\subref{fig:8-d}, \subref{fig:8-e} features a real-life example, observe
the edge of the section shown.) The general behaviour of the model
across multiple sections can be tracked more easily with front faces of
the original section on display. The presence of both representations
accelerates QC.

\hypertarget{geodesic-distances-for-mesh-painting}{%
\subsection{Geodesic distances for mesh
painting}\label{geodesic-distances-for-mesh-painting}}

We also implemented a VR-based mesh painting facility, mostly based on
MeshLab code base \citep{km39}. In this mode the colour spheres, which
our user can place with the controller, produce a geodesically coloured
region on the mesh instead of an annotation. These two functions,
annotations and mesh painting, are conveyed to be clearly different to
the user.

The selected colour is imposed on all vertices inside the geodesic
radius from the centre of the sphere. We would like to paint on, for
example, a part of a blood vessel that has a specific property. At the
same time, we would like \emph{not} to colour other blood vessels that
might be inside the painting sphere, but are not immediately related to
the selected blood vessel. This is facilitated with geodesic distances,
as Figures~\ref{fig:geo} shows.

The markings from mesh painting lead to the final separation of the
entities (such as blood vessel types, kinds of capillary sheaths, etc.)
in the visualisation.

\hypertarget{front-plane-clipping}{%
\subsection{Front plane clipping}\label{front-plane-clipping}}

\label{sec:method-clipping}

The classic view frustum in computer graphics consists of six planes,
four \enquote{display edges} building the frustum sides, a back plane
(the farthest visible boundary) and the front plane, the closest
boundary. The clipping planes are recomputed when the camera (\ie the
user) is moving. In the cases when there are too many self-occluding
objects in the scene, the observer cannot \enquote{pierce through}
further than few closest objects. In other words, the observer can only
see the closest objects. (This fact motivates occlusion culling.)

Such an occlusion was the case with our denser data sets. With a simple
change in the code, we moved the front plane of the view frustum further
away, in an adjustable manner. Basically, the user \enquote{cuts} parts
of the reconstruction in front of their eyes, allowing for the detailed
inspection of the \emph{inside} of the reconstruction.

This adjustment is very minor from the computer graphics point of view,
but it was very much welcomed by our actual users, the medical experts.
With appropriate front plane clipping set at about \SI{60}{\centi\meter}
from the camera, it becomes possible to inspect very dense medical data
sets from \enquote{inside}. (Figs.~\ref{fig:7}, \ref{fig:clipping},
\ref{fig:complex-150}, \ref{fig:sep-150} demonstrate this effect.) The
user \enquote{cuts away} the currently unneeded layers with their
movements.

\hypertarget{results}{%
\section{Results}\label{results}}

\hypertarget{hardware}{%
\subsection{Hardware}\label{hardware}}

We conducted most of our investigations on a 64-bit Intel machine with
i7-6700K CPU at \SI{4}{GHz}, \SI{16}{GB} RAM, and Windows~10. We used
NVidia GTX~1070 with \SI{8}{GB} VRAM, and HTC Vive.

Our VR application was initially developed with HTC Vive in mind; it
performed well on other headsets, such as HTC Vive Pro Eye and Valve
Index. We observed convincing performance on Intel i7-9750H at
\SI{2.6}{GHz}, \SI{64}{GB} RAM (MacBook Pro~\mbox{16\hspace{1pt}$''$)}
and NVidia RTX~2080 Ti with \SI{11}{GB} VRAM in Razor Core~X eGPU with
HTC Vive Pro Eye, as well as on AMD Ryzen 2700X, \SI{32}{GB} RAM, NVidia
RTX~2070 Super with \SI{8}{GB} VRAM with Valve Index. Our application
also should perform well with further headsets such as Oculus Rift. It
was possible to use previous-generation GPUs, we also tested our
application with NVidia GTX~960. Overall, it is possible to work with
our application using an inexpensive setup.

The largest limitation factor seems to be the VRAM used by the
uncompressed original image stack. The second largest limitation is the
number of vertices of the visualised meshes and the rasteriser
performance in case of very large, undecimated reconstructions.

\hypertarget{bone-marrow-data-set}{%
\subsection{\texorpdfstring{\enquote{Bone marrow} data set
}{``Bone marrow'' data set }}\label{bone-marrow-data-set}}

We have reconstructed the 3D shape of smaller and larger microvessels in
hard, undecalcified, methacylate-embedded human bone serial sections
\akm. Shape diameter function on the reconstructed mesh allows to
distinguish capillaries from sinuses. Figure~\ref{fig:4} shows
\subref{fig:4-a} a single section (part of the input data to
reconstruction), \subref{fig:4-b} a volume rendering of all 21 sections,
and \subref{fig:4-c} our 3D reconstruction. In \citet{km16own} we did
not use VR. Here, we use the same data set to showcase some features of
our VR-based method. It took us much more manpower and time to validate
the reconstructions then, without VR, as Section~\ref{sec:impact}
details (Fig.~\ref{fig:volren}).

The process of annotation is demonstrated in
Fig.~\ref{fig:4},~\subref{fig:4-d}. The next subfigures show further
investigation of the annotated area in VR either in combined
mesh-section view~\subref{fig:4-e}, or showing the corresponding section
only~\subref{fig:4-f}. To discriminate between capillaries (smaller,
coloured red) and sinuses (larger, coloured green), we computed shape
diameter function on the reconstructed meshes and colour-coded resulting
values on the mesh, as shown in \subref{fig:4-c}--\subref{fig:4-e}. The
handling of the reconstruction and serial section data in VR showcases
the annotation process.

\hypertarget{follicle-double-data-set}{%
\subsection{\texorpdfstring{\enquote{Follicle-double} data set
}{``Follicle-double'' data set }}\label{follicle-double-data-set}}

The human spleen contains accumulations of special migratory
lymphocytes, the so-termed follicles. We reconstructed the capillaries
inside and outside the follicles \afoll. We show some results from this
work in this section and in the next. Fig.~\ref{fig:5} presents one of
three ROIs that were quality controlled.

Our 3D reconstruction demonstrates that follicles are embedded in a
superficial capillary meshwork resembling a basketball basket.
Figure~\ref{fig:5} shows that our VR tool enables easy annotation and
projection of the original data leading to further results \afoll. In
Fig.~\ref{fig:5}, \subref{fig:5-e}, some brown dots have been marked
inside a follicle. The 3D model shows, that the dots indeed represent
capillaries cut orthogonally to their long axis. Thus, we additionally
find that infrequent capillaries also occur inside the follicles. The
superficial capillary network of the follicles is thus connected to very
few internal capillaries and to an external network of capillaries in
the red pulp. We observed the latter two networks to have a shape which
is totally different from the superficial follicular network. The
external network is partially covered by capillary sheaths stained in
blue colour. In total, we examined three \enquote{follicle-double} data
sets in VR.

\hypertarget{follicle-single-data-set}{%
\subsection{\texorpdfstring{\enquote{Follicle-single} data
set}{``Follicle-single'' data set}}\label{follicle-single-data-set}}

To continue the investigation of capillaries inside and outside the
follicles, Fig.~\ref{fig:6} shows that the annotated elongated
structures in the follicles and in the T-cell zone at least partially
belong to long capillaries, which accompany the outside of larger
arteries, so-termed \emph{vasa vasorum}. With our VR-based method, we
investigated this \(4k \times 4k\) ROI at
\SI{0.3}{\micro\meter\per\pixel} and three further ROIs (not shown) with
\(1600 \times 1600\) pixels at \SI{0.6}{\micro\meter\per\pixel} \afoll.
Fig.~\ref{fig:6} also shows a Vive controller tracing one of the longer
capillaries with the annotation ball as a form of communication of the
findings to spectators outside VR.

\hypertarget{red-pulp-data-set}{%
\subsection{\texorpdfstring{\enquote{Red pulp} data
set}{``Red pulp'' data set}}\label{red-pulp-data-set}}

The location of capillary sheaths in human spleens has not been
clarified in detail until recently~\asheath. Our 3D reconstructions
indicate that sheaths primarily occur in a post-arteriolar position in
the part of the organ, which does not contain lymphocyte accumulations
(so-termed red pulp), although length and diameter of the sheaths are
variable. Many sheaths are interrupted by the boundaries of the ROI.
(The remedy was a longer series of sections, as presented in
Section~\ref{sec:explain-150}.) For this reason it makes sense to
collect only sheaths which are completely included in the
reconstruction. Such a selection was done with our VR classification
tool.

Figure~\ref{fig:8}, \subref{fig:8-a} shows an overview of the
annotations. In Figs.~\ref{fig:8}, \subref{fig:8-b}--\subref{fig:8-d} it
becomes clear, that the sheaths indeed end at the marked positions.
Notice the enabled front face culling on the section cuboid in the
closeups. Figure~\ref{fig:8}, \subref{fig:8-e} additionally shows the
reconstructed meshes for the sheaths. We show a single ROI at
\(2k\times 2k\) pixels. We have inspected 11 such ROIs in VR.

\hypertarget{sheaths-alternating-data-set-and-clipping}{%
\subsection{\texorpdfstring{\enquote{Sheaths alternating} data set and
clipping}{``Sheaths alternating'' data set and clipping}}\label{sheaths-alternating-data-set-and-clipping}}

\label{sec:explain-150}

The \enquote{sheaths alternating} data set with up to 150 sections was
created to further investigate the morphology and (to some extent) the
function of capillary sheaths \citep{steiniger_150}. The resulting 3D
data set was extremely dense. The increased amount of \enquote{channels}
and the nature of the study (tracking the blood vessels) was a big
challenge. The amount of the reconstructed blood vessels and their
self-occlusion prohibited any possible insight when viewing them from
the outside. Here we utilised front plane clipping
(Section~\ref{sec:method-clipping}). Figures~\ref{fig:7-b},
\ref{fig:clipping}, \ref{fig:complex-150} (and also
Fig.~\ref{fig:sinus-cut} for \enquote{sinus} data set) showcase this
minor, but important adjustment. Figs.~\ref{fig:complex-150},
\ref{fig:sep-150} further demonstrate the complexity of the
\enquote{sheaths alternating} data set.

\hypertarget{mesh-painting-and-obtaining-insights}{%
\subsection{Mesh painting and obtaining
insights}\label{mesh-painting-and-obtaining-insights}}

\label{sec:geo}

As already seen in Figs.~\ref{fig:7}, \ref{fig:5}, \subref{fig:5-a},
\subref{fig:5-b}, \ref{fig:6}, we can point to anatomical structures
with the Valve Index controller. Similarly, annotations can be placed
and mesh fragments can be painted in different colours. An~example of
real-life mesh painting with geodesic distances is in
Figure~\ref{fig:sinus-geo}. The arrows show a part of a structure
already painted by user in red in~\subref{fig:sinus-geo-a}. It is
painted back to blue in~\subref{fig:sinus-geo-b}.

\newcommand{\dofigurecrop}[1]{%
\begin{figure}[#1]
\centering
\subfloat[][]{\label{fig:sinus-geo-a}\includegraphics[width=1\linewidth]{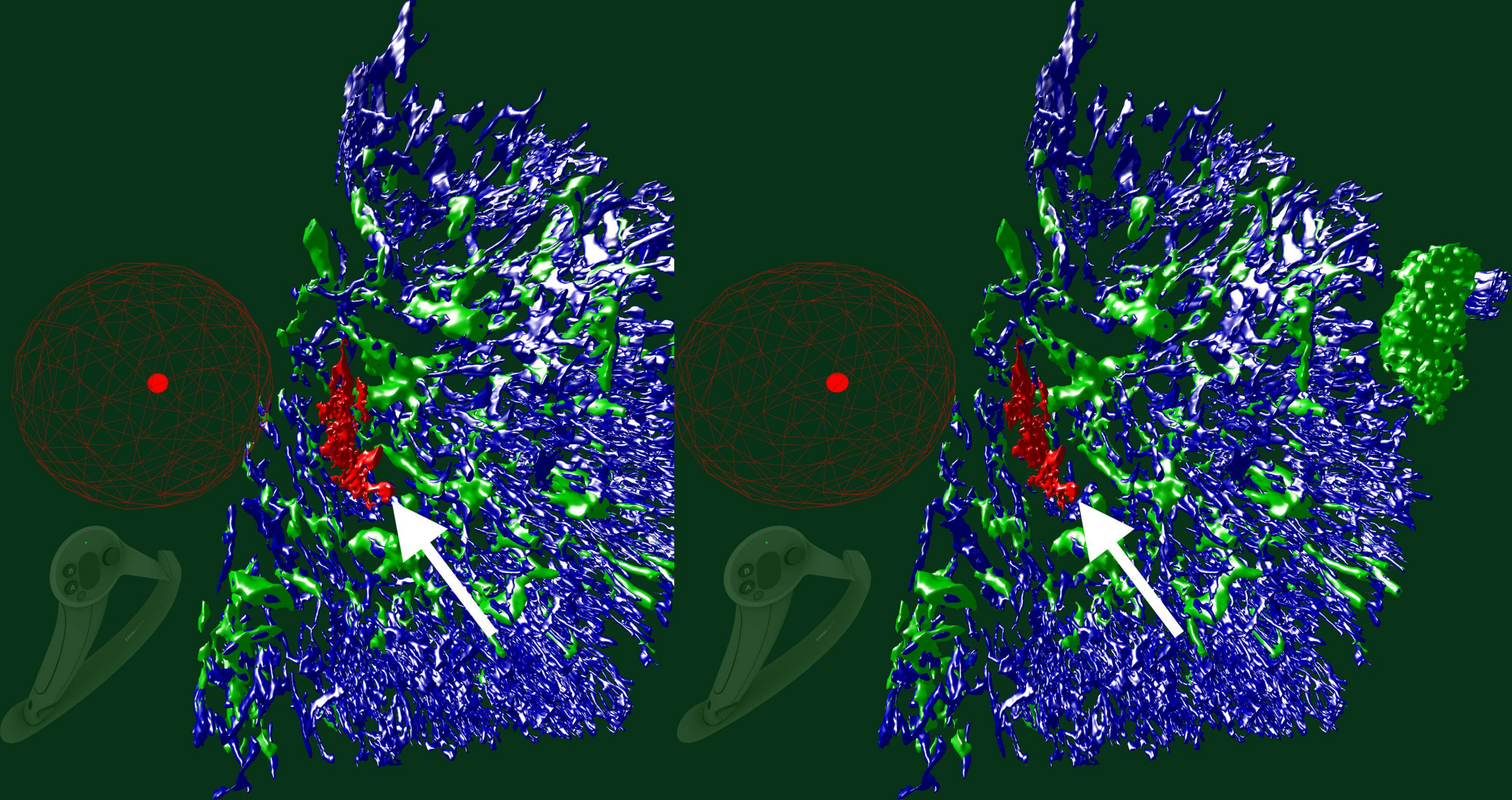}}

\subfloat[][]{\label{fig:sinus-geo-b}\includegraphics[width=0.498\linewidth]{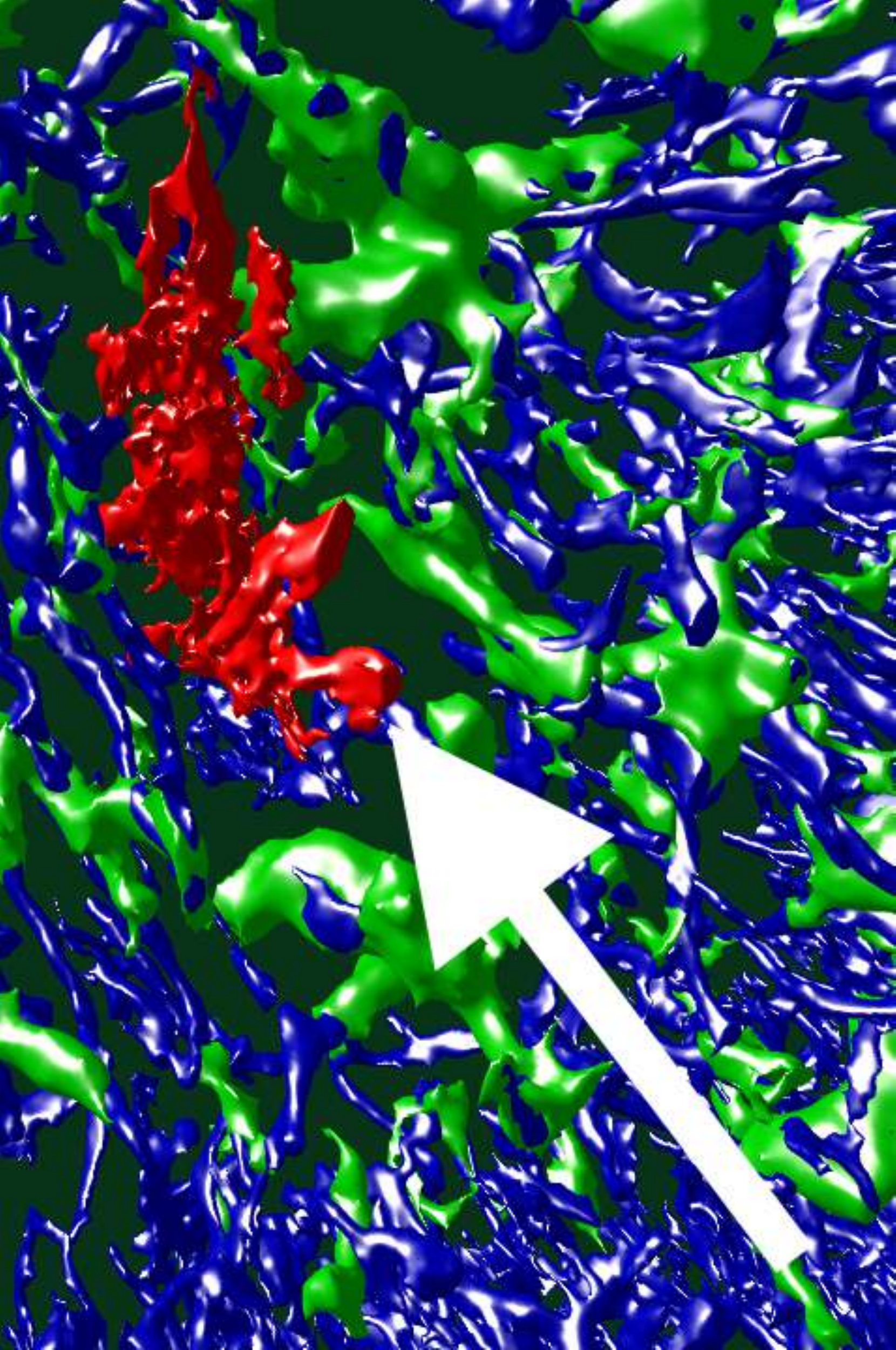}}\hfill%
\subfloat[][]{\label{fig:sinus-geo-c}\includegraphics[width=0.498\linewidth]{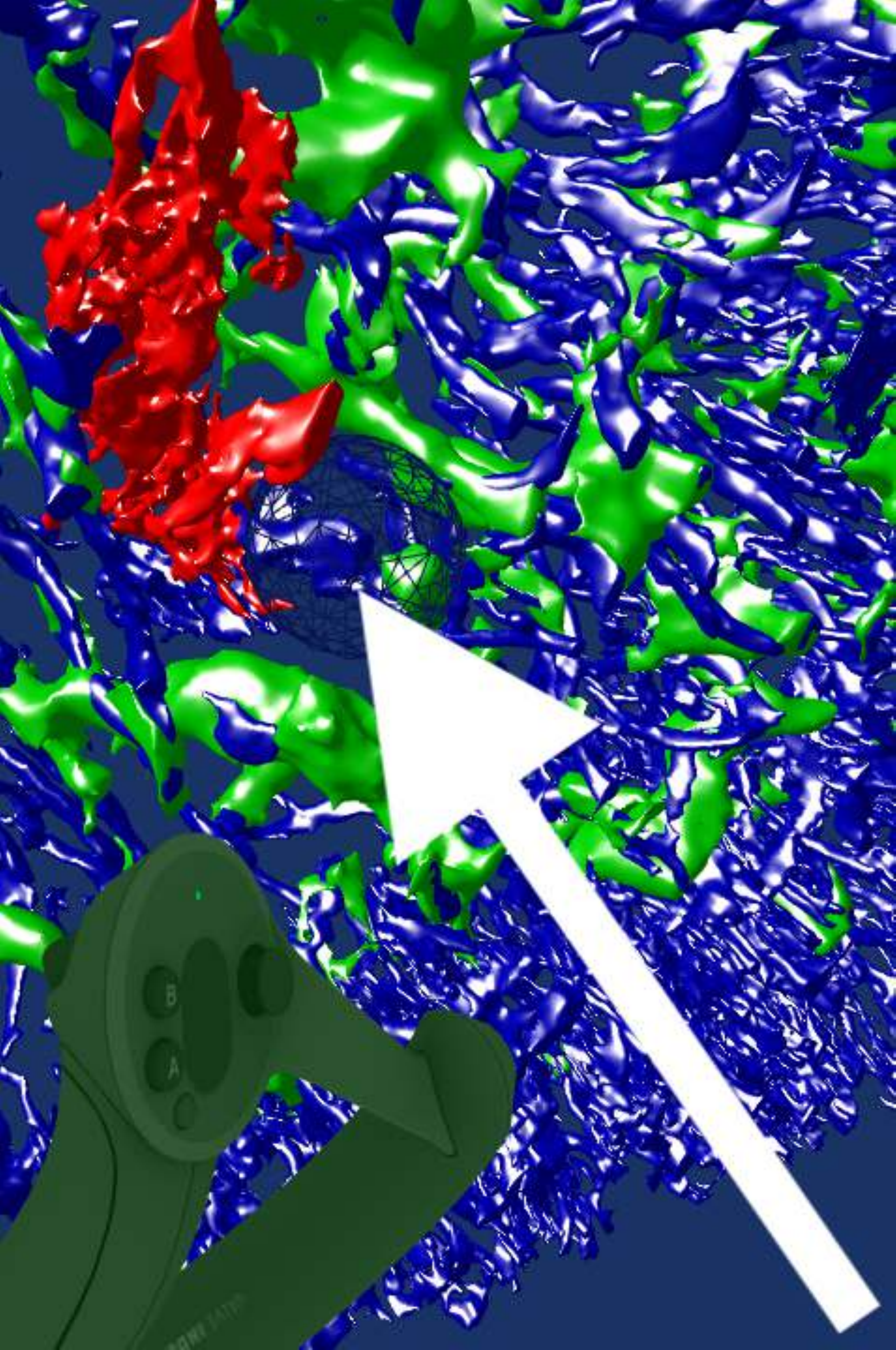}}

\caption{An ongoing session of mesh painting with geodesic distances as VR screenshots. We use the \enquote{sinus} data set. \protect\subref{fig:sinus-geo-a}: Notice the huge annotation ball on the controller, the bright red dot is its centre. This centre is the starting point of the geodesic computation, initiated by the trigger on the controller.
The large radius of the marking tool is bounded by the connectivity: the vertices which are within the radius, but are not connected to the starting point or are \enquote{too far} geodesically, are not painted.\newline
\protect\subref{fig:sinus-geo-b}: For better visibility, we show a crop from the left eye view of \protect\subref{fig:sinus-geo-a}. The white arrow shows a point of interest.\newline
\protect\subref{fig:sinus-geo-c}: An~excessive marking (white arrow), is removed with a repeated painting operation. On the bottom left in \protect\subref{fig:sinus-geo-a}, \protect\subref{fig:sinus-geo-c} a Valve Index controller is visible. The background colour in this figure signifies the highlighting mode used.}
\label{fig:sinus-geo}
\ifthenelse{\boolean{review} \OR \boolean{arxivpreprint}}{%
\thisfloatpagestyle{empty}
}{}
\end{figure}
}
\ifthenelse{\boolean{review} \OR \boolean{arxivpreprint}}{%
\begin{figure}[p]
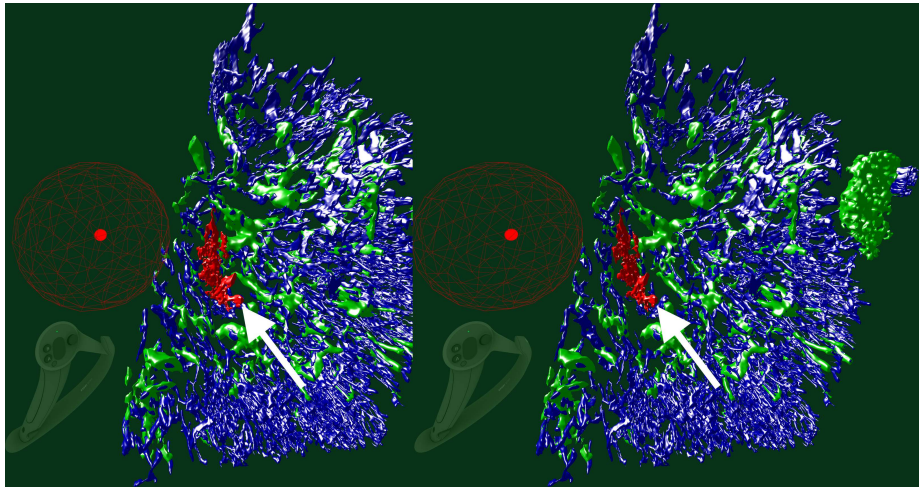
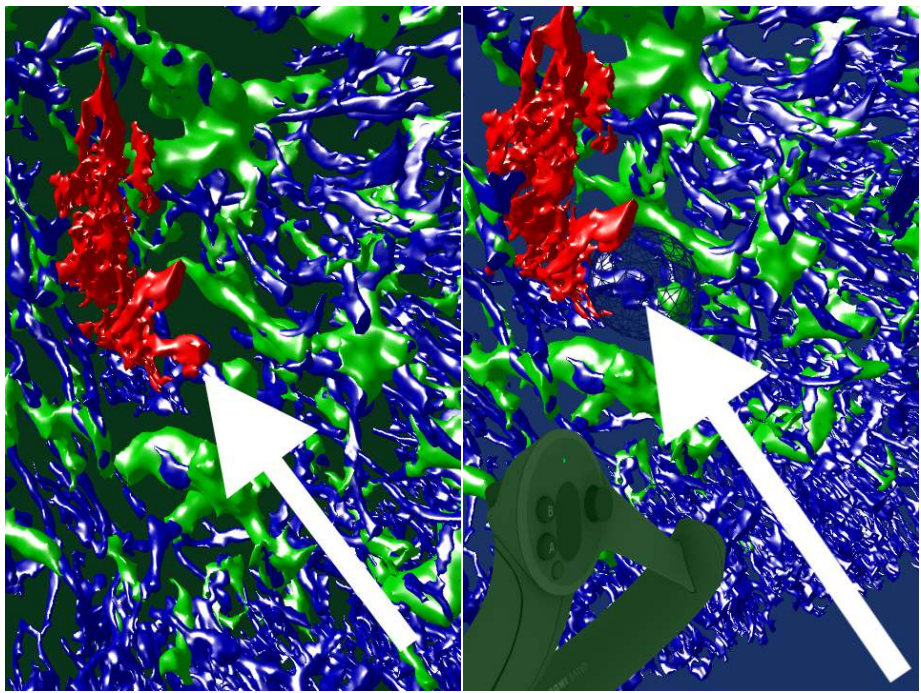

\centering
\subfloat[][]{\label{fig:sinus-geo-a}\includegraphics[width=1\linewidth]{images/sinus/geo_2020-07-31-PM_08_21_16-mark.pdf}}

\subfloat[][]{\label{fig:sinus-geo-b}\includegraphics[width=0.498\linewidth]{images/sinus/geo_2020-07-31-PM_08_21_16-mark+crop.pdf}}\hfill%
\subfloat[][]{\label{fig:sinus-geo-c}\includegraphics[width=0.498\linewidth]{images/sinus/geo_2020-07-31-PM_08_21_42-mark+crop.pdf}}

\caption{An ongoing session of mesh painting with geodesic distances as VR screenshots. We use the \enquote{sinus} data set. \protect\subref{fig:sinus-geo-a}: Notice the huge annotation ball on the controller, the bright red dot is its centre. This centre is the starting point of the geodesic computation, initiated by the trigger on the controller.
The large radius of the marking tool is bounded by the connectivity: the vertices which are within the radius, but are not connected to the starting point or are \enquote{too far} geodesically, are not painted.\newline
\protect\subref{fig:sinus-geo-b}: For better visibility, we show a crop from the left eye view of \protect\subref{fig:sinus-geo-a}. The white arrow shows a point of interest.\newline
\protect\subref{fig:sinus-geo-c}: An~excessive marking (white arrow), is removed with a repeated painting operation. On the bottom left in \protect\subref{fig:sinus-geo-a}, \protect\subref{fig:sinus-geo-c} a Valve Index controller is visible. The background colour in this figure signifies the highlighting mode used.}
\label{fig:sinus-geo}
\ifthenelse{\boolean{review} \OR \boolean{arxivpreprint}}{%
\thisfloatpagestyle{empty}
}{}
\end{figure}

}{%
\begin{figure}[!t]
\centering
\subfloat[][]{\label{fig:sinus-geo-a}\includegraphics[width=1\linewidth]{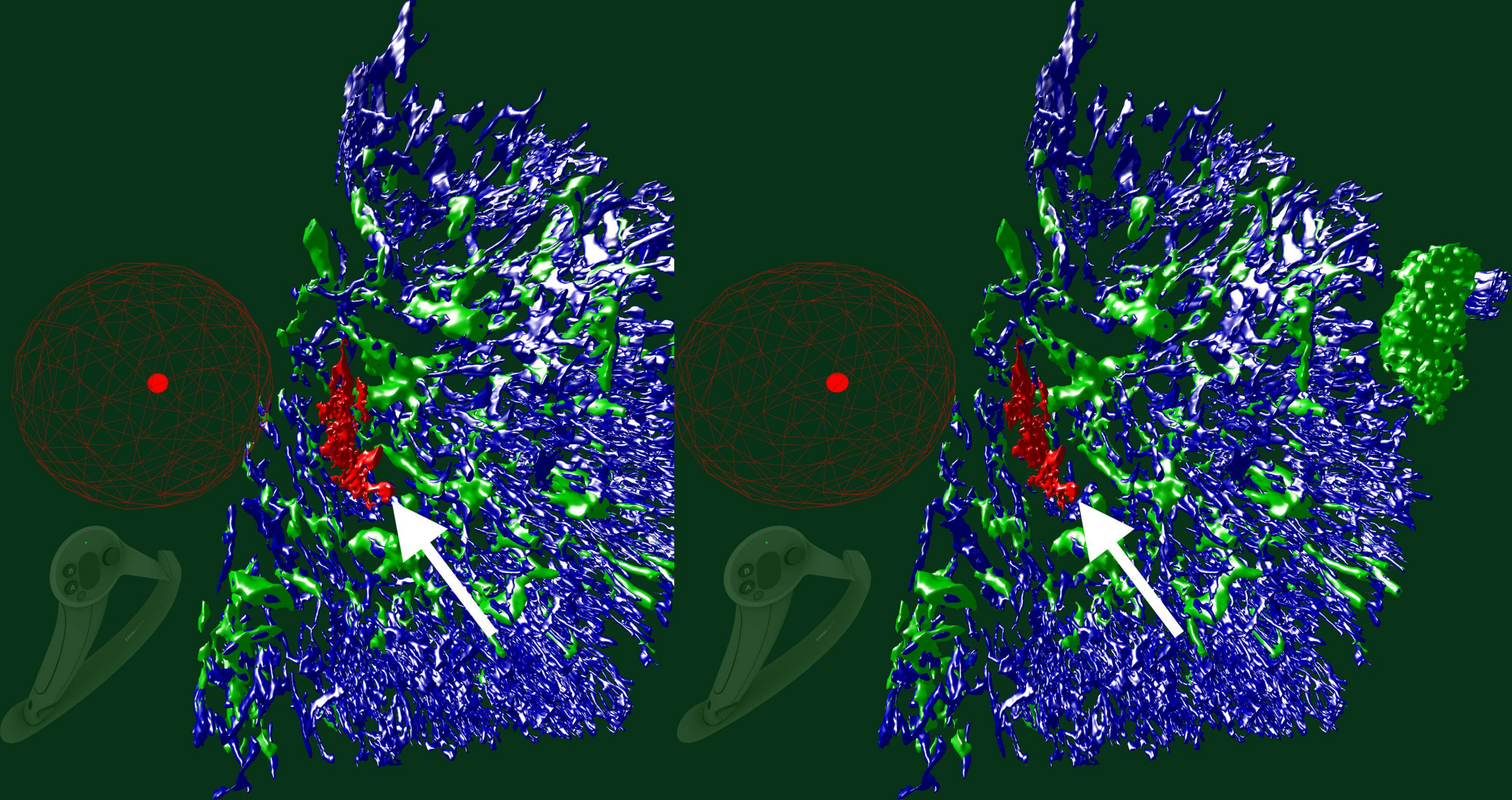}}

\subfloat[][]{\label{fig:sinus-geo-b}\includegraphics[width=0.498\linewidth]{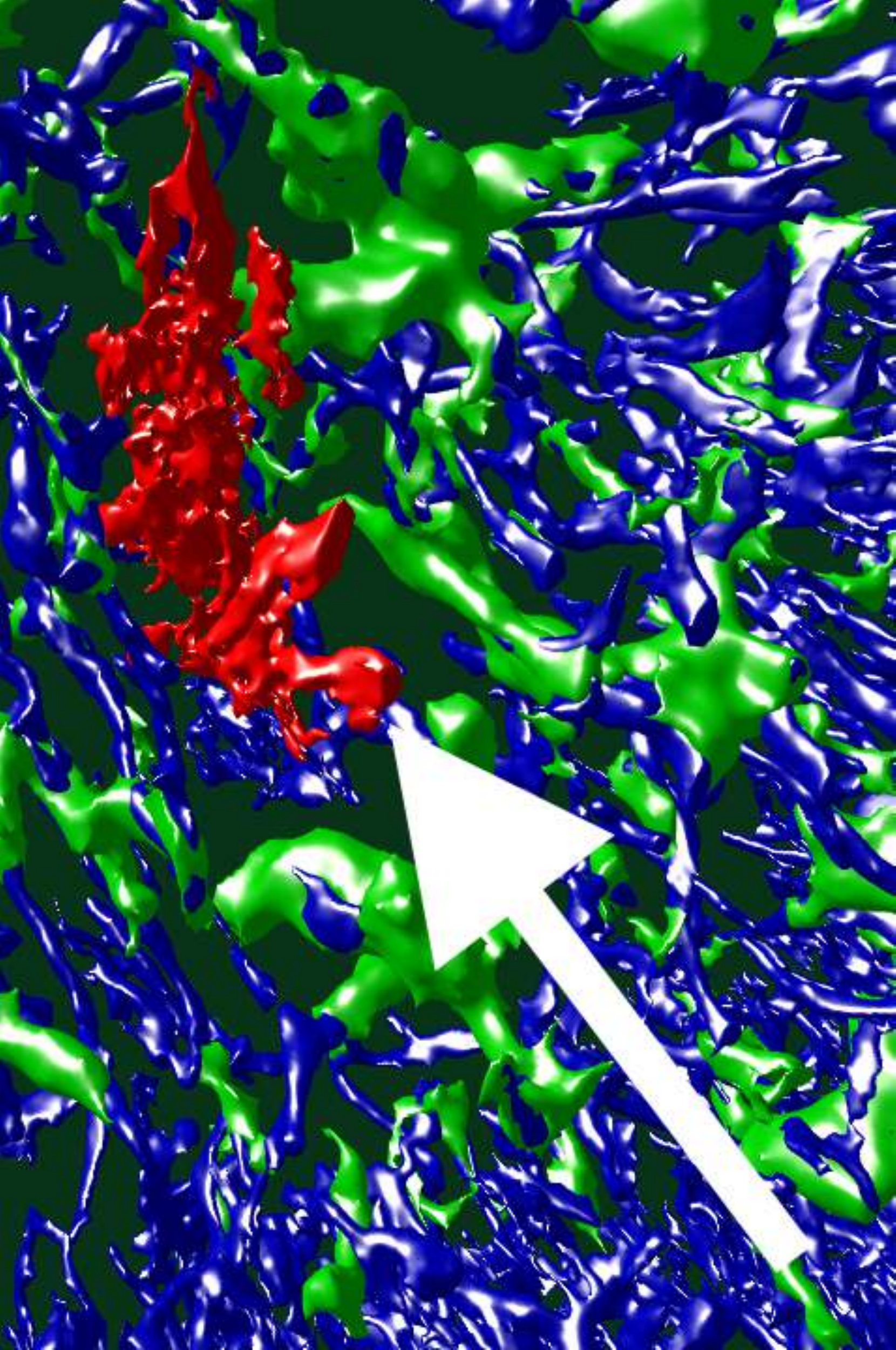}}\hfill%
\subfloat[][]{\label{fig:sinus-geo-c}\includegraphics[width=0.498\linewidth]{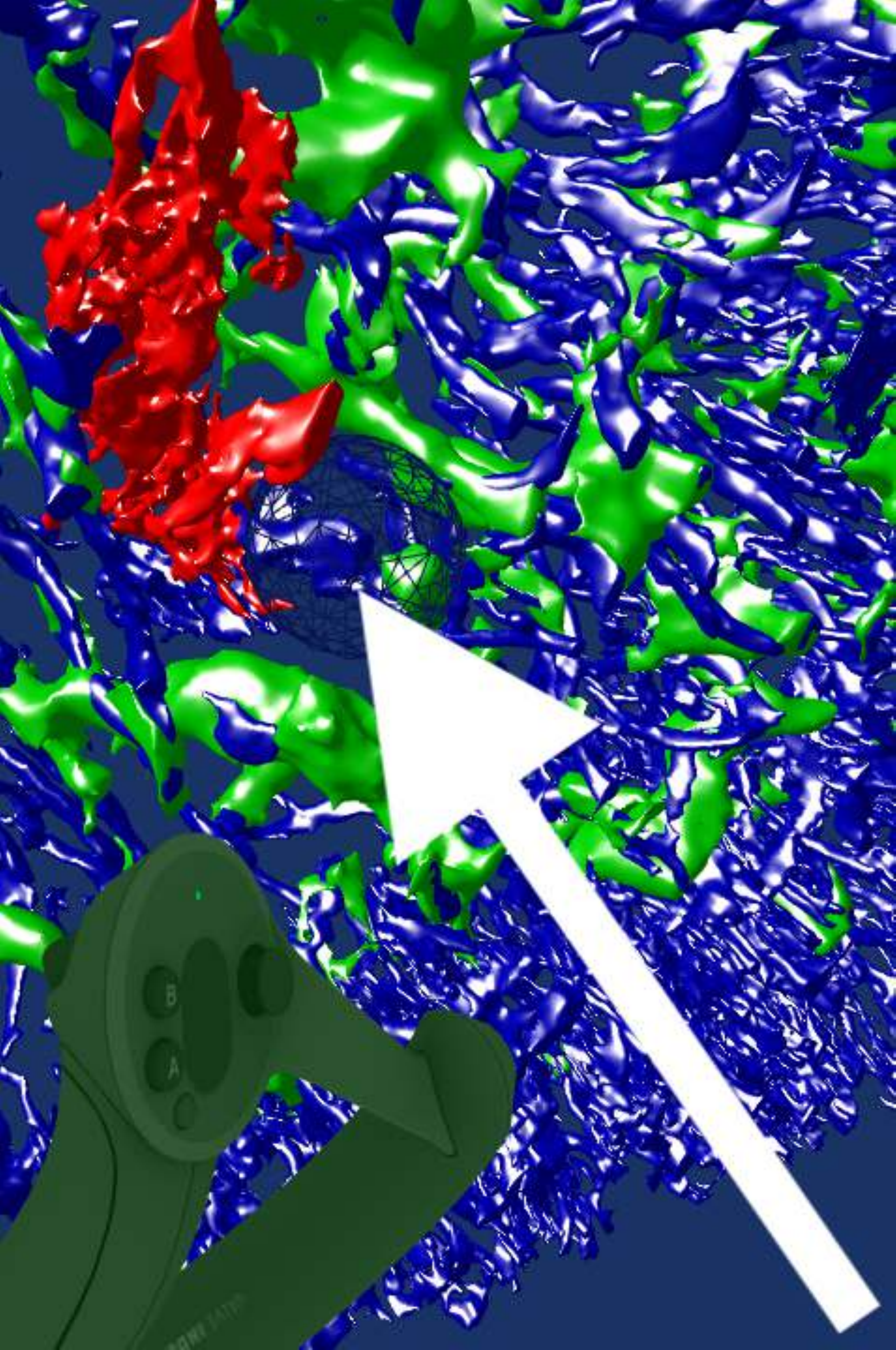}}

\caption{An ongoing session of mesh painting with geodesic distances as VR screenshots. We use the \enquote{sinus} data set. \protect\subref{fig:sinus-geo-a}: Notice the huge annotation ball on the controller, the bright red dot is its centre. This centre is the starting point of the geodesic computation, initiated by the trigger on the controller.
The large radius of the marking tool is bounded by the connectivity: the vertices which are within the radius, but are not connected to the starting point or are \enquote{too far} geodesically, are not painted.\newline
\protect\subref{fig:sinus-geo-b}: For better visibility, we show a crop from the left eye view of \protect\subref{fig:sinus-geo-a}. The white arrow shows a point of interest.\newline
\protect\subref{fig:sinus-geo-c}: An~excessive marking (white arrow), is removed with a repeated painting operation. On the bottom left in \protect\subref{fig:sinus-geo-a}, \protect\subref{fig:sinus-geo-c} a Valve Index controller is visible. The background colour in this figure signifies the highlighting mode used.}
\label{fig:sinus-geo}
\ifthenelse{\boolean{review} \OR \boolean{arxivpreprint}}{%
\thisfloatpagestyle{empty}
}{}
\end{figure}

}

In this manner we have refined an automatic heuristics for arterioles
(larger blood vessels, red in Fig.~\ref{fig:sheaths}) to be always
correct and to lead up to the capillary sheaths.

With a similar tool, working on unconnected components, we changed the
colour of the sheaths in the \enquote{red pulp} and \enquote{sheaths
alternating} data sets. The colour change effected the classification of
the sheaths. The sheaths were initially all blue in our visualisations
of the \enquote{red pulp} data set. Sheaths around capillaries,
following known arterioles, were then coloured green. We also annotated
very few capillaries that \emph{should} have a sheath, but did not
(white). Figure~\ref{fig:sheaths} shows one of the final results, a
connected vascular component with accompanying sheaths, follicle
structures, and smooth muscle actin.

Figure~\ref{fig:complex-150} underlines the complexity of the
\enquote{sheaths alternating} data set. Both images in this figure show
the final results of mesh painting, the actual work is already done.
Still, they convey how interwoven and obscure the original situation is.
CD271\textsuperscript{+} cells, mostly present in capillary sheaths, are
in various shades of green in this figure. Figure~\ref{fig:complex-150}
is highly complex; the fact that something \emph{can} be seen in still
images is the merit of applied annotations and mesh painting, of an
active front plane clipping, and of a proper choice of a view point.
A~viewer in VR has no problem navigating such data sets because of the
active control of view direction and of movement dynamics. The latter
also means a control of the front plane clipping dynamics.

Figure~\ref{fig:sep-150} shows a further development: a separated
arteriole with its further capillary branches and capillary sheaths
\citep{steiniger_150}. The sheath in the figure was cut open by front
plane clipping. This \enquote{separation} was generated from user inputs
in VR, similar to the previous figure. Now, however, the complexity is
reduced to a degree, that allows showing the still image in a medical
research publication.

Summarising, our visualisations in VR were used to obtain insights
\citep{steiniger_locating_2018, steiniger_150} on the position of the
capillary sheaths---a problem that was initially discussed
\citep{Schweigger-Seidel1862} more than 150 years ago!

\hypertarget{discussion}{%
\section{Discussion}\label{discussion}}

\label{sec:impact}

The novelty of this work stems from how VR streamlines and facilitates
better QC and VA of our reconstructions. The presented VR-based tool
plays an important role in our pipeline. \enquote{Usual} 3D
reconstructions from serial histological sections are known, but are
quite rare, because they involve cumbersome tasks. The reasons for this
are threefold: technical difficulties to create them; struggles to QC
the reconstructions; investigation and comprehension problems in dense,
self-occluding reconstructions. A~proper registration, correct
reconstruction options, and possibly also inter-slice interpolation are
necessary for creating a satisfying reconstruction. For QC we need to
visually ensure the correctness of processing, identify what the
reconstruction actually shows, keep artefacts at bay by creating a
better reconstruction if needed. Finding a good reconstruction from
previously unexplored data with a new staining is an iterative process.
While we create the reconstructions quite efficiently, QC was a lot of
work in the past. With an immersive VR application, QC is much easier
and faster, in our experience.

Annotations and mesh colouring provide for visual analytics abilities
and these facilitate better distinction between various aspects of the
reconstruction. To give some example, capillary sheaths, surely
following arterioles, can be separated from capillary sheaths, the
origins or ends of which lie outside of the ROI. Such distinctions allow
for better understanding in microanatomical research.

\label{page:volren} Our experience emphasises the importance of VR-based
QC. Our older 3D reconstruction study \akm featured
\(3500 \times 3500 \times 21\) voxels in four regions. From each
reconstruction a further version was derived. They did not need to be
quality controlled again, but their inspection was crucial to produce a
diagnosis for medical research. We used both a non-VR visualisation tool
for QC and pre-rendered videos for inspection. It took a group of 3 to 5
experts multiple day-long meetings to QC these reconstructions with the
non-VR tool (Fig.~\ref{fig:volren}). Deducing the anatomical findings
from pre-rendered videos was also not easy for the domain experts.

\label{page:free-mov} We found free user movement essential for
long-term usability of our application---our users spend hours immersed
in consecutive sessions. Basically, the model is not otherwise
translated, rotated, or scaled in the productive use, but only in
response to tracking and reacting to user's own movements. Such free
user movement allows the immersed user to utilise their brain's systems
for spatial orientation and spatial memory. In their turn, the
recognition and annotation of structures become easier. Free user
movement also distinguishes our application from
\citet{egger_comprehensive_2020}: they used a VR flight mode on 3D
models from~CT.

We first found the benefits of VR-based visualisation during the
preparation of \citet{steiniger_capillary_2018}. Unlike the bone marrow
data set \citep{km16own}, in our next work
\citep{steiniger_locating_2018}, the total number of voxels was slightly
larger, and QC was much faster with the new method. Our domain expert
alone quality controlled with our VR-based method \emph{eleven} regions
with \(2000 \times 2000 \times 24\) voxels per day in one instance
\citep{steiniger_locating_2018} and two to four
\(\approx 2000\times 2000\times 84\) regions per day in another instance
\citep{steiniger_150}. These sum up to slightly more than \(10^9\)
voxels per day in the first case and up to \(1.36 \cdot 10^9\) voxels
per day in the second case. We would like to highlight, that these
amounts of data were routinely quality controlled by a single person in
a single day. Thus VR immersion saved an order of magnitude of man-hours
for QC of our medical research 3D reconstructions
\citep{steiniger_capillary_2018, steiniger_locating_2018, steiniger_150}.

Our immersive application also enabled VA of the same reconstructions.
Without immersive VA and (later on) interactive \enquote{cutting} into
the reconstructions with front plane clipping in~VR, it would be
exorbitantly harder or even impossible for us to obtain the research
results, summarised in Figs.~\ref{fig:sheaths}--\ref{fig:sep-150}
\citep{steiniger_locating_2018, steiniger_150}.

\hypertarget{conclusions}{%
\section{Conclusions}\label{conclusions}}

3D reconstructions from histological serial sections require quality
control (QC) and further investigations. Domain experts were not
satisfied by previously existing QC methods. We present a VR-based
solution to explore mesh data. Our application also allows to
superimpose the original serial sections. Such display is essential for
QC. In our experience, immersion accelerates QC by an order of
magnitude. Our users can annotate areas of interest and communicate the
annotations. VR-powered VA allowed for a more exact and fast distinction
and classification of various microanatomical entities, such as
post-arteriolar capillaries and other kinds of capillaries. The
classification of arterial blood vessels in its turn facilitated the
classification of capillary sheaths. Summarising, our VR tool greatly
enhances productivity and allows for more precise reconstructions that
enable new insights
\citep{steiniger_capillary_2018, steiniger_locating_2018, steiniger_150}
in microanatomical research.

\hypertarget{future-work}{%
\subsection{Future work}\label{future-work}}

Making our application an even a better visual analytics tool is always
viable. Minor improvements at user input handling include more input
combinations and gestures. A~planned feature is to record spoken
annotations for every annotation marker. Recorded memos would facilitate
better explanation of markings at their revision. The application has a
potential to evolve in the direction of a non-medical 3D sculpting
utility. A~better maintainability of the code base through an excessive
use of software product lines \citep{apel2016feature} is an important
goal. Not all builds need all features and software product lines can
accommodate this point.

Improvements of the rendering performance are both important and viable.
Possible points of interest are better occlusion culling
\citep[\egX ][]{mattausch_chc_2008, hasselgren_masked_2016} and
progressive meshes \citep[\egX ][]{derzapf_dependency-free_2012}. There
are further ways to improve the anti-aliasing and thus even further
improve the immersive user experience. A~possibility to consider is an
advanced interpolation for higher internal frame rates.

A~promising idea is to learn better view angles \citep[similar
to][]{Burns:2007:FEC:2384179.2384223} from the transformation matrices
saved as parts of annotations. Better pre-rendered videos might be
produced in this manner. \citep[ have a similar
motivation.]{gutierrez_toolbox_2018} Texture compression in general and
volume compression techniques in particular,
{[}\egX \citet{1183757};\citet{Guthe:2016:VLC:3056901.3056915};\citet{guarda_method_2017}\},
would help to reduce the GPU memory consumption caused by data for
original slices.

VR might be \emph{the} pivotal instrument for better understanding in
teaching complex 3D structures \citep{philippe_multimodal_2020}, \eg in
medicine or in machine engineering. An~effect of VR in training and
education in such professions \citep[and also in other areas,
\egX ][]{calvert_impact_2020} might need a more detailed assessment.

Of course, viable future work includes applications of our
visualisations to reconstructions of further organs and tissues
(\eg future bone marrow, lung, heart, or tonsil probes) and expansion to
further modalities of medical (such as MRI or CT) or non-medical data.
Recently, we experimented with VR presentation of serial block face
electron microscopy data. Multi-modality is an interesting topic, too
\citep{tang_perceptual_2020}. Possible examples of further applications
include materials science, computational fluid dynamics, and, most
surely, computer graphics.

\hypertarget{acknowledgements}{%
\subsection*{Acknowledgements}\label{acknowledgements}}
\addcontentsline{toc}{subsection}{Acknowledgements}

\ifthenelse{\boolean{anonymous}}{Omitted for review.}{Most of this work was done when the first two authors were members of University of Bayreuth. We would like to thank Vitus Stachniss, Verena Wilhelmi, and Christine Ulrich (Philipps-University of Marburg) for their efforts with non-VR QC tool. We thank Christian M\"uhlfeld (Hannover Medical School) for the possibility to test our application on a MacBook Pro \mbox{16\hspace{1pt}$''$.} Paul Schmiedel (then: University of Bayreuth) worked on the codebase of the VR tool in Bayreuth.  Figure~\ref{fig:new-user} depicts Lena Vo\ss, we would like to thank her for the permission to use this image. 
}

{
\bibliographystyle{elsarticle-harv}\biboptions{authoryear}
\linespread{1}\selectfont
\bibliography{flow.bib,km16.bib,own.bib,newmain.bib,vr.bib}

\providecommand{\noopsort}[1]{}
\begin{thebibliography}{82}
\expandafter\ifx\csname natexlab\endcsname\relax\def\natexlab#1{#1}\fi
\providecommand{\url}[1]{\texttt{#1}}
\providecommand{\href}[2]{#2}
\providecommand{\path}[1]{#1}
\providecommand{\DOIprefix}{doi:}
\providecommand{\ArXivprefix}{arXiv:}
\providecommand{\URLprefix}{URL: }
\providecommand{\Pubmedprefix}{pmid:}
\providecommand{\doi}[1]{\href{http://dx.doi.org/#1}{\path{#1}}}
\providecommand{\Pubmed}[1]{\href{pmid:#1}{\path{#1}}}
\providecommand{\bibinfo}[2]{#2}
\ifx\xfnm\relax \def\xfnm[#1]{\unskip,\space#1}\fi
\bibitem[{Ahrens et~al.(2005)Ahrens, Geveci and Law}]{ahrens_paraview:_2005}
\bibinfo{author}{Ahrens, J.}, \bibinfo{author}{Geveci, B.},
  \bibinfo{author}{Law, C.}, \bibinfo{year}{2005}.
\newblock \bibinfo{title}{{ParaView}: {An} end-user tool for large data
  visualization}.
\newblock \bibinfo{journal}{The visualization handbook} \bibinfo{volume}{717}.
\newblock \URLprefix
  \url{http://datascience.dsscale.org/wp-content/uploads/2016/06/ParaView.pdf}.
  \bibinfo{note}{{LA-UR-03-1560}}.
\bibitem[{Apel et~al.(2016)Apel, Batory, K{\"a}stner and
  Saake}]{apel2016feature}
\bibinfo{author}{Apel, S.}, \bibinfo{author}{Batory, D.},
  \bibinfo{author}{K{\"a}stner, C.}, \bibinfo{author}{Saake, G.},
  \bibinfo{year}{2016}.
\newblock \bibinfo{title}{Feature-oriented software product lines}.
\newblock \bibinfo{publisher}{Springer}.
\newblock \DOIprefix\doi{10.1007/978-3-642-37521-7}.
\bibitem[{Ayachit(2015)}]{ayachit_paraview_2015}
\bibinfo{author}{Ayachit, U.}, \bibinfo{year}{2015}.
\newblock \bibinfo{title}{The {ParaView} guide: a parallel visualization
  application}.
\newblock \bibinfo{publisher}{Kitware, Inc.}
\bibitem[{Berg and Vance(2017)}]{berg_industry_2017}
\bibinfo{author}{Berg, L.P.}, \bibinfo{author}{Vance, J.M.},
  \bibinfo{year}{2017}.
\newblock \bibinfo{title}{Industry use of virtual reality in product design and
  manufacturing: a survey}.
\newblock \bibinfo{journal}{Virtual Real.} \bibinfo{volume}{21},
  \bibinfo{pages}{1--17}.
\newblock \DOIprefix\doi{10.1007/s10055-016-0293-9}.
\bibitem[{Bouaoud et~al.(2020)Bouaoud, El~Beheiry, Jablon, Schouman, Bertolus,
  Picard, Masson and Khonsari}]{bouaoud_diva_2020}
\bibinfo{author}{Bouaoud, J.}, \bibinfo{author}{El~Beheiry, M.},
  \bibinfo{author}{Jablon, E.}, \bibinfo{author}{Schouman, T.},
  \bibinfo{author}{Bertolus, C.}, \bibinfo{author}{Picard, A.},
  \bibinfo{author}{Masson, J.B.}, \bibinfo{author}{Khonsari, R.H.},
  \bibinfo{year}{2020}.
\newblock \bibinfo{title}{{DIVA}, a {3D} virtual reality platform, improves
  undergraduate craniofacial trauma education}.
\newblock \bibinfo{journal}{Journal of Stomatology, Oral and Maxillofacial
  Surgery} \URLprefix
  \url{http://www.sciencedirect.com/science/article/pii/S2468785520302214},
  \DOIprefix\doi{10.1016/j.jormas.2020.09.009}.
\bibitem[{{Brooks Jr.}(1999)}]{799723}
\bibinfo{author}{{Brooks Jr.}, F.P.}, \bibinfo{year}{1999}.
\newblock \bibinfo{title}{What's real about virtual reality?}
\newblock \bibinfo{journal}{{IEEE} Comput. Graph.} \bibinfo{volume}{19},
  \bibinfo{pages}{16--27}.
\newblock \DOIprefix\doi{10.1109/38.799723}.
\bibitem[{Burns et~al.(2007)Burns, Haidacher, Wein, Viola and
  {Gr\"{o}ller}}]{Burns:2007:FEC:2384179.2384223}
\bibinfo{author}{Burns, M.}, \bibinfo{author}{Haidacher, M.},
  \bibinfo{author}{Wein, W.}, \bibinfo{author}{Viola, I.},
  \bibinfo{author}{{Gr\"{o}ller}, M.E.}, \bibinfo{year}{2007}.
\newblock \bibinfo{title}{Feature emphasis and contextual cutaways for
  multimodal medical visualization}, in: \bibinfo{booktitle}{Proceedings of the
  {9\xth} Joint {Eurographics} / {IEEE} {VGTC} Conference on Visualization},
  \bibinfo{publisher}{EG}. pp. \bibinfo{pages}{275--282}.
\newblock \DOIprefix\doi{10.2312/VisSym/EuroVis07/275-282}.
\bibitem[{Calvert and Abadia(2020)}]{calvert_impact_2020}
\bibinfo{author}{Calvert, J.}, \bibinfo{author}{Abadia, R.},
  \bibinfo{year}{2020}.
\newblock \bibinfo{title}{Impact of immersing university and high school
  students in educational linear narratives using virtual reality technology}.
\newblock \bibinfo{journal}{Computers \& Education} \bibinfo{volume}{159},
  \bibinfo{pages}{104005}.
\newblock \DOIprefix\doi{10.1016/j.compedu.2020.104005}.
\bibitem[{Calì et~al.(2019)Calì, Kare, Agus, Veloz~Castillo, Boges, Hadwiger
  and Magistretti}]{cali_method_2019}
\bibinfo{author}{Calì, C.}, \bibinfo{author}{Kare, K.}, \bibinfo{author}{Agus,
  M.}, \bibinfo{author}{Veloz~Castillo, M.F.}, \bibinfo{author}{Boges, D.},
  \bibinfo{author}{Hadwiger, M.}, \bibinfo{author}{Magistretti, P.},
  \bibinfo{year}{2019}.
\newblock \bibinfo{title}{A method for {3D} reconstruction and virtual reality
  analysis of glial and neuronal cells}.
\newblock \bibinfo{journal}{JoVE} \bibinfo{volume}{151},
  \bibinfo{pages}{e59444}.
\newblock \DOIprefix\doi{10.3791/59444}.
\bibitem[{Chan et~al.(2013)Chan, Conti, Salisbury and
  Blevins}]{doi:10.1227/NEU.0b013e3182750d26}
\bibinfo{author}{Chan, S.}, \bibinfo{author}{Conti, F.},
  \bibinfo{author}{Salisbury, K.}, \bibinfo{author}{Blevins, N.H.},
  \bibinfo{year}{2013}.
\newblock \bibinfo{title}{Virtual reality simulation in neurosurgery:
  Technologies and evolution}.
\newblock \bibinfo{journal}{Neurosurgery} \bibinfo{volume}{72},
  \bibinfo{pages}{A154--A164}.
\newblock \DOIprefix\doi{10.1227/NEU.0b013e3182750d26}.
\bibitem[{Checa and Bustillo(2020)}]{checa_review_2020}
\bibinfo{author}{Checa, D.}, \bibinfo{author}{Bustillo, A.},
  \bibinfo{year}{2020}.
\newblock \bibinfo{title}{A review of immersive virtual reality serious games
  to enhance learning and training}.
\newblock \bibinfo{journal}{Multimed. Tools Appl.} \bibinfo{volume}{79},
  \bibinfo{pages}{5501--5527}.
\newblock \DOIprefix\doi{10.1007/s11042-019-08348-9}.
\bibitem[{Chen et~al.(2015)Chen, Xu, Wang, Wang, Wang, Zeng, Wang and
  Egger}]{Chen2015124}
\bibinfo{author}{Chen, X.}, \bibinfo{author}{Xu, L.}, \bibinfo{author}{Wang,
  Y.}, \bibinfo{author}{Wang, H.}, \bibinfo{author}{Wang, F.},
  \bibinfo{author}{Zeng, X.}, \bibinfo{author}{Wang, Q.},
  \bibinfo{author}{Egger, J.}, \bibinfo{year}{2015}.
\newblock \bibinfo{title}{Development of a surgical navigation system based on
  augmented reality using an optical see-through head-mounted display}.
\newblock \bibinfo{journal}{J. Biomed. Inform.} \bibinfo{volume}{55},
  \bibinfo{pages}{124--131}.
\newblock \DOIprefix\doi{10.1016/j.jbi.2015.04.003}.
\bibitem[{Choi et~al.(2016)Choi, Chan, Leung and Chui}]{7814821}
\bibinfo{author}{Choi, K.S.}, \bibinfo{author}{Chan, S.T.},
  \bibinfo{author}{Leung, C.H.M.}, \bibinfo{author}{Chui, Y.P.},
  \bibinfo{year}{2016}.
\newblock \bibinfo{title}{Stereoscopic three-dimensional visualization for
  immersive and intuitive anatomy learning}, in: \bibinfo{booktitle}{IEEE 8\xth
  International Conference on Technology for Education},
  \bibinfo{publisher}{IEEE}. pp. \bibinfo{pages}{184--187}.
\newblock \DOIprefix\doi{10.1109/T4E.2016.046}.
\bibitem[{Cignoni et~al.(2008)Cignoni, Callieri, Corsini, Dellepiane, Ganovelli
  and Ranzuglia}]{km39}
\bibinfo{author}{Cignoni, P.}, \bibinfo{author}{Callieri, M.},
  \bibinfo{author}{Corsini, M.}, \bibinfo{author}{Dellepiane, M.},
  \bibinfo{author}{Ganovelli, F.}, \bibinfo{author}{Ranzuglia, G.},
  \bibinfo{year}{2008}.
\newblock \bibinfo{title}{Meshlab: an open-source mesh processing tool.}, in:
  \bibinfo{booktitle}{Eurographics Italian Chapter Conference}, pp.
  \bibinfo{pages}{129--136}.
\bibitem[{Daly(2018)}]{daly_future_2018}
\bibinfo{author}{Daly, C.J.}, \bibinfo{year}{2018}.
\newblock \bibinfo{title}{The future of education? {Using} {3D} animation and
  virtual reality in teaching physiology}.
\newblock \bibinfo{journal}{Physiology News} \bibinfo{volume}{111},
  \bibinfo{pages}{43}.
\newblock \URLprefix \url{http://www.physoc.org/magazine}.
\bibitem[{Daly(2019a)}]{daly_confocal_2019}
\bibinfo{author}{Daly, C.J.}, \bibinfo{year}{2019}a.
\newblock \bibinfo{title}{From confocal microscope to virtual reality and
  computer games; technical and educational considerations}.
\newblock \bibinfo{journal}{Infocus Magazine} \bibinfo{volume}{54},
  \bibinfo{pages}{51--59}.
\newblock \URLprefix \url{http://eprints.gla.ac.uk/188105/}.
\bibitem[{Daly(2019b)}]{daly_imaging_2019}
\bibinfo{author}{Daly, C.J.}, \bibinfo{year}{2019}b.
\newblock \bibinfo{title}{Imaging the vascular wall: {From} microscope to
  virtual reality}, in: \bibinfo{editor}{Touyz, R.M.}, \bibinfo{editor}{Delles,
  C.} (Eds.), \bibinfo{booktitle}{Textbook of {Vascular} {Medicine}}.
  \bibinfo{publisher}{Springer}, \bibinfo{address}{Cham}, pp.
  \bibinfo{pages}{59--66}.
\newblock \DOIprefix\doi{10.1007/978-3-030-16481-2_6}.
\bibitem[{van Dam et~al.(2000)van Dam, Forsberg, Laidlaw, LaViola and
  Simpson}]{888006}
\bibinfo{author}{van Dam, A.}, \bibinfo{author}{Forsberg, A.S.},
  \bibinfo{author}{Laidlaw, D.H.}, \bibinfo{author}{LaViola, J.J.},
  \bibinfo{author}{Simpson, R.M.}, \bibinfo{year}{2000}.
\newblock \bibinfo{title}{Immersive {VR} for scientific visualization: a
  progress report}.
\newblock \bibinfo{journal}{{IEEE} Comput. Graph.} \bibinfo{volume}{20},
  \bibinfo{pages}{26--52}.
\newblock \DOIprefix\doi{10.1109/38.888006}.
\bibitem[{Derzapf and Guthe(2012)}]{derzapf_dependency-free_2012}
\bibinfo{author}{Derzapf, E.}, \bibinfo{author}{Guthe, M.},
  \bibinfo{year}{2012}.
\newblock \bibinfo{title}{Dependency-free parallel progressive meshes}.
\newblock \bibinfo{journal}{Comput. Graph. Forum} \bibinfo{volume}{31},
  \bibinfo{pages}{2288--2302}.
\newblock \DOIprefix\doi{10.1111/j.1467-8659.2012.03154.x}.
\bibitem[{Dorweiler et~al.(2019)Dorweiler, Vahl and
  Ghazy}]{dorweiler_zukunftsperspektiven_2019}
\bibinfo{author}{Dorweiler, B.}, \bibinfo{author}{Vahl, C.F.},
  \bibinfo{author}{Ghazy, A.}, \bibinfo{year}{2019}.
\newblock \bibinfo{title}{Zukunftsperspektiven digitaler
  {Visualisierungstechnologien} in der {Gefäßchirurgie}}.
\newblock \bibinfo{journal}{Gefässchirurgie} \bibinfo{volume}{24},
  \bibinfo{pages}{531--538}.
\newblock \DOIprefix\doi{10.1007/s00772-019-00570-x}.
\bibitem[{Duarte et~al.(2020)Duarte, Santos, Guimarães~Júnior and
  Peccin}]{duarte_learning_2020}
\bibinfo{author}{Duarte, M.L.}, \bibinfo{author}{Santos, L.R.},
  \bibinfo{author}{Guimarães~Júnior, J.B.}, \bibinfo{author}{Peccin, M.S.},
  \bibinfo{year}{2020}.
\newblock \bibinfo{title}{Learning anatomy by virtual reality and augmented
  reality. {A} scope review}.
\newblock \bibinfo{journal}{Morphologie}
  \DOIprefix\doi{10.1016/j.morpho.2020.08.004}.
\bibitem[{Egger et~al.(2017)Egger, Gall, Wallner, Boechat, Hann, Li, Chen and
  Schmalstieg}]{10.1371/journal.pone.0173972}
\bibinfo{author}{Egger, J.}, \bibinfo{author}{Gall, M.},
  \bibinfo{author}{Wallner, J.}, \bibinfo{author}{Boechat, P.},
  \bibinfo{author}{Hann, A.}, \bibinfo{author}{Li, X.}, \bibinfo{author}{Chen,
  X.}, \bibinfo{author}{Schmalstieg, D.}, \bibinfo{year}{2017}.
\newblock \bibinfo{title}{{HTC} {Vive} {MeVisLab} integration via {OpenVR} for
  medical applications}.
\newblock \bibinfo{journal}{PLOS ONE} \bibinfo{volume}{12},
  \bibinfo{pages}{1--14}.
\newblock \DOIprefix\doi{10.1371/journal.pone.0173972}.
\bibitem[{Egger et~al.(2020)Egger, Gunacker, Pepe, Melito, Gsaxner, Li,
  Ellermann and Chen}]{egger_comprehensive_2020}
\bibinfo{author}{Egger, J.}, \bibinfo{author}{Gunacker, S.},
  \bibinfo{author}{Pepe, A.}, \bibinfo{author}{Melito, G.M.},
  \bibinfo{author}{Gsaxner, C.}, \bibinfo{author}{Li, J.},
  \bibinfo{author}{Ellermann, K.}, \bibinfo{author}{Chen, X.},
  \bibinfo{year}{2020}.
\newblock \bibinfo{title}{A comprehensive workflow and framework for immersive
  virtual endoscopy of dissected aortae from {CTA} data}, in:
  \bibinfo{editor}{Fei, B.}, \bibinfo{editor}{Linte, C.A.} (Eds.),
  \bibinfo{booktitle}{Medical {Imaging} 2020: {Image}-{Guided} {Procedures},
  {Robotic} {Interventions}, and {Modeling}}, \bibinfo{publisher}{SPIE}. pp.
  \bibinfo{pages}{774--779}.
\newblock \DOIprefix\doi{10.1117/12.2559239}.
\bibitem[{El~Beheiry et~al.(2019)El~Beheiry, Doutreligne, Caporal, Ostertag,
  Dahan and Masson}]{el_beheiry_virtual_2019}
\bibinfo{author}{El~Beheiry, M.}, \bibinfo{author}{Doutreligne, S.},
  \bibinfo{author}{Caporal, C.}, \bibinfo{author}{Ostertag, C.},
  \bibinfo{author}{Dahan, M.}, \bibinfo{author}{Masson, J.B.},
  \bibinfo{year}{2019}.
\newblock \bibinfo{title}{Virtual reality: {Beyond} visualization}.
\newblock \bibinfo{journal}{J. Mol. Biol.} \bibinfo{volume}{431},
  \bibinfo{pages}{1315--1321}.
\newblock \DOIprefix\doi{10.1016/j.jmb.2019.01.033}.
\bibitem[{Esfahlani et~al.(2018)Esfahlani, Thompson, Parsa, Brown and
  Cirstea}]{esfahlani_rehabgame:_2018}
\bibinfo{author}{Esfahlani, S.S.}, \bibinfo{author}{Thompson, T.},
  \bibinfo{author}{Parsa, A.D.}, \bibinfo{author}{Brown, I.},
  \bibinfo{author}{Cirstea, S.}, \bibinfo{year}{2018}.
\newblock \bibinfo{title}{{ReHabgame}: {A} non-immersive virtual reality
  rehabilitation system with applications in neuroscience}.
\newblock \bibinfo{journal}{Heliyon} \bibinfo{volume}{4},
  \bibinfo{pages}{e00526}.
\newblock \DOIprefix\doi{10.1016/j.heliyon.2018.e00526}.
\bibitem[{Faludi et~al.(2019)Faludi, Zoller, Gerig, Zam, Rauter and
  Cattin}]{faludi_direct_2019}
\bibinfo{author}{Faludi, B.}, \bibinfo{author}{Zoller, E.I.},
  \bibinfo{author}{Gerig, N.}, \bibinfo{author}{Zam, A.},
  \bibinfo{author}{Rauter, G.}, \bibinfo{author}{Cattin, P.C.},
  \bibinfo{year}{2019}.
\newblock \bibinfo{title}{Direct visual and haptic volume rendering of medical
  data sets for an immersive exploration in virtual reality}, in:
  \bibinfo{editor}{Shen, D.}, \bibinfo{editor}{Liu, T.},
  \bibinfo{editor}{Peters, T.M.}, \bibinfo{editor}{Staib, L.H.},
  \bibinfo{editor}{Essert, C.}, \bibinfo{editor}{Zhou, S.},
  \bibinfo{editor}{Yap, P.T.}, \bibinfo{editor}{Khan, A.} (Eds.),
  \bibinfo{booktitle}{Medical {Image} {Computing} and {Computer} {Assisted}
  {Intervention} – {MICCAI} 2019}, \bibinfo{publisher}{Springer},
  \bibinfo{address}{Cham}. pp. \bibinfo{pages}{29--37}.
\newblock \DOIprefix\doi{10.1007/978-3-030-32254-0_4}.
\bibitem[{Forsberg et~al.(2000)Forsberg, Laidlaw, van Dam, Kirby, Karniadakis
  and Elion}]{Forsberg:2000:IVR:375213.375297}
\bibinfo{author}{Forsberg, A.S.}, \bibinfo{author}{Laidlaw, D.H.},
  \bibinfo{author}{van Dam, A.}, \bibinfo{author}{Kirby, R.M.},
  \bibinfo{author}{Karniadakis, G.E.}, \bibinfo{author}{Elion, J.L.},
  \bibinfo{year}{2000}.
\newblock \bibinfo{title}{Immersive virtual reality for visualizing flow
  through an artery}, in: \bibinfo{booktitle}{Visualization},
  \bibinfo{publisher}{IEEE Computer Society Press}. pp.
  \bibinfo{pages}{457--460}.
\bibitem[{Guarda et~al.(2017)Guarda, Santos, da~Silva~Cruz, Assunção,
  Rodrigues and de~Faria}]{guarda_method_2017}
\bibinfo{author}{Guarda, A.F.R.}, \bibinfo{author}{Santos, J.M.},
  \bibinfo{author}{da~Silva~Cruz, L.A.}, \bibinfo{author}{Assunção, P.A.A.},
  \bibinfo{author}{Rodrigues, N.M.M.}, \bibinfo{author}{de~Faria, S.M.M.},
  \bibinfo{year}{2017}.
\newblock \bibinfo{title}{A method to improve {HEVC} lossless coding of
  volumetric medical images}.
\newblock \bibinfo{journal}{Signal Process.-Image} \bibinfo{volume}{59},
  \bibinfo{pages}{96--104}.
\newblock \DOIprefix\doi{10.1016/j.image.2017.02.002}.
\bibitem[{Guthe and Goesele(2016)}]{Guthe:2016:VLC:3056901.3056915}
\bibinfo{author}{Guthe, S.}, \bibinfo{author}{Goesele, M.},
  \bibinfo{year}{2016}.
\newblock \bibinfo{title}{Variable length coding for {GPU}-based direct volume
  rendering}, in: \bibinfo{booktitle}{Vision, Modeling \& Visualization},
  \bibinfo{publisher}{EG}. pp. \bibinfo{pages}{77--84}.
\newblock \DOIprefix\doi{10.2312/vmv.20161345}.
\bibitem[{Guthe et~al.(2002)Guthe, Wand, Gonser and Strasser}]{1183757}
\bibinfo{author}{Guthe, S.}, \bibinfo{author}{Wand, M.},
  \bibinfo{author}{Gonser, J.}, \bibinfo{author}{Strasser, W.},
  \bibinfo{year}{2002}.
\newblock \bibinfo{title}{Interactive rendering of large volume data sets}, in:
  \bibinfo{booktitle}{IEEE Visualization}, pp. \bibinfo{pages}{53--60}.
\newblock \DOIprefix\doi{10.1109/VISUAL.2002.1183757}.
\bibitem[{Gutiérrez et~al.(2018)Gutiérrez, David, Rai and
  Callet}]{gutierrez_toolbox_2018}
\bibinfo{author}{Gutiérrez, J.}, \bibinfo{author}{David, E.},
  \bibinfo{author}{Rai, Y.}, \bibinfo{author}{Callet, P.L.},
  \bibinfo{year}{2018}.
\newblock \bibinfo{title}{Toolbox and dataset for the development of saliency
  and scanpath models for {omnidirectional/\SI{360}{\degree}~still} images}.
\newblock \bibinfo{journal}{Signal Process.-Image} \bibinfo{volume}{69},
  \bibinfo{pages}{35--42}.
\newblock \DOIprefix\doi{10.1016/j.image.2018.05.003}.
\bibitem[{Hasselgren et~al.(2016)Hasselgren, Andersson and
  Akenine-Möller}]{hasselgren_masked_2016}
\bibinfo{author}{Hasselgren, J.}, \bibinfo{author}{Andersson, M.},
  \bibinfo{author}{Akenine-Möller, T.}, \bibinfo{year}{2016}.
\newblock \bibinfo{title}{Masked software occlusion culling}, in:
  \bibinfo{booktitle}{Proceedings of {High} {Performance} {Graphics}},
  \bibinfo{publisher}{{EG}}. pp. \bibinfo{pages}{23--31}.
\bibitem[{Inoue et~al.(2016)Inoue, Ikeda, Yatabe and
  Oikawa}]{doi:10.1121/2.0000381}
\bibinfo{author}{Inoue, A.}, \bibinfo{author}{Ikeda, Y.},
  \bibinfo{author}{Yatabe, K.}, \bibinfo{author}{Oikawa, Y.},
  \bibinfo{year}{2016}.
\newblock \bibinfo{title}{Three-dimensional sound-field visualization system
  using head mounted display and stereo camera}.
\newblock \bibinfo{journal}{Proceedings of Meetings on Acoustics}
  \bibinfo{volume}{29}, \bibinfo{pages}{025001--1--13}.
\newblock \DOIprefix\doi{10.1121/2.0000381}.
\bibitem[{Ju(2004)}]{km37}
\bibinfo{author}{Ju, T.}, \bibinfo{year}{2004}.
\newblock \bibinfo{title}{Robust repair of polygonal models}.
\newblock \bibinfo{journal}{ACM T.\ Graphic.} \bibinfo{volume}{23},
  \bibinfo{pages}{888--895}.
\bibitem[{Kikinis et~al.(2014)Kikinis, Pieper and Vosburgh}]{kikinis_3d_2014}
\bibinfo{author}{Kikinis, R.}, \bibinfo{author}{Pieper, S.D.},
  \bibinfo{author}{Vosburgh, K.G.}, \bibinfo{year}{2014}.
\newblock \bibinfo{title}{{3D} {Slicer}: {A} platform for subject-specific
  image analysis, visualization, and clinical support}, in:
  \bibinfo{editor}{Jolesz, F.A.} (Ed.), \bibinfo{booktitle}{Intraoperative
  {Imaging} and {Image}-{Guided} {Therapy}}, \bibinfo{publisher}{Springer}. pp.
  \bibinfo{pages}{277--289}.
\newblock \DOIprefix\doi{10.1007/978-1-4614-7657-3_19}.
\bibitem[{Knodel et~al.(2018)Knodel, Lemke, Lampe, Hoffer, Gillmann, Uder,
  Hillengaß, Wittum and Bäuerle}]{knodel_virtual_2018}
\bibinfo{author}{Knodel, M.M.}, \bibinfo{author}{Lemke, B.},
  \bibinfo{author}{Lampe, M.}, \bibinfo{author}{Hoffer, M.},
  \bibinfo{author}{Gillmann, C.}, \bibinfo{author}{Uder, M.},
  \bibinfo{author}{Hillengaß, J.}, \bibinfo{author}{Wittum, G.},
  \bibinfo{author}{Bäuerle, T.}, \bibinfo{year}{2018}.
\newblock \bibinfo{title}{Virtual reality in advanced medical immersive
  imaging: a workflow for introducing virtual reality as a supporting tool in
  medical imaging}.
\newblock \bibinfo{journal}{Comput. Visual Sci.} \bibinfo{volume}{18},
  \bibinfo{pages}{203--212}.
\newblock \DOIprefix\doi{10.1007/s00791-018-0292-3}.
\bibitem[{Liimatainen et~al.(2020)Liimatainen, Latonen, Valkonen, Kartasalo and
  Ruusuvuori}]{liimatainen_virtual_2020}
\bibinfo{author}{Liimatainen, K.}, \bibinfo{author}{Latonen, L.},
  \bibinfo{author}{Valkonen, M.}, \bibinfo{author}{Kartasalo, K.},
  \bibinfo{author}{Ruusuvuori, P.}, \bibinfo{year}{2020}.
\newblock \bibinfo{title}{Virtual reality for {3D} histology: multi-scale
  visualization of organs with interactive feature exploration}.
\newblock \bibinfo{journal}{arXiv:2003.11148 [cs, eess]} \URLprefix
  \url{http://arxiv.org/abs/2003.11148}. \bibinfo{note}{arXiv: 2003.11148}.
\bibitem[{Lobachev(2018)}]{habil}
\bibinfo{author}{Lobachev, O.}, \bibinfo{year}{2018}.
\newblock \bibinfo{title}{On Three-dimensional reconstruction}.
\newblock \bibinfo{type}{Habilitation thesis}. University of Bayreuth.
\bibitem[{Lobachev(2020)}]{lobachev_tempest_2020}
\bibinfo{author}{Lobachev, O.}, \bibinfo{year}{2020}.
\newblock \bibinfo{title}{The tempest in a cubic millimeter: {Image}-based
  refinements necessitate the reconstruction of {3D} microvasculature from a
  large series of damaged alternately-stained histological sections}.
\newblock \bibinfo{journal}{IEEE Access} \bibinfo{volume}{8},
  \bibinfo{pages}{13489--13506}.
\newblock \DOIprefix\doi{10.1109/ACCESS.2020.2965885}.
\bibitem[{Lobachev et~al.(2019)Lobachev, Pfeffer, Guthe and Steiniger}]{poster}
\bibinfo{author}{Lobachev, O.}, \bibinfo{author}{Pfeffer, H.},
  \bibinfo{author}{Guthe, M.}, \bibinfo{author}{Steiniger, B.S.},
  \bibinfo{year}{2019}.
\newblock \bibinfo{title}{Inspecting human {3D} histology in virtual reality.
  {Mesoscopic} models of the splenic red pulp microvasculature computed from
  immunostained serial sections}, in: \bibinfo{booktitle}{114\xth Annual
  Meeting}, \bibinfo{organization}{Anatomische Gesellschaft},
  \bibinfo{address}{Institut für Anatomie und Zellbiologie der Universität
  Würzburg}.
\newblock \bibinfo{note}{Poster}.
\bibitem[{Lobachev et~al.(2017a)Lobachev, Steiniger and Guthe}]{sccg17}
\bibinfo{author}{Lobachev, O.}, \bibinfo{author}{Steiniger, B.S.},
  \bibinfo{author}{Guthe, M.}, \bibinfo{year}{2017}a.
\newblock \bibinfo{title}{Compensating anisotropy in histological serial
  sections with optical flow-based interpolation}, in:
  \bibinfo{booktitle}{Proceedings of the 33\xrd Spring Conference on Computer
  Graphics}, \bibinfo{publisher}{{ACM}}. pp. \bibinfo{pages}{14:1--14:11}.
\newblock \DOIprefix\doi{10.1145/3154353.3154366}.
\bibitem[{Lobachev et~al.(2017b)Lobachev, Ulrich, Steiniger, Wilhelmi,
  Stachniss and Guthe}]{media16}
\bibinfo{author}{Lobachev, O.}, \bibinfo{author}{Ulrich, C.},
  \bibinfo{author}{Steiniger, B.S.}, \bibinfo{author}{Wilhelmi, V.},
  \bibinfo{author}{Stachniss, V.}, \bibinfo{author}{Guthe, M.},
  \bibinfo{year}{2017}b.
\newblock \bibinfo{title}{Feature-based multi-resolution registration of
  immunostained serial sections}.
\newblock \bibinfo{journal}{Med.\ Image Anal.} \bibinfo{volume}{35},
  \bibinfo{pages}{288--302}.
\newblock \DOIprefix\doi{10.1016/j.media.2016.07.010}.
\bibitem[{Lorensen and Cline(1987)}]{Lorensen1987}
\bibinfo{author}{Lorensen, W.E.}, \bibinfo{author}{Cline, H.E.},
  \bibinfo{year}{1987}.
\newblock \bibinfo{title}{Marching cubes: A high resolution {3D} surface
  construction algorithm}.
\newblock \bibinfo{journal}{ACM SIGGRAPH Computer Graphics}
  \bibinfo{volume}{21}, \bibinfo{pages}{163--169}.
\newblock \DOIprefix\doi{10.1145/37402.37422}.
\bibitem[{López~Chávez et~al.(2020)López~Chávez, Rodríguez and
  Gutierrez-Garcia}]{lopez_chavez_comparative_2020}
\bibinfo{author}{López~Chávez, O.}, \bibinfo{author}{Rodríguez, L.F.},
  \bibinfo{author}{Gutierrez-Garcia, J.O.}, \bibinfo{year}{2020}.
\newblock \bibinfo{title}{A comparative case study of {2D}, {3D} and
  immersive-virtual-reality applications for healthcare education}.
\newblock \bibinfo{journal}{International Journal of Medical Informatics}
  \bibinfo{volume}{141}, \bibinfo{pages}{104226}.
\newblock \DOIprefix\doi{10.1016/j.ijmedinf.2020.104226}.
\bibitem[{Mann et~al.(2018)Mann, Furness, Yuan, Iorio and Wang}]{mann2018all}
\bibinfo{author}{Mann, S.}, \bibinfo{author}{Furness, T.},
  \bibinfo{author}{Yuan, Y.}, \bibinfo{author}{Iorio, J.},
  \bibinfo{author}{Wang, Z.}, \bibinfo{year}{2018}.
\newblock \bibinfo{title}{All reality: Virtual, augmented, mixed (x), mediated
  (x, y), and multimediated reality}.
\newblock \href{http://arxiv.org/abs/1804.08386}{{\tt arXiv:1804.08386}}.
  \bibinfo{note}{arXiv preprint}.
\bibitem[{Mathur(2015)}]{7223437}
\bibinfo{author}{Mathur, A.S.}, \bibinfo{year}{2015}.
\newblock \bibinfo{title}{Low cost virtual reality for medical training}, in:
  \bibinfo{booktitle}{2015 IEEE Virtual Reality (VR)}, pp.
  \bibinfo{pages}{345--346}.
\newblock \DOIprefix\doi{10.1109/VR.2015.7223437}.
\bibitem[{Mattausch et~al.(2008)Mattausch, Bittner and
  Wimmer}]{mattausch_chc_2008}
\bibinfo{author}{Mattausch, O.}, \bibinfo{author}{Bittner, J.},
  \bibinfo{author}{Wimmer, M.}, \bibinfo{year}{2008}.
\newblock \bibinfo{title}{{CHC}++: {Coherent} hierarchical culling revisited}.
\newblock \bibinfo{journal}{Comput. Graph. Forum} \bibinfo{volume}{27},
  \bibinfo{pages}{221--230}.
\newblock \DOIprefix\doi{10.1111/j.1467-8659.2008.01119.x}.
\bibitem[{Misiak et~al.(2018)Misiak, Schreiber, Fuhrmann, Zur, Seider and
  Nafeie}]{misiak_islandviz:_2018}
\bibinfo{author}{Misiak, M.}, \bibinfo{author}{Schreiber, A.},
  \bibinfo{author}{Fuhrmann, A.}, \bibinfo{author}{Zur, S.},
  \bibinfo{author}{Seider, D.}, \bibinfo{author}{Nafeie, L.},
  \bibinfo{year}{2018}.
\newblock \bibinfo{title}{{IslandViz}: {A} tool for visualizing modular
  software systems in virtual reality}, in: \bibinfo{booktitle}{2018 {IEEE}
  {Working} {Conference} on {Software} {Visualization} ({VISSOFT})},
  \bibinfo{publisher}{IEEE}. pp. \bibinfo{pages}{112--116}.
\newblock \DOIprefix\doi{10.1109/VISSOFT.2018.00020}.
\bibitem[{Moro et~al.(2017)Moro, Štromberga, Raikos and
  Stirling}]{moro_effectiveness_2017}
\bibinfo{author}{Moro, C.}, \bibinfo{author}{Štromberga, Z.},
  \bibinfo{author}{Raikos, A.}, \bibinfo{author}{Stirling, A.},
  \bibinfo{year}{2017}.
\newblock \bibinfo{title}{The effectiveness of virtual and augmented reality in
  health sciences and medical anatomy}.
\newblock \bibinfo{journal}{Anat Sci Educ} \bibinfo{volume}{10},
  \bibinfo{pages}{549--559}.
\newblock \DOIprefix\doi{10.1002/ase.1696}.
\bibitem[{Philippe et~al.(2020)Philippe, Souchet, Lameras, Petridis, Caporal,
  Coldeboeuf and Duzan}]{philippe_multimodal_2020}
\bibinfo{author}{Philippe, S.}, \bibinfo{author}{Souchet, A.D.},
  \bibinfo{author}{Lameras, P.}, \bibinfo{author}{Petridis, P.},
  \bibinfo{author}{Caporal, J.}, \bibinfo{author}{Coldeboeuf, G.},
  \bibinfo{author}{Duzan, H.}, \bibinfo{year}{2020}.
\newblock \bibinfo{title}{Multimodal teaching, learning and training in virtual
  reality: a review and case study}.
\newblock \bibinfo{journal}{Virtual Reality \& Intelligent Hardware}
  \bibinfo{volume}{2}, \bibinfo{pages}{421--442}.
\newblock \DOIprefix\doi{10.1016/j.vrih.2020.07.008}.
\bibitem[{Pieper et~al.(2004)Pieper, Halle and Kikinis}]{pieper_3d_2004}
\bibinfo{author}{Pieper, S.}, \bibinfo{author}{Halle, M.},
  \bibinfo{author}{Kikinis, R.}, \bibinfo{year}{2004}.
\newblock \bibinfo{title}{{3D} {Slicer}}, in: \bibinfo{booktitle}{{IEEE}
  {International} {Symposium} on {Biomedical} {Imaging}: {Nano} to {Macro}},
  pp. \bibinfo{pages}{632--635}.
\newblock \DOIprefix\doi{10.1109/ISBI.2004.1398617}.
\bibitem[{Pieterse et~al.(2020)Pieterse, Hierck, de~Jong, Kroese, Willems and
  Reinders}]{pieterse_design_2020}
\bibinfo{author}{Pieterse, A.D.}, \bibinfo{author}{Hierck, B.P.},
  \bibinfo{author}{de~Jong, P.G.M.}, \bibinfo{author}{Kroese, J.},
  \bibinfo{author}{Willems, L.N.A.}, \bibinfo{author}{Reinders, M.E.J.},
  \bibinfo{year}{2020}.
\newblock \bibinfo{title}{Design and implementation of \enquote{{AugMedicine}:
  Lung Cases,} an augmented reality application for the medical curriculum on
  the presentation of dyspnea}.
\newblock \bibinfo{journal}{Front. Virtual Real.} \bibinfo{volume}{1}.
\newblock \DOIprefix\doi{10.3389/frvir.2020.577534}. \bibinfo{note}{publisher:
  Frontiers}.
\bibitem[{Preim and Saalfeld(2018)}]{preim_survey_2018}
\bibinfo{author}{Preim, B.}, \bibinfo{author}{Saalfeld, P.},
  \bibinfo{year}{2018}.
\newblock \bibinfo{title}{A survey of virtual human anatomy education systems}.
\newblock \bibinfo{journal}{Comput. Graph.} \bibinfo{volume}{71},
  \bibinfo{pages}{132--153}.
\newblock \DOIprefix\doi{10.1016/j.cag.2018.01.005}.
\bibitem[{Quam et~al.(2015)Quam, Gundert, Ellwein, Larkee, Hayden, Migrino,
  Otake and LaDisa}]{vr-fluid}
\bibinfo{author}{Quam, D.J.}, \bibinfo{author}{Gundert, T.J.},
  \bibinfo{author}{Ellwein, L.}, \bibinfo{author}{Larkee, C.E.},
  \bibinfo{author}{Hayden, P.}, \bibinfo{author}{Migrino, R.Q.},
  \bibinfo{author}{Otake, H.}, \bibinfo{author}{LaDisa, Jr., J.F.},
  \bibinfo{year}{2015}.
\newblock \bibinfo{title}{Immersive visualization for enhanced computational
  fluid dynamics analysis}.
\newblock \bibinfo{journal}{J. Biomech. Eng.-T. {ASME}} \bibinfo{volume}{137},
  \bibinfo{pages}{031004--031004--12}.
\newblock \DOIprefix\doi{10.1115/1.4029017}.
\bibitem[{Ritter et~al.(2011)Ritter, Boskamp, Homeyer, Laue, Schwier, Link and
  Peitgen}]{ritter_medical_2011}
\bibinfo{author}{Ritter, F.}, \bibinfo{author}{Boskamp, T.},
  \bibinfo{author}{Homeyer, A.}, \bibinfo{author}{Laue, H.},
  \bibinfo{author}{Schwier, M.}, \bibinfo{author}{Link, F.},
  \bibinfo{author}{Peitgen, H.O.}, \bibinfo{year}{2011}.
\newblock \bibinfo{title}{Medical image analysis}.
\newblock \bibinfo{journal}{IEEE Pulse} \bibinfo{volume}{2},
  \bibinfo{pages}{60--70}.
\newblock \DOIprefix\doi{10.1109/MPUL.2011.942929}.
\bibitem[{Rizvic et~al.(2019)Rizvic, Boskovic, Okanovic, Kihic and
  Sljivo}]{rizvic_virtual_2019}
\bibinfo{author}{Rizvic, S.}, \bibinfo{author}{Boskovic, D.},
  \bibinfo{author}{Okanovic, V.}, \bibinfo{author}{Kihic, I.I.},
  \bibinfo{author}{Sljivo, S.}, \bibinfo{year}{2019}.
\newblock \bibinfo{title}{Virtual reality experience of {Sarajevo} war
  heritage}, in: \bibinfo{editor}{Rizvic, S.},
  \bibinfo{editor}{Rodriguez~Echavarria, K.} (Eds.),
  \bibinfo{booktitle}{Eurographics {Workshop} on {Graphics} and {Cultural}
  {Heritage}}, \bibinfo{publisher}{EG}.
\newblock \DOIprefix\doi{10.2312/gch.20191340}.
\bibitem[{Scholl et~al.(2018)Scholl, Suder and Schiffer}]{scholl_direct_2018}
\bibinfo{author}{Scholl, I.}, \bibinfo{author}{Suder, S.},
  \bibinfo{author}{Schiffer, S.}, \bibinfo{year}{2018}.
\newblock \bibinfo{title}{Direct volume rendering in virtual reality}, in:
  \bibinfo{editor}{Maier, A.}, \bibinfo{editor}{Deserno, T.M.},
  \bibinfo{editor}{Handels, H.}, \bibinfo{editor}{Maier-Hein, K.H.},
  \bibinfo{editor}{Palm, C.}, \bibinfo{editor}{Tolxdorff, T.} (Eds.),
  \bibinfo{booktitle}{Bildverarbeitung für die {Medizin} 2018},
  \bibinfo{publisher}{Springer}, \bibinfo{address}{Berlin, Heidelberg}. pp.
  \bibinfo{pages}{297--302}.
\newblock \DOIprefix\doi{10.1007/978-3-662-56537-7_79}.
\bibitem[{Schweigger-Seidel(1862)}]{Schweigger-Seidel1862}
\bibinfo{author}{Schweigger-Seidel, F.}, \bibinfo{year}{1862}.
\newblock \bibinfo{title}{{Untersuchungen} {\"u}ber die {Milz}}.
\newblock \bibinfo{journal}{Arch. Pathol. Anat. Ph.} \bibinfo{volume}{23},
  \bibinfo{pages}{526--570}.
\newblock \DOIprefix\doi{10.1007/BF01939038}.
\bibitem[{Shen et~al.(2008)Shen, Boulanger and Noga}]{4618615}
\bibinfo{author}{Shen, R.}, \bibinfo{author}{Boulanger, P.},
  \bibinfo{author}{Noga, M.}, \bibinfo{year}{2008}.
\newblock \bibinfo{title}{{MedVis}: A real-time immersive visualization
  environment for the exploration of medical volumetric data}, in:
  \bibinfo{booktitle}{Information Visualization in Medical and Biomedical
  Informatics}, pp. \bibinfo{pages}{63--68}.
\newblock \DOIprefix\doi{10.1109/MediVis.2008.10}.
\bibitem[{Shimabukuro and Minghim(1998)}]{694195}
\bibinfo{author}{Shimabukuro, M.H.}, \bibinfo{author}{Minghim, R.},
  \bibinfo{year}{1998}.
\newblock \bibinfo{title}{Visualisation and reconstruction in dentistry}, in:
  \bibinfo{booktitle}{Information Visualization}, pp. \bibinfo{pages}{25--31}.
\newblock \DOIprefix\doi{10.1109/IV.1998.694195}.
\bibitem[{Silva et~al.(2009)Silva, Santos, Madeira and
  Silva}]{silva_processing_2009}
\bibinfo{author}{Silva, S.}, \bibinfo{author}{Santos, B.S.},
  \bibinfo{author}{Madeira, J.}, \bibinfo{author}{Silva, A.},
  \bibinfo{year}{2009}.
\newblock \bibinfo{title}{Processing, visualization and analysis of medical
  images of the heart: {An} example of fast prototyping using {MeVisLab}}, in:
  \bibinfo{booktitle}{2009 2\xnd {International} {Conference} in
  {Visualisation}}, pp. \bibinfo{pages}{165--170}.
\newblock \DOIprefix\doi{10.1109/VIZ.2009.40}.
\bibitem[{Slater et~al.(2020)Slater, Gonzalez-Liencres, Haggard, Vinkers,
  Gregory-Clarke, Jelley, Watson, Breen, Schwarz, Steptoe, Szostak, Halan, Fox
  and Silver}]{slater_ethics_2020}
\bibinfo{author}{Slater, M.}, \bibinfo{author}{Gonzalez-Liencres, C.},
  \bibinfo{author}{Haggard, P.}, \bibinfo{author}{Vinkers, C.},
  \bibinfo{author}{Gregory-Clarke, R.}, \bibinfo{author}{Jelley, S.},
  \bibinfo{author}{Watson, Z.}, \bibinfo{author}{Breen, G.},
  \bibinfo{author}{Schwarz, R.}, \bibinfo{author}{Steptoe, W.},
  \bibinfo{author}{Szostak, D.}, \bibinfo{author}{Halan, S.},
  \bibinfo{author}{Fox, D.}, \bibinfo{author}{Silver, J.},
  \bibinfo{year}{2020}.
\newblock \bibinfo{title}{The ethics of realism in virtual and augmented
  reality}.
\newblock \bibinfo{journal}{Front. virtual real.} \bibinfo{volume}{1},
  \bibinfo{pages}{1}.
\newblock \DOIprefix\doi{10.3389/frvir.2020.00001}.
\bibitem[{Stefani et~al.(2018)Stefani, Lacy-Hulbert and
  Skillman}]{stefani_confocalvr_2018}
\bibinfo{author}{Stefani, C.}, \bibinfo{author}{Lacy-Hulbert, A.},
  \bibinfo{author}{Skillman, T.}, \bibinfo{year}{2018}.
\newblock \bibinfo{title}{{ConfocalVR}: Immersive visualization for confocal
  microscopy}.
\newblock \bibinfo{journal}{J. Mol. Biol.} \bibinfo{volume}{430},
  \bibinfo{pages}{4028--4035}.
\newblock \DOIprefix\doi{10.1016/j.jmb.2018.06.035}.
\bibitem[{Steiniger et~al.(2007)Steiniger, Stachniss, Schwarzbach and
  Barth}]{steiniger_phenotypic_2007}
\bibinfo{author}{Steiniger, B.}, \bibinfo{author}{Stachniss, V.},
  \bibinfo{author}{Schwarzbach, H.}, \bibinfo{author}{Barth, P.J.},
  \bibinfo{year}{2007}.
\newblock \bibinfo{title}{Phenotypic differences between red pulp capillary and
  sinusoidal endothelia help localizing the open splenic circulation in
  humans}.
\newblock \bibinfo{journal}{Histochem. Cell. Biol.} \bibinfo{volume}{128},
  \bibinfo{pages}{391--398}.
\newblock \DOIprefix\doi{10.1007/s00418-007-0320-8}.
\bibitem[{Steiniger et~al.(2020)Steiniger, Pfeffer, Guthe and
  Lobachev}]{steiniger_150}
\bibinfo{author}{Steiniger, B.S.}, \bibinfo{author}{Pfeffer, H.},
  \bibinfo{author}{Guthe, M.}, \bibinfo{author}{Lobachev, O.},
  \bibinfo{year}{2020}.
\newblock \bibinfo{title}{Exploring human splenic red pulp vasculature in
  virtual reality. {Details} of sheathed capillaries and the open capillary
  network}.
\newblock \bibinfo{journal}{Histochem. Cell Biol.} \URLprefix
  \url{https://rdcu.be/b8KgZ}, \DOIprefix\doi{10.1007/s00418-020-01924-3}.
\bibitem[{Steiniger et~al.(2016)Steiniger, Stachniss, Wilhelmi, Seiler, Lampp,
  Neff, Guthe and Lobachev}]{km16own}
\bibinfo{author}{Steiniger, B.S.}, \bibinfo{author}{Stachniss, V.},
  \bibinfo{author}{Wilhelmi, V.}, \bibinfo{author}{Seiler, A.},
  \bibinfo{author}{Lampp, K.}, \bibinfo{author}{Neff, A.},
  \bibinfo{author}{Guthe, M.}, \bibinfo{author}{Lobachev, O.},
  \bibinfo{year}{2016}.
\newblock \bibinfo{title}{Three-dimensional arrangement of human bone marrow
  microvessels revealed by immunohistology in undecalcified sections}.
\newblock \bibinfo{journal}{PLOS ONE} \bibinfo{volume}{11},
  \bibinfo{pages}{1--25}.
\newblock \DOIprefix\doi{10.1371/journal.pone.0168173}.
\bibitem[{Steiniger et~al.(2018a)Steiniger, Ulrich, Berthold, Guthe and
  Lobachev}]{steiniger_capillary_2018}
\bibinfo{author}{Steiniger, B.S.}, \bibinfo{author}{Ulrich, C.},
  \bibinfo{author}{Berthold, M.}, \bibinfo{author}{Guthe, M.},
  \bibinfo{author}{Lobachev, O.}, \bibinfo{year}{2018}a.
\newblock \bibinfo{title}{Capillary networks and follicular marginal zones in
  human spleens. {Three}-dimensional models based on immunostained serial
  sections}.
\newblock \bibinfo{journal}{PLOS ONE} \bibinfo{volume}{13},
  \bibinfo{pages}{1--21}.
\newblock \DOIprefix\doi{10.1371/journal.pone.0191019}.
\bibitem[{Steiniger et~al.(2018b)Steiniger, Wilhelmi, Berthold, Guthe and
  Lobachev}]{steiniger_locating_2018}
\bibinfo{author}{Steiniger, B.S.}, \bibinfo{author}{Wilhelmi, V.},
  \bibinfo{author}{Berthold, M.}, \bibinfo{author}{Guthe, M.},
  \bibinfo{author}{Lobachev, O.}, \bibinfo{year}{2018}b.
\newblock \bibinfo{title}{Locating human splenic capillary sheaths in virtual
  reality}.
\newblock \bibinfo{journal}{Sci. Rep.} \bibinfo{volume}{8},
  \bibinfo{pages}{15720}.
\newblock \DOIprefix\doi{10.1038/s41598-018-34105-3}.
\bibitem[{Stets et~al.(2017)Stets, Sun, Corning and
  Greenwald}]{stets_visualization_2017}
\bibinfo{author}{Stets, J.D.}, \bibinfo{author}{Sun, Y.},
  \bibinfo{author}{Corning, W.}, \bibinfo{author}{Greenwald, S.W.},
  \bibinfo{year}{2017}.
\newblock \bibinfo{title}{Visualization and labeling of point clouds in virtual
  reality}, in: \bibinfo{booktitle}{{SIGGRAPH} {Asia} 2017 {Posters} --- {SA}
  '17}, \bibinfo{publisher}{{ACM}}, \bibinfo{address}{Bangkok, Thailand}. pp.
  \bibinfo{pages}{1--2}.
\newblock \DOIprefix\doi{10.1145/3145690.3145729}.
\bibitem[{Stone et~al.(2010)Stone, Kohlmeyer, Vandivort and
  Schulten}]{Stone2010}
\bibinfo{author}{Stone, J.E.}, \bibinfo{author}{Kohlmeyer, A.},
  \bibinfo{author}{Vandivort, K.L.}, \bibinfo{author}{Schulten, K.},
  \bibinfo{year}{2010}.
\newblock \bibinfo{title}{Immersive molecular visualization and interactive
  modeling with commodity hardware}, in: \bibinfo{editor}{Bebis, G.},
  \bibinfo{editor}{Boyle, R.}, \bibinfo{editor}{Parvin, B.},
  \bibinfo{editor}{Koracin, D.}, \bibinfo{editor}{Chung, R.},
  \bibinfo{editor}{Hammound, R.}, \bibinfo{editor}{Hussain, M.},
  \bibinfo{editor}{Kar-Han, T.}, \bibinfo{editor}{Crawfis, R.},
  \bibinfo{editor}{Thalmann, D.}, \bibinfo{editor}{Kao, D.},
  \bibinfo{editor}{Avila, L.} (Eds.), \bibinfo{booktitle}{Advances in Visual
  Computing}, \bibinfo{publisher}{Springer}. pp. \bibinfo{pages}{382--393}.
\newblock \DOIprefix\doi{10.1007/978-3-642-17274-8_38}.
\bibitem[{Sutherland(1965)}]{sutherland_ultimate_1965}
\bibinfo{author}{Sutherland, I.E.}, \bibinfo{year}{1965}.
\newblock \bibinfo{title}{The ultimate display}, in:
  \bibinfo{booktitle}{Proceedings of {IFIP} {Congress},}, pp.
  \bibinfo{pages}{506--508}.
\bibitem[{Sutherland(1968)}]{sutherland_head-mounted_1968}
\bibinfo{author}{Sutherland, I.E.}, \bibinfo{year}{1968}.
\newblock \bibinfo{title}{A head-mounted three dimensional display}, in:
  \bibinfo{booktitle}{Proceedings of the {December} 9-11, 1968, fall joint
  computer conference, part {I}}, \bibinfo{publisher}{{ACM}}. pp.
  \bibinfo{pages}{757--764}.
\newblock \DOIprefix\doi{10.1145/1476589.1476686}.
\bibitem[{Sutherland(1970)}]{sutherland_computer_1970}
\bibinfo{author}{Sutherland, I.E.}, \bibinfo{year}{1970}.
\newblock \bibinfo{title}{Computer displays}.
\newblock \bibinfo{journal}{Sci. Am.} \bibinfo{volume}{222},
  \bibinfo{pages}{56--81}.
\newblock \URLprefix \url{https://www.jstor.org/stable/24925827}.
\bibitem[{Tang et~al.(2020)Tang, Tian, Li, Hu, Yu and
  Xu}]{tang_perceptual_2020}
\bibinfo{author}{Tang, L.}, \bibinfo{author}{Tian, C.}, \bibinfo{author}{Li,
  L.}, \bibinfo{author}{Hu, B.}, \bibinfo{author}{Yu, W.}, \bibinfo{author}{Xu,
  K.}, \bibinfo{year}{2020}.
\newblock \bibinfo{title}{Perceptual quality assessment for multimodal medical
  image fusion}.
\newblock \bibinfo{journal}{Signal Process.-Image} \bibinfo{volume}{85},
  \bibinfo{pages}{115852}.
\newblock \DOIprefix\doi{10.1016/j.image.2020.115852}.
\bibitem[{Tomikawa et~al.(2010)Tomikawa, Hong, Shiotani, Tokunaga, Konishi,
  Ieiri, Tanoue, Akahoshi, Maehara and Hashizume}]{tomikawa_real-time_2010}
\bibinfo{author}{Tomikawa, M.}, \bibinfo{author}{Hong, J.},
  \bibinfo{author}{Shiotani, S.}, \bibinfo{author}{Tokunaga, E.},
  \bibinfo{author}{Konishi, K.}, \bibinfo{author}{Ieiri, S.},
  \bibinfo{author}{Tanoue, K.}, \bibinfo{author}{Akahoshi, T.},
  \bibinfo{author}{Maehara, Y.}, \bibinfo{author}{Hashizume, M.},
  \bibinfo{year}{2010}.
\newblock \bibinfo{title}{Real-time 3-dimensional virtual reality navigation
  system with open {MRI} for breast-conserving surgery}.
\newblock \bibinfo{journal}{J. Am. Coll. Surgeons} \bibinfo{volume}{210},
  \bibinfo{pages}{927--933}.
\newblock \DOIprefix\doi{10.1016/j.jamcollsurg.2010.01.032}.
\bibitem[{Ulrich et~al.(2014)Ulrich, Grund, Derzapf, Lobachev and Guthe}]{mc14}
\bibinfo{author}{Ulrich, C.}, \bibinfo{author}{Grund, N.},
  \bibinfo{author}{Derzapf, E.}, \bibinfo{author}{Lobachev, O.},
  \bibinfo{author}{Guthe, M.}, \bibinfo{year}{2014}.
\newblock \bibinfo{title}{Parallel iso-surface extraction and simplification},
  in: \bibinfo{editor}{Skala, V.} (Ed.), \bibinfo{booktitle}{WSCG
  Communications proceedings}.
\newblock \URLprefix \url{http://hdl.handle.net/11025/26435}.
\bibitem[{Uppot et~al.(2019)Uppot, Laguna, McCarthy, De~Novi, Phelps, Siegel
  and Courtier}]{uppot_implementing_2019}
\bibinfo{author}{Uppot, R.N.}, \bibinfo{author}{Laguna, B.},
  \bibinfo{author}{McCarthy, C.J.}, \bibinfo{author}{De~Novi, G.},
  \bibinfo{author}{Phelps, A.}, \bibinfo{author}{Siegel, E.},
  \bibinfo{author}{Courtier, J.}, \bibinfo{year}{2019}.
\newblock \bibinfo{title}{Implementing virtual and augmented reality tools for
  radiology education and training, communication, and clinical care}.
\newblock \bibinfo{journal}{Radiology} \bibinfo{volume}{291},
  \bibinfo{pages}{570--580}.
\newblock \DOIprefix\doi{10.1148/radiol.2019182210}.
\bibitem[{Uruthiralingam and Rea(2020)}]{rea_augmented_2020}
\bibinfo{author}{Uruthiralingam, U.}, \bibinfo{author}{Rea, P.M.},
  \bibinfo{year}{2020}.
\newblock \bibinfo{title}{Augmented and virtual reality in anatomical education
  – a systematic review}, in: \bibinfo{editor}{Rea, P.M.} (Ed.),
  \bibinfo{booktitle}{Biomedical {Visualisation}}.
  \bibinfo{publisher}{Springer}, \bibinfo{address}{Cham}. volume
  \bibinfo{volume}{1235} of \textit{\bibinfo{series}{Advances in Experimental
  Medicine and Biology}}, pp. \bibinfo{pages}{89--101}.
\newblock \DOIprefix\doi{10.1007/978-3-030-37639-0_5}.
\bibitem[{Walsh et~al.(2012)Walsh, Sherlock, Ling and
  Carnahan}]{walsh_virtual_2012}
\bibinfo{author}{Walsh, C.M.}, \bibinfo{author}{Sherlock, M.E.},
  \bibinfo{author}{Ling, S.C.}, \bibinfo{author}{Carnahan, H.},
  \bibinfo{year}{2012}.
\newblock \bibinfo{title}{Virtual reality simulation training for health
  professions trainees in gastrointestinal endoscopy}.
\newblock \bibinfo{journal}{Cochrane Database of Systematic Reviews} \URLprefix
  \url{http://doi.wiley.com/10.1002/14651858.CD008237.pub2},
  \DOIprefix\doi{10.1002/14651858.CD008237.pub2}.
\bibitem[{Wißmann et~al.(2020)Wißmann, Mišiak, Fuhrmann and
  Latoschik}]{wismann_accelerated_2020}
\bibinfo{author}{Wißmann, N.}, \bibinfo{author}{Mišiak, M.},
  \bibinfo{author}{Fuhrmann, A.}, \bibinfo{author}{Latoschik, M.E.},
  \bibinfo{year}{2020}.
\newblock \bibinfo{title}{Accelerated stereo rendering with hybrid
  reprojection-based rasterization and adaptive ray-tracing}, in:
  \bibinfo{booktitle}{2020 {IEEE} {Conference} on {Virtual} {Reality} and {3D}
  {User} {Interfaces} ({VR})}, pp. \bibinfo{pages}{828--835}.
\newblock \DOIprefix\doi{10.1109/VR46266.2020.00107}. \bibinfo{note}{iSSN:
  2642-5254}.
\bibitem[{Xia et~al.(2013)Xia, Lopes and Restivo}]{xia_virtual_2013}
\bibinfo{author}{Xia, P.}, \bibinfo{author}{Lopes, A.M.},
  \bibinfo{author}{Restivo, M.T.}, \bibinfo{year}{2013}.
\newblock \bibinfo{title}{Virtual reality and haptics for dental surgery: a
  personal review}.
\newblock \bibinfo{journal}{Vis Comput} \bibinfo{volume}{29},
  \bibinfo{pages}{433--447}.
\newblock \DOIprefix\doi{10.1007/s00371-012-0748-2}.
\bibitem[{Zoller et~al.(2020)Zoller, Faludi, Gerig, Jost, Cattin and
  Rauter}]{zoller_force_2020}
\bibinfo{author}{Zoller, E.I.}, \bibinfo{author}{Faludi, B.},
  \bibinfo{author}{Gerig, N.}, \bibinfo{author}{Jost, G.F.},
  \bibinfo{author}{Cattin, P.C.}, \bibinfo{author}{Rauter, G.},
  \bibinfo{year}{2020}.
\newblock \bibinfo{title}{Force quantification and simulation of pedicle screw
  tract palpation using direct visuo-haptic volume rendering}.
\newblock \bibinfo{journal}{Int. J. Comput. Ass. Rad.}
  \DOIprefix\doi{10.1007/s11548-020-02258-0}.

\end{thebibliography}
}

\appendix

\hypertarget{short-glossary-of-medical-terms}{%
\section{Short glossary of medical
terms}\label{short-glossary-of-medical-terms}}

\begin{itemize}
\tightlist
\item
  \emph{Histology}: the science studying biological tissues at
  microscopic level.
\item
  \emph{Histological section}: a thin slice of a tissue that can be
  inspected under a microscope. The sectioning happens on a device
  called \enquote{microtome}. Sections should be \emph{stained}
  (coloured) for better results.
\item
  \emph{Serial sections}: a consecutive series of histological sections.
  After registration the series forms volume data.
\item
  \emph{Vasculature}: a network of blood vessels.
\item
  \emph{Capillary}: the smallest type of blood vessels.
\item
  \emph{Sinus}: a capillary with large diameter and specialised wall
  structure in specific organs, such as spleen or bone marrow. Depending
  on the organ, sinuses differ in wall structure and function. Sinuses
  only occur in bone marrow regions where blood cells are generated
  (\enquote{hematopoietic regions}). Sinuses are ubiquitous in the
  spleen. Whether (venous) spleen sinuses are connected to capillaries
  of the arterial side, is an open question.
\item
  \emph{Follicle}: generally, a round structure. In spleen and tonsils,
  for example, follicles are important dynamic structures where specific
  white blood cells often appear. This makes follicles important for
  understanding immune system functions.
\item
  \emph{Spleen}: one of the organs which essentially function as a blood
  filter. The spleen has a unique vasculature, where blood flow also
  occurs outside blood vessels. It features some unique structures both
  with respect to the vasculature and to the arrangement of two major
  types of migratory white blood cells called lymphocytes, which either
  occupy follicles or T-cell zones.
\item
  \emph{Bone marrow}: the inside of bones not only contains fatty
  tissue, but also regions responsible for generation of new blood
  cells. Bone marrow features a unique vasculature with capillaries and
  sinuses.
\item
  \emph{Staining}: for better visual inspection, specific parts of the
  tissue sections can be coloured.
\item
  \emph{Immunohistology}: specific detection of different molecules in
  tissue sections using antibody solutions. Binding of an antibody to a
  tissue component is finally visualised by deposition of a coloured
  insoluble polymerisation product of a previously uncoloured soluble
  stain. The stainings can, for example, detect membrane glycoproteins
  in the innermost cells of blood vessels, so-termed endothelial cells.
  Membrane glycoproteins have been numbered in the order of their
  discovery using the CD (cluster of differentiation) nomenclature.
\item
  \emph{MRI}, magnetic resonance imaging, and \emph{CT}, computed
  tomography from a series of X-ray images are non-invasive imaging
  techniques that revolutionised the diagnostics. Unfortunately, the
  spatial resolution and selectivity of these techniques are not enough
  for our goals.
\item
  \emph{Anti-CD34 staining}: primarily stains endothelial cells of
  arterial vessels and capillaries in human spleen and bone marrow.
  Typically used colour is brown. Some stem cells in bone marrow are
  also stained. It is also weakly present in sinus endothelia in the
  proximity of follicles.
\item
  \emph{Anti-CD141 staining}: stains sinus endothelial cells in human
  bone marrow and in human spleen. Typically used colour is brown.
\item
  \emph{Anti-SMA staining}: stains smooth muscle alpha-actin. It is
  present, \eg  in walls of larger blood vessels on the \enquote{input}
  arterial side, the so-called \enquote{arterioles.} Typically used
  staining colour is brown.
\item
  \emph{Anti-CD271 staining}: stains capillary sheath cells and
  additional fibroblast-like cells in human spleen. The sheaths are
  multi-cellular structures around the initial segment of human splenic
  capillaries. Sheath cells obviously represent the sessile
  fibroblast-derived part of capillary sheaths. Typically used staining
  colour is blue or red. Specialised fibroblasts inside the follicles
  are more weakly stained with this antibody.
\item
  \emph{Anti-CD20 staining}: stains B-lymphocytes. Typically used colour
  is red.
\end{itemize}

\section{Supplementary Material}
The supplementary video for this paper is available under \url{https://zenodo.org/record/4268535}. It shows the details of our visualisations in dynamics. The video also showcases various approaches towards visual analytics. All imagery in the video is a real-time feed from VR headset.

The datasets analyzed in this study can be found in the Zenodo repositories
\url{https://zenodo.org/record/1039241}, \url{https://zenodo.org/record/1229434}, \url{https://zenodo.org/record/4059595}.

\end{document}